\documentclass[aps,twocolumn,prd,reprint,nofootinbib,floatfix]{revtex4-1}
\RequirePackage[l2tabu, orthodox]{nag}

\usepackage[utf8x]{inputenc}
\usepackage{microtype} 

\usepackage{amsmath}
\usepackage{amstext, amssymb, amsthm, amsfonts, slashed}
\usepackage{physics}
\usepackage{mathtools}
\usepackage{mhchem}
\usepackage{graphicx}
\usepackage{booktabs}
\usepackage{dcolumn}% Align table columns on decimal point
\usepackage{bm}% bold math
\usepackage{dsfont}
\usepackage[usenames,dvipsnames]{xcolor}
\usepackage[normalem]{ulem}
\usepackage{footmisc}

\usepackage{url}
\usepackage[colorlinks=true,urlcolor=blue,linkcolor=blue,citecolor=blue,hypertexnames]{hyperref}
\usepackage[nameinlink]{cleveref}
\usepackage{braket}
\usepackage{simplewick}
\usepackage{float}

\hypersetup{
	colorlinks,
	linkcolor={blue!75!black},
	citecolor={blue!75!black},
	urlcolor={blue!75!black}
}

\newcommand{\pderiv}[2]{\frac{\partial #1}{\partial #2}}
\newcommand{\dderiv}[2]{\frac{\dd #1}{\dd #2}}

\newcommand{\nf}{N_f}
\newcommand{\nc}{N_c}

\crefname{section}{Sec.\!}{Secs.\!}
\crefname{figure}{Fig.\!}{Figs.\!}
\crefname{equation}{}{}
\crefname{table}{Tab.\!}{Tabs.\!}
\crefname{appendix}{App.\!}{Apps.\!}

\def\s0#1#2{\mbox{\small{$ \frac{#1}{#2} $}}}
\def\0#1#2{\frac{#1}{#2}}

\usepackage{xcolor,colortbl}

\definecolor{Gray}{gray}{0.85}
\definecolor{LightGreen}{rgb}{0.88, 1, 0.88}
\definecolor{Blue}{rgb}{0,1,1}
\definecolor{Lime}{rgb}{0,1,0}
\definecolor{LightCyan}{rgb}{0.88,1,1}
\definecolor{LightRed}{rgb}{1, 0.85, 0.85}
\definecolor{Red}{rgb}{1, 0, 0}
\definecolor{LightYellow}{rgb}{1, 1, 0.85}
\definecolor{Yellow}{rgb}{1,1,0.05}
\definecolor{LightBlue}{rgb}{0.87, 0.94, 1}
\definecolor{white}{gray}{1}
\definecolor{black}{gray}{0}

\definecolor{LightGray}{gray}{0.93}

\newcolumntype{?}{!{\vrule width 1pt}}
\newcolumntype{`}{!{\vrule width 1.5pt}}

\AtBeginDocument{
\heavyrulewidth=.16em
\lightrulewidth=.1em
\cmidrulewidth=.03em
\belowrulesep=.4ex
\belowbottomsep=0pt
\aboverulesep=.4ex
\abovetopsep=0pt
\cmidrulesep=\doublerulesep
\cmidrulekern=.5em
\defaultaddspace=.5em
}

\graphicspath{{figures/}, {}}
\makeatletter
    \def\CT@@do@color{%
      \global\let\CT@do@color\relax
            \@tempdima\wd\z@
            \advance\@tempdima\@tempdimb
            \advance\@tempdima\@tempdimc
    \advance\@tempdimb\tabcolsep
    \advance\@tempdimc\tabcolsep
    \advance\@tempdima2\tabcolsep
            \kern-\@tempdimb
            \leaders\vrule
                    \hskip\@tempdima\@plus  1fill
            \kern-\@tempdimc
            \hskip-\wd\z@ \@plus -1fill }
    \makeatother

\begin{document}

%%%%%%%%%%%%%%%%%%%%%%%%%%%%%%%%%%%%%%%%%%%
% opening
%%%%%%%%%%%%%%%%%%%%%%%%%%%%%%%%%%%%%%%%%%%
\title{Spectral Functions of Gauge Theories with Banks-Zaks Fixed Points}

\author{Yannick~Kluth}
\affiliation{Department  of  Physics  and  Astronomy,  University  of  Sussex,  Brighton,  BN1  9QH,  U.K.}
\author{Daniel~F.~Litim}
\affiliation{Department  of  Physics  and  Astronomy,  University  of  Sussex,  Brighton,  BN1  9QH,  U.K.}
\author{Manuel~Reichert}
\affiliation{Department  of  Physics  and  Astronomy,  University  of  Sussex,  Brighton,  BN1  9QH,  U.K.}

\begin{abstract}
We investigate spectral functions of matter-gauge theories that are asymptotically free in the ultraviolet and display a Banks-Zaks conformal fixed point in the infrared. Using perturbation theory, Callan-Symanzik resummations, and UV-IR connecting renormalisation group trajectories,  we analytically determine the gluon, quark, and ghost propagators in the entire complex momentum plane. At weak coupling, we find that a  K\"all\'en-Lehmann spectral representation of propagators is achieved for all fields, and determine suitable ranges for gauge-fixing parameters. 
 At strong coupling, a proliferation of complex conjugated branch cuts renders a causal representation impossible. We also derive relations for scaling exponents that determine the presence or absence of propagator non-analyticities. Further results include spectral functions for all fields up to five loop order, bounds on the conformal window, and an algorithm to find running gauge coupling analytically at higher loops. Implications of our findings and extensions to other theories are discussed.
\end{abstract}

\maketitle

\tableofcontents

%%%%%%%%%%%%%%%%%%%%%%%%%%%%
\section{Introduction}
Connected two-point correlation functions are key objects in quantum field theory. They describe the propagation of particles and encode important physical information such as the particle masses, bound states, or decay widths. Gaining access to field propagators for timelike and spacelike momenta is crucial for the understanding of spectra and the unitarity in any given quantum field theory. The latter is often accessed via the K\"all\'en-Lehmann (KL) spectral representation \cite{Kallen:1952zz, Lehmann:1954xi}. For stable physical particles, the KL representation is a positive-definite and normalisable function, which can be understood as a probability density for the transition to an excited state with a given energy. It then becomes an important task to clarify how viable spectral functions arise within renormalisable quantum field theories. 

On a different tack, it is widely appreciated that ultraviolet (UV) fixed points such as in asymptotic freedom \cite{Gross:1973id, Politzer:1973fx} or asymptotic safety  \cite{Litim:2014uca, Bond:2016dvk, Bond:2022xvr} are mandatory for particle theories to stay well-defined and predictive up to highest energies. UV-complete theories then arise from small perturbations in the vicinity of UV fixed points. The latter trigger the renormalisation group (RG) flow of couplings which, in principle, encode all physical information. In four  dimensions, all known UV-complete theories  involve non-abelian gauge fields \cite{Coleman:1973sx, Bond:2018oco}, and, possibly, gravity \cite{Weinberg:1980gg}. Hence, we are facing the intriguing conundrum that spectral information related to the causality and unitarity of theories is encoded in the correlation functions of primarily gauge-variant quantities, e.g.~\cite{Gardi:1998ch}. 

An interesting step forward has been achieved recently in the context of quantum gravity \cite{Fehre:2021eob}, where an interacting UV fixed point has been found in Lorentzian signature. Small perturbations trigger an RG flow which connects the high energy fixed point with classical general relativity in the infrared (IR),  thereby providing the graviton propagator at all scales. Curiously, the graviton is found to admit a causal KL representation, opening a  door to address aspects of causality and unitarity of quantised gravity in a field-theoretical setting.

\begin{figure}[t]
	\includegraphics[width=\linewidth]{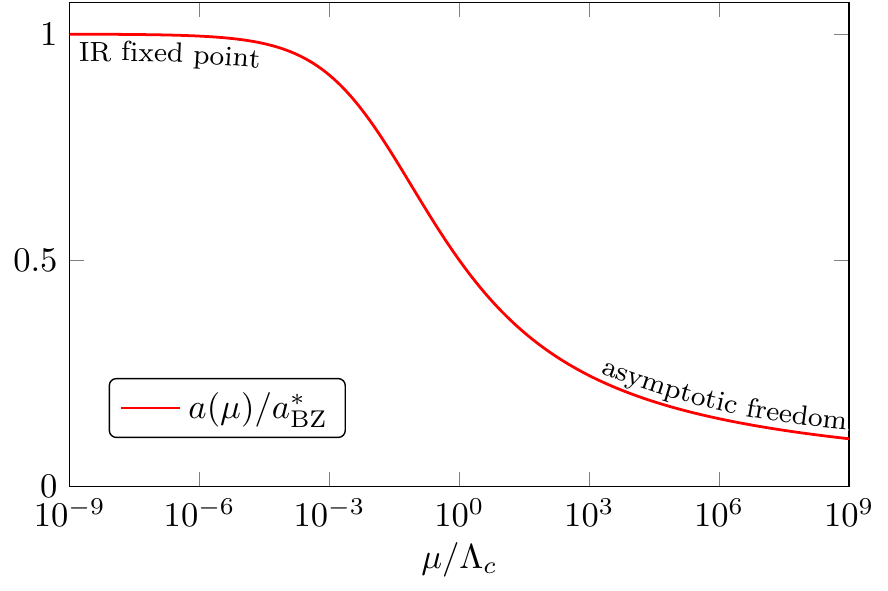}
	\caption{RG trajectory of a weakly coupled matter-gauge theory with coupling $a(\mu)$, interpolating between an asymptotically free fixed point in the UV and an interacting conformal fixed point  $a_*$ in the IR with cross-over scale $\Lambda_c$.}
	\label{fig:BZ}
\end{figure}

In this work, motivated by these findings, we revisit spectral functions of gauge theories with matter from first principles, without gravity. We concentrate on theories which asymptote into conformal fixed points in the UV and the IR. The role models for this are $SU(N)$ gauge theories coupled to $\nf$ fermions in the fundamental representation.
For small $\nf$, the theory is asymptotically free and confining, such as in QCD, while
for large $\nf$, asymptotic freedom is lost and viable UV-completions have not been found. For intermediate $\nf$, however, the theory remains asymptotically free and develops the interacting Banks-Zaks (BZ)  fixed point in the IR \cite{Belavin:1974gu, Caswell:1974gg, Banks:1981nn},  whose running coupling is illustrated in \cref{fig:BZ}.
For our purposes, the latter offers several benefits and features: 
\begin{itemize}
	\item The theory is a perturbatively renormalisable and unitary quantum field theory, courtesy of asymptotic freedom, while the intricacies of confinement and chiral symmetry breaking are avoided, courtesy of the BZ fixed point.
	\item The fundamental fields, gluons and quarks, remain good degrees of freedom to parametrise the physics both in the UV and the IR. 
	\item The theory can be brought under rigorous perturbative control at all scales.
\end{itemize}
Our set-up is conceptually similar to the scenario for gravity \cite{Fehre:2021eob} where the metric field is taken as the primary carrier of the gravitational force at all scales, except that interactions are parametrically small throughout. We expect that structural insights achieved here may also be of relevance in gravity.

In this spirit, we investigate  gluon, quark, and ghost propagators and their spectral functions at weak coupling, both analytically and numerically. To achieve accurate results, we additionally exploit findings from perturbation theory up to five loop for the running gauge coupling, and up to four loop to account for self-energy corrections \cite{Chetyrkin:2000dq, Baikov:2016tgj, Herzog:2017ohr, Luthe:2017ttg, Chetyrkin:2017bjc}.  Another  key step is the Callan-Symanzik resummation of self-energy logarithms, which provides  analytical access to the entire complex plane of propagators. 
 In places, we employ the large-$N$ Veneziano limit, which gives rise to a controlled expansion in the small conformal parameter $\epsilon\ll 1$. At stronger coupling the BZ fixed point disappears, and we investigate how the loss of conformality correlates with the loss of a KL spectral representation. We also derive general conditions, solely expressed in terms of universal scaling exponents, which determine the presence or absence of propagator non-analyticities.

%%%%%%%%%
%%%%%%%%

This paper is structured as follows. In \cref{sec:BZ}, we introduce the BZ fixed point and the explicit analytic coupling solution at two-loop order. We use the Callan-Symanzik equation to resum the two-loop results and obtain full analytical access to the complex momentum plane of the field propagators. In \cref{sec:spectral-function}, we derive the corresponding spectral functions, discuss the conditions for their existence, and discuss their properties and dependence on the gauge parameter. In \cref{sec:higher-orders}, we extend our analysis to higher-loop order and investigate the size of the conformal BZ window from the perspective of existing spectral representations of the field propagators. We furthermore compare the perturbative results to ones obtained with the functional renormalisation group. In \cref{sec:general}, we study general perturbative and resummed $\beta$-functions and derive conditions for the absence of non-analyticities. We conclude in \cref{sec:conluctions}.

%%%%%%%%%%%%%%%%%%%%%%%%%%%%
\section{Banks-Zaks Phase and Propagator}
\label{sec:BZ}
In this section, we first recap the known properties of the BZ fixed point and the analytic solution of the two-loop $\beta$-function in terms of the $W$-Lambert function. We then proceed to use the Callan-Symanzik equation for the resummation of large logarithms and study all field propagators analytically in the entire complex plane.

\subsection{Setup}
\label{sec:Setup}
We are interested in four-dimensional Yang-Mills theories with gauge group $SU(\nc)$  coupled to $\nf$ massless Dirac fermions $\psi$ in the fundamental representation. Modulo gauge-fixing and ghost terms, the perturbatively renormalisable Lagrangian is given by
\begin{align}\label{eq:L}
	L &= -\frac{1}{2} \text{Tr}\! \left( F_{\mu \nu} F^{\mu \nu} \right) + \text{Tr}\! \left( \bar{\psi} i \slashed D \psi \right),
\end{align}
where $F_{\mu \nu}$ is the field strength of the gauge bosons, and the trace runs over the colour and flavour indices. The theory has a global $U(\nf)$ flavour symmetry and is otherwise characterised by the gauge and matter field multiplicities, and by the  gauge coupling $g$, which we scale with a perturbative loop factor 
\begin{align}\label{eq:coupling}
	a = g^2/(16 \pi^2)\,.
\end{align}
The dependence of the gauge coupling on the energy  scale $\mu$ is expressed via the $\beta$-function $\beta (a)\equiv\mu^2 {\dd a(\mu^2)}/{\dd \mu^2}$. In perturbation theory, it is given by
\begin{align}
	\label{eq:runningcoupling1}
	\beta (a) = \sum_{n=1}\beta_n\,a^{n+1}\,,
\end{align}
with loop coefficients $\beta_n$  known up to five-loop order in the $\overline{\textrm{MS}}$ scheme \cite{Herzog:2017ohr}. 

Free or interacting renormalisation group fixed points with $\beta(a_*)=0$ are of particular interest \cite{Polchinski:1987dy}, the reason being that scale-invariance for any relativistic and unitary four-dimensional theory that remains perturbative in the UV or IR asymptotes into a conformal field theory \cite{Komargodski:2011vj, Komargodski:2011xv, Luty:2012ww}. The theory \cref{eq:L} always displays a free Gaussian fixed point $a_*=0$. For sufficiently few matter fields, the one-loop coefficient $\beta_1$ is negative leading to asymptotic freedom such as in QCD \cite{Gross:1973id, Politzer:1973fx}. Conversely, adding too many matter fields implies that asymptotic freedom is lost and the theory becomes IR free such as in QED. 

The competition of gauge and matter field fluctuations may also lead to interacting quantum fixed points $(a_*> 0)$. At weak coupling, interacting fixed points are either of the Banks-Zaks (BZ) or the gauge-Yukawa (GY) type \cite{Bond:2016dvk, Bond:2018oco}. BZ fixed points \cite{Belavin:1974gu, Caswell:1974gg, Banks:1981nn} are always IR in any quantum field theory \cite{Bond:2016dvk}, while fixed points involving Yukawa interactions may be either IR or UV \cite{Bond:2016dvk, Bond:2018oco}, see \cite{Litim:2014uca, Litim:2015iea, Bond:2016dvk, Bond:2017wut, Bond:2017tbw, Bond:2017lnq, Bond:2017suy, Bond:2019npq}. It is also well-established that conformal windows with BZ or GY fixed points exist at strong coupling, see e.g.~\cite{Gies:2005as, Dietrich:2006cm, Jarvinen:2011qe, Kusafuka:2011fd,  Alvares:2012kr, DeGrand:2015zxa, Gukov:2016tnp,  Simmons-Duffin:2016gjk, Poland:2018epd,Ryttov:2016ner, Ryttov:2017lkz,  Kuipers:2018lux, Hasenfratz:2018wpq, Fodor:2018uih,  Antipin:2018asc, Hasenfratz:2019dpr, DiPietro:2020jne, Kim:2020yvr, Bond:2021tgu, Bond:2022xvr}. For a recent conjecture of a BZ phase with spontaneously broken scale symmetry, see~\cite{DelDebbio:2021xwu}.

Much less is known about the fixed points outside the BZ or GY conformal windows. In pure Yang-Mills theory, it has been suggested that its IR limit relates to a non-perturbative fixed point, e.g.~\cite{Aguilar:2002tc, Gies:2002af, Pawlowski:2003hq}. In a different vein, it has also been speculated that a new strongly interacting UV fixed points may arise in the many fermion limit \cite{PalanquesMestre:1983zy, Gracey:1996he, Holdom:2010qs}, though the viability for this has been called into question as of late \cite{Martin:2000cr, Ryttov:2019aux, Alanne:2019vuk, Leino:2019qwk, Dondi:2019ivp, Dondi:2020qfj,Bond:2021tgu}.

In this work, we focus on settings with a BZ fixed point and exploit results from perturbation theory up to five loops and suitable resummations thereof.

\subsection{Banks-Zaks}
\label{sec:BZ-FP}
Our starting point is a regime with a BZ fixed point  \cite{Belavin:1974gu, Caswell:1974gg, Banks:1981nn}, which, furthermore, is under strict perturbative control. If so, the running gauge coupling remains small along the entire renormalisation group (RG) trajectory interpolating between asymptotic freedom in the UV and a perturbatively controlled BZ fixed point in the IR. Most notably perturbative methods are sufficient to analyse the properties of the theory.

In this spirit, we notice that the $\beta$-function \cref{eq:runningcoupling1} features a non-trivial BZ fixed point, which takes the form
\begin{align}\label{eq:FP}
	a_* = - \frac{\beta_1}{\beta_2} + \order{\beta_1^2} 
\end{align}
to the leading orders in perturbation theory. The one- and two-loop coefficients are given by
\begin{align} \label{eq:b1b2}
	\beta _1&=-\frac{2  \nc}{3} \varepsilon \,, \notag \\
	\beta_2 &= \frac{25 \nc^2-11}{2} + \frac{3-13 \nc^2}{3} \varepsilon \,.
\end{align}
Here, we introduced the Veneziano parameter $\varepsilon$ 
\begin{align}
	\label{eq:def-epsilon}
	\varepsilon = \frac{11}{2} - \frac{\nf}{\nc} \,,
\end{align}
to replace the free parameters $(\nc,\nf)$ by $(\nc,\varepsilon)$. The parameter \cref{eq:def-epsilon}  may take values between $[-\infty,\frac{11}{2}]$. For $\varepsilon>0$, the theory is asymptotically free, corresponding to $\nf < \frac{11}{2} \nc$. Further, the one loop gauge coefficient is parametrically small provided that
\begin{align}
	\label{eq:vareps}
	0<\varepsilon\ll1\,.
\end{align}
Consequently, interacting fixed points in the regime \cref{eq:vareps} are under strict perturbative control \cite{Banks:1981nn, Bond:2019npq}.

The fixed point \cref{eq:FP} stems from a cancellation between the one-loop term and the remainder of the $\beta$-function, starting with the two-loop coefficient. Such a cancellation leads to a reliable fixed point within perturbation theory if $\beta_1/\beta_2$ in \cref{eq:FP} is parametrically small. This is precisely the case when $\varepsilon\ll1$, and provided that the two-loop term $\beta_2$ remains of order unity and positive. The latter holds true in general: for any $4d$  quantum gauge theory coupled to matter with a parametrically small one loop coefficient $\beta_1$, the two-loop coefficient $\beta_2$ is of order unity, and strictly positive \cite{Bond:2016dvk}. Hence, BZ fixed points are invariably IR and never UV. 

A regime with arbitrarily small $\varepsilon$ can always be achieved in the large-$N$ Veneziano limit where $\nc, \nf \to \infty$ with $\nf/\nc$ fixed, and where the parameter \cref{eq:def-epsilon} becomes continuous, also reducing the number of free parameters to one. In this work, we follow two strategies to determine fixed points: Firstly, we use the perturbative loop expansion to determine fixed points from order to order. This provides the fixed point as a rational function of $\varepsilon$, and will mostly be used when $\nc, \nf$ are finite. Alternatively, we may determine the fixed point as a strict power series in $\varepsilon$. The latter mixes contributions from different loop orders owing to the fact that loop coefficients $\beta_i/(\nc)^i$ contain terms of different order in $\varepsilon$, see \cref{eq:b1b2}. In the Veneziano limit, this is sometimes referred to as the conformal expansion.

Beginning with the loop expansion and neglecting three- and higher-loop corrections, we can find an analytical solution for the running coupling to study its properties in the complex plane. At two-loop order, the value for the BZ fixed point reads
\begin{align}
	\label{eq:BZ2loop}
	a_* = \frac{4 \varepsilon}{\nc (75 - 26 \varepsilon) - \tfrac{6}{\nc} \left(\frac{11}{2} - \varepsilon\right)} \,.
\end{align}
Further, its universal scaling exponent 
\begin{align}\label{eq:theta}
	\theta = \left.\frac{\partial \beta_a}{\partial a} \right|_{a = a_*},
\end{align}
is given by
\begin{align}
	\label{eq:BZ2loopeigv}
	\theta = \frac{8}{3} \frac{\varepsilon^2}{\left( 75 - 26 \varepsilon \right) - \tfrac{6}{\nc^2} \left(\frac{11}{2} - \varepsilon\right)} \,,
\end{align}
at two-loop accuracy. The corresponding RG trajectory for the running gauge coupling that connects the asymptotically free UV fixed point with the BZ fixed point in the IR, as shown in \cref{fig:BZ}, can be found analytically (see \cref{app:analyticsol}). It reads \cite{Corless:1996zz, Gardi:1998qr}
\begin{align}
	a(\mu^2) = \frac{a_*}{1 + W_0(z)} \, , 
	\label{eq:afullsol}
\end{align}
in terms of the principal branch of the $W$-Lambert function, with
\begin{align}
	z(\mu^2) &= \omega_0 e^{\omega_0} \left(\frac{\mu ^2}{\mu_0^2}\right)^{\!\theta} \, , 
	&
	\omega_0 &= \frac{a_* - a_0}{a_0} \, ,
	\label{eq:zdef}
\end{align}
and initial condition $a(\mu_0^2)=a_0$. As such, the expressions \cref{eq:BZ2loop,eq:BZ2loopeigv,eq:afullsol,eq:zdef} are the two-loop results for the BZ fixed point and the running coupling for all scales.

The  gauge coupling  interpolates between asymptotic freedom in the UV and the Banks-Zaks fixed point in the IR. 
The transition between the two scaling regimes is characterised by an RG invariant cross-over scale $\Lambda_c$,  indicated in  \cref{fig:BZ}. 
It can be written as
\begin{equation}\label{eq:Lambdac}
\Lambda_c=
\mu\cdot
 \exp(\frac{1}{\beta_1\,\delta\alpha(\mu)})\,,
\end{equation}
where $0<\delta\alpha(\mu)\ll \alpha_*$ denotes the  initial deviation of the gauge coupling from its free UV fixed point at the high scale $\mu$, while $\beta_1<0$ denotes  the one-loop gauge coefficient, see \cref{eq:b1b2}. It is readily confirmed that $\mathrm d\Lambda_c/\mathrm d\ln \mu=0$. Notice that the expression \cref{eq:Lambdac}  is parametrically the same as for any asymptotically free gauge theory and coincides with the definition for $\Lambda_{\rm QCD}$ in perturbative QCD, as it must, because the UV initial condition $\delta\alpha$ does not know that the theory achieves  an interacting fixed point rather than  confinement in the  IR.

Alternatively, we can perform a conformal expansion organised in powers of the Veneziano parameter $\varepsilon$. We expand the BZ fixed point and its critical exponent up to the appropriate power of $\varepsilon$. From two-loop perturbation theory, we can obtain leading order expressions in $\varepsilon$ for the BZ fixed point
\begin{align}
	\label{eq:BZVeneziano}
	\nc a_* = \frac{4}{75} \varepsilon 	+ \mathcal{O}\! \left( \varepsilon^2,  \frac{1}{\nc^2} \right) ,
\end{align}
as well as its eigenvalue
\begin{align}
	\label{eq:BZVenezianoeigv}
	\theta = \frac{8}{225} \varepsilon ^2 + \mathcal{O}\! \left( \varepsilon^3, \frac{1}{\nc} \right)  .
\end{align}
Higher-order expressions for the fixed point and the scaling exponent up to five loop order in the Veneziano limit are given in \cref{sec:AppBZ}. Inserted into \cref{eq:afullsol} leads to the running coupling at leading order in the Veneziano expansion. 

For small $\varepsilon\ll1$, the difference between loop expansion and the Veneziano expansion is parametrically small. For larger values of $\varepsilon$, there is a quantitative difference between the loop expansion and the Veneziano expansion due to the fact that loop coefficients may contain different orders in $\varepsilon$. In this work, we display results in both expansions. In the conformal expansion, we restrict ourselves to small $\varepsilon$ and work exclusively in the $\nc\to\infty$ limit. In the loop expansion, we also explore larger values of $\varepsilon$ as well as small values of $\nc$.

\subsection{Gauge Coupling in the Complex Plane}
\label{sec:complex-plane-2-loop}
We now discuss the properties of the running coupling in the complex $\mu^2/\mu_0^2$ plane. While $a$ is uniquely defined for $\mu^2/\mu_0^2 > 0$, uniqueness is lost in the complex plane provided there are branch cuts. There are two sources of branching points:
\begin{itemize}
	\item[(i)] the branching point at $\mu^2/\mu_0^2 = 0$ originating from the power law in the definition of $z$ in \cref{eq:zdef},
	\item[(ii)] further branching points originating through the $W$-Lambert function.
\end{itemize}
The branching point of the first type (i) is always present and continues along $\mu^2/\mu_0^2 < 0$ until $- \infty$. This branch cut is important for the physics as its discontinuity later results in the spectral functions of the fields.

In contrast to that, the branching points of the second type (ii) may be absent or present including in larger numbers, depending on the values for $\varepsilon$ and $\nc$. As we will see, these types of branching points spoil the existence of a standard KL spectral representation. 

%%%%%%%%%
\begin{figure}[t]
	\includegraphics[width=.95\linewidth]{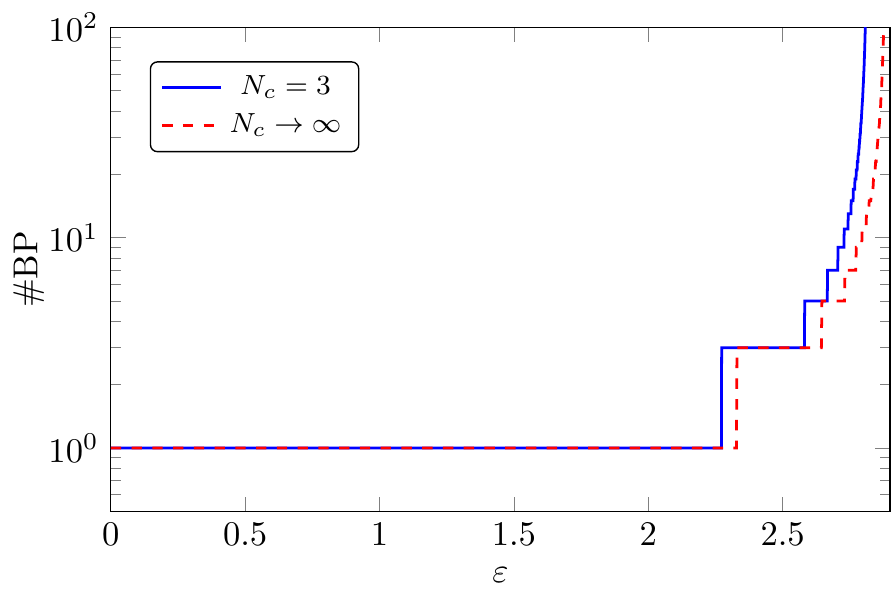}
	\caption{Shown are the number of branching points  in the complexified $\mu^2$-plane of the running gauge coupling at two loop, as a function of the Veneziano parameter $\varepsilon$.  A single branch cut at weak coupling ($0<\varepsilon <\varepsilon_\text{branch cut}$) proliferates rapidly into many more at stronger coupling ($\varepsilon_\text{branch cut}<\varepsilon <\varepsilon_{\rm max}$) before the fixed point disappears $(\varepsilon=\varepsilon_{\rm max})$.}
	\label{fig:branching}
\end{figure}
%%%%%%%%

%%%%%%%%
\begin{figure*}[t]
	\includegraphics[width=0.41\linewidth]{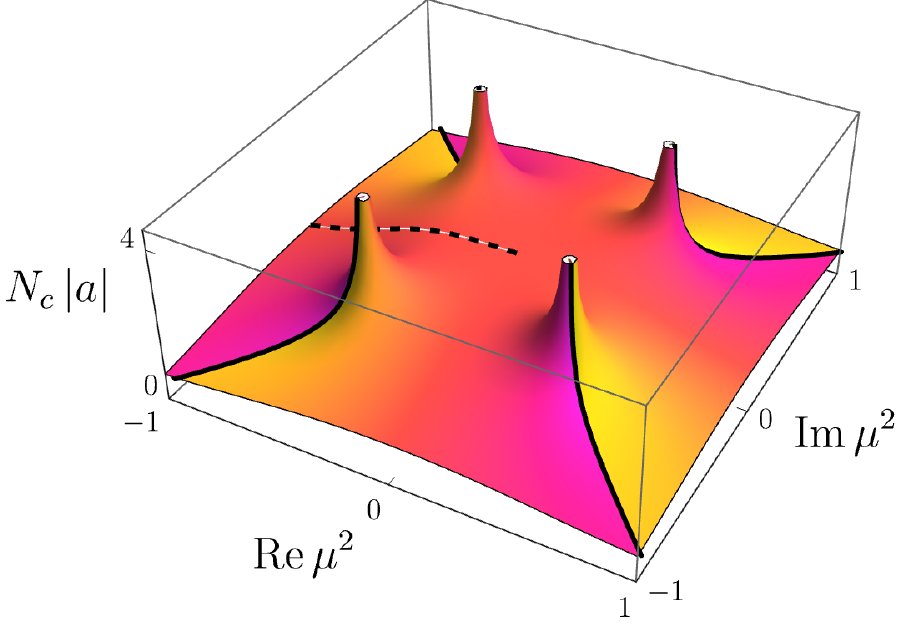}\hspace{1cm}
	\includegraphics[width=0.41\linewidth]{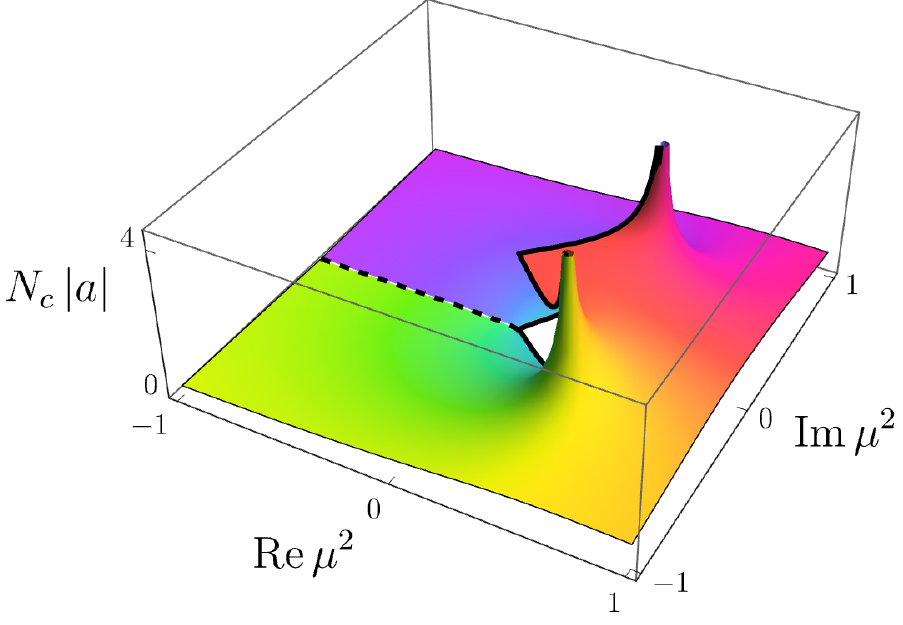}
	\hfill
	\includegraphics[width=0.07\linewidth]{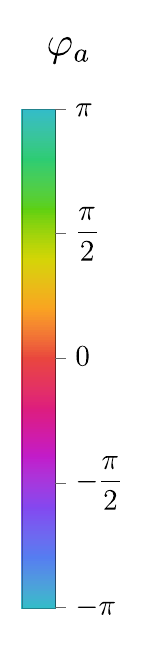}
	\caption{Magnitude $|a|$ and phase $\varphi_a$ of the gauge coupling $a$ in the complex plane of the RG scale $\mu^2$ at two-loop order. We use $\varepsilon = 2.7$,  $\mu_0^2 = 1$, and $\nc \rightarrow \infty$. In the left panel, we use the principal branch \cref{eq:afullsol}, and in the right panel, the modified prescription from \cref{eq:amod}. Due to the properties of the $W$-Lambert function two branch points have disappeared in this modified version of the running coupling. Branch cuts originating from the $W$-Lambert function (power law of $\mu^2$) are drawn as black solid (dashed) lines.} 
	\label{fig:a_branchcutmods}
\end{figure*}
%%%%%%%%%

The principal branch of the $W$-Lambert function $W_0 (z)$ has its branch cut starting at $z = -1/e$ and is chosen to continue along the negative $z$ axis towards $z= - \infty$. Thus, to obtain the branch cut of $W_0 (z)$ we must have
\begin{align}
	z(\mu^2) < - \frac{1}{e} \, .
	\label{eq:cond-BP}
\end{align}
In order for this equation to have a solution, the phase of $\mu^2/\mu_0^2$ must be given by
\begin{align}
	\varphi (\mu^2/\mu_0^2) = -\frac{\pi}{\theta} (2n + 1) \, ,
	\label{eq:astdbranchstart}
\end{align}
with $n$ an integer and $\theta$ the eigenvalue of the BZ fixed point given in \cref{eq:BZ2loopeigv} and \cref{eq:BZVenezianoeigv}. Using the principal branch, only those solutions in \cref{eq:astdbranchstart} exist which contain a phase $\varphi$ with $- \pi < \varphi \leq \pi$. This leads us to the requirement
\begin{align}
	- \frac{1}{|\theta|} < 2n + 1 \leq \frac{1}{|\theta|} \, ,
\end{align}
which means the number of branching points originating from the $W$-Lambert function is given by twice the amount of odd numbers less than $| \theta |$. For the total number of branch cuts, this means
\begin{align}
	\label{eq:BP-two-loop}
	\# \text{bp} = 1 + 2 \left\lfloor \frac{\left| \theta \right| + 1}{2} \right\rfloor .
\end{align}
The existence of non-analyticities for $|\theta|>1$ was first noticed in \cite{Gardi:1998ch}. In \cref{fig:branching}, we show the resulting number of branching points as a function of $\varepsilon$ using the two-loop expression of the BZ eigenvalue \cref{eq:BZ2loopeigv}. We observe a single branch cut at $\mu^2/\mu^2_0=0$ at weak and moderate coupling ($0<\varepsilon <\varepsilon_{\rm branch\,cut}$), with 
\begin{align}\label{eq:epscrit}
	\varepsilon_{\rm branch\,cut}=\left\{
	\begin{array}{lcr}
		2.2723&{\rm for}& \nc=3\,,\\[1ex]
		2.32850&{\rm for}&1/\nc =0\,.
	\end{array}\right.
\end{align}
At stronger coupling ($\varepsilon_\text{branch cut}<\varepsilon <\varepsilon_{\rm max}$), and beyond the threshold $\varepsilon_\text{branch cut}$, their number rapidly  proliferates into many more branching points with increasing $\varepsilon$. Their number diverges just when the fixed point ceases to exist $(\varepsilon=\varepsilon_{\rm max})$, with 
\begin{align}
	\varepsilon_{\rm max}=\left\{
	\begin{array}{lcr}
		2.8158&{\rm for}& \nc=3\,,\\[1ex]
		2.8846&{\rm for}&1/\nc =0\,,
	\end{array}\right. 
\end{align}
at two-loop accuracy.

The presence of branch cuts signifies that the analytic continuation of the running gauge coupling into the complex plane is not unique, simply because the location of branch cuts is ambiguous. In fact, even the number of branching points may become ambiguous due to \emph{e.g.}~the properties of the $W$-Lambert function. In \cref{fig:a_branchcutmods}, we illustrate this ambiguity at two-loop, using $\varepsilon = 2.7$. In the standard prescription in the left panel of \cref{fig:a_branchcutmods}, we follow \cref{eq:afullsol} and observe in total five branching points, four of which are obtained from the $W$-Lambert function. The plot in the right panel of \cref{fig:a_branchcutmods} instead assumes the running coupling to be given by
\begin{align}
	\label{eq:amod}
	a(\mu^2) = \frac{a_*}{1 + W_i (z)} \,,
\end{align}
with
\begin{align}
	i =  \begin{cases}
		-2 & \phantom{\text{for} \;\varphi (\mu^2/\mu_0^2)} < -3 \frac{\pi}{\theta} \\[1ex]
		-1 	& \phantom{\text{for} \;\varphi (\mu^2/\mu_0^2)} < -1 \frac{\pi}{\theta} \\[1ex]
		\ \ 0 &\text{for} \;\varphi (\mu^2/\mu_0^2) < \ \ 1 \frac{\pi}{\theta} \\[1ex]
		\ \ 1&\phantom{ \text{for} \;\varphi (\mu^2/\mu_0^2)} < \ \ 3 \frac{\pi}{\theta} \\[1ex]
		\ \ 2 &\phantom{\text{for} \;\varphi (\mu^2/\mu_0^2)} > \ \ 3 \frac{\pi}{\theta}
	\end{cases} \, .
\end{align}
Since only $W_0 (z)$ has a branching point at $z = -1/e$ and all other branches at $z = 0$, two branching points have moved due to this prescription to the origin and their branch cuts go along the negative axis towards $- \infty$. Thus, the presence of additional branch cuts turns out to make the running coupling in the complex plane ambiguous.

\subsection{Callan-Symanzik Resummation}
\label{sec:CS-prop}
At this point, we want to compute the propagator for all momenta using perturbative methods. We consider all species of particles included in the theory, i.e., gluons, quarks, and ghosts. Starting from corresponding two-point functions, we follow, e.g., \cite{Chetyrkin:2000dq} and define
\begin{align}
	D_{\mu \nu}^{a b} (p) &\equiv i \int \!\dd^dx \, e^{i p  x} \langle T A_\mu^a (x) A_\nu^b (0) \rangle \, ,\notag\\
	\Delta^{a b} (p) &\equiv i \int \!\dd^dx \, e^{i p x} \langle T c^a (x) \bar{c}^b (0) \rangle \, ,\notag \\
	S_{i j} (p) &\equiv i \int \!\dd^dx \, e^{i p x} \langle T \psi_i (x) \bar{\psi}_j (0) \rangle \, .
\end{align}
The tensor structures of the propagators follow from general arguments such as Ward identities with only the self energies being undetermined,
\begin{align}
	D_{\mu \nu}^{a b} (p) &= \frac{\delta^{a b}}{- p^2} \left[ \left( - g_{\mu \nu} + \frac{p_\mu p_\nu}{p^2} \right) \frac{1}{1 + \Pi_A (p^2,\mu^2)} - \xi \frac{p_\mu p_\nu}{p^2} \right] ,\notag \\
	\Delta^{a b} (p) &= \frac{\delta^{a b}}{- p^2} \frac{1}{1 + \Pi_c (p^2,\mu^2)} \,,\notag \\
	S_{i j} (p) &= \frac{\delta^{i j}}{- p^2} \frac{\slashed{p}}{1 + \Pi_\psi (p^2,\mu^2)} \,.
\end{align}
Perturbative results for the self-energies $\Pi_\phi$ for all the propagators $\phi \in \{A,c,\psi\}$ have been obtained in \cite{Chetyrkin:2000dq, Ruijl:2017eht}. These contain logarithmic contributions of the form $\log (-p^2/\mu^2)$ which diverge in the UV as well as in the IR. These large logarithms lead to a breakdown of the perturbative expansion and it is required to resum them in order to obtain the correct large and small momentum behaviour. 

A practical tool to resum logarithmic contributions lies in the perturbative renormalisation group for $n$-point functions. It stems from the property of a bare $n$-point function being independent of the renormalisation scale $\mu^2$. We apply it here to the scalar parts of the propagators defined with 
\begin{align}
	G_\phi (p^2,\mu^2) &= \frac{1}{-p^2} \frac{1}{1 + \Pi_\phi (p^2,\mu^2)} \, .
\end{align}
For the bare propagators $G_{\phi, b}$, the independence of $\mu^2$ implies
\begin{align}
	\mu^2 \dderiv{}{\mu^2} G_{\phi, b} = \mu^2 \dderiv{}{\mu^2} \left( Z_\phi \, G_\phi \right) = 0 \, .
	\label{eq:dmu-bare-G}
\end{align}
Here, we denote the renormalised propagator as $G_{\phi}$ and $Z_\phi$ is the renormalisation constant of the field $\phi$.
The perturbative propagator depends on the renormalisation scale only through the running coupling and the logarithmic terms that we want to resum. Therefore, we obtain the Callan-Symanzik (CS) equation
\begin{align}
	\left( \mu^2 \pderiv{}{\mu^2} + \beta (a) \pderiv{}{a} - \gamma_\phi \right) G_\phi = 0 \, ,
	\label{eq:callansymanzik}
\end{align}
with the anomalous dimension defined by
\begin{align}
	\gamma_\phi = - \frac{1}{Z_\phi} \dderiv{Z_\phi}{\log \mu^2} \, .
\end{align}
Solving \cref{eq:callansymanzik} results in the general form of the propagator up to an integration constant. This integration constant can be obtained from the perturbative result at the point where the logarithmic corrections vanish, i.e.\ at $-p^2 = \mu^2$. 

To solve \cref{eq:callansymanzik} we first note that 
\begin{align}
	G_\phi (a, p^2, \mu^2) = \frac{f_\phi (a, p^2 / \mu^2)}{p^2} \, ,
\end{align}
with some function $f_\phi$ that only depends on the ratio $p^2 / \mu^2$.
This allows us to trade $\mu^2$-derivatives for $p^2$-derivatives,
\begin{align}
	\left( - p^2 \pderiv{}{p^2} + \beta (a) \pderiv{}{a} - \gamma_\phi -1\right) G_\phi (a, p^2, \mu^2) = 0 \, .
\end{align}
This differential equation can be solved by the method of characteristics. To that end, we introduce a new momentum dependent running coupling $\bar{a} (-p^2)$
\begin{align}
	\int_a^{\bar{a}} \! \frac{\mathrm d a'}{\beta (a')} = \int_{\mu^2}^{-p^2} \! \frac{\mathrm d \bar{p}^2}{\bar{p}^2} \, .
\end{align}
The two-loop solution for $\bar{a}$ is given by the solution for $a$ upon the replacement of $\mu^2$ by $-p^2$ and $\mu_0^2$ by $\mu^2$,.
\begin{align}
	\bar{a} (-p^2) = \frac{a_*}{1 + W_0 (\bar{z})} \, ,
	\label{eq:def-abar}
\end{align}
with
\begin{align}
	\bar{z} = \left( \frac{a_*}{a} - 1 \right) e^{a_*/a - 1} \left( \frac{-p^2}{\mu^2} \right)^{\!\theta} \, .
	\label{eq:zbardef}
\end{align}
This allows the elimination of the variable $p^2$ in favour of $\bar{a}$. The CS equation for $G_\phi (\bar{a}, a, \mu^2)$ then takes the form of an ordinary first-order differential equation
\begin{align}
	\left( \beta (a) \pderiv{}{a} - \gamma_\phi - 1 \right) G_\phi =0\, .
\end{align}
Its solution $G_\phi (\bar{a}, a, \mu^2) $ is given by
\begin{align}
	G_\phi &= \frac{ \hat{C}_\phi (\bar{a}, \mu^2)}{-p^2} \exp( - \int_{a}^{\bar{a}} \!\frac{\text{d} a'}{\beta (a')} \, \gamma_\phi (a') ) \,,
	\label{eq:callimpprop}
\end{align}
with $\hat{C}_\phi (\bar{a}, \mu^2)$ an integration constant independent of $a$. At two loop accuracy, and writing $\gamma_\phi (a) = \gamma_\phi^{(1)} a + {\cal O}(a^2)$, the integral in \cref{eq:callimpprop} is given by
\begin{align}
	\int_{a}^{\bar{a}}\! \frac{\text{d} a'}{\beta (a')} \, \gamma_\phi (a')
	&= \frac{\gamma_\phi^{(1)}}{\beta_1} \log\frac{\bar{a}}{a} + \frac{\gamma_\phi (a_*)}{\theta} \log \frac{\bar{a} - a_*}{a - a_*}.
\end{align}
The only remaining task is the determination of the integration constant $\hat{C}_\phi (\bar{a}, \mu^2)$. This can be obtained by comparing to the perturbative two-loop result at $p^2 = - \mu^2$, leading to
\begin{align}
	\label{eq:chat-2loop}
	\hat{C}_\phi (\bar{a}, \mu^2) = - \frac{\mathcal{N}_\phi}{1 + \Pi_\phi^{(1)} \bar{a} + \Pi_\phi^{(2)} \bar{a}^2}  \,,
\end{align}
where the  coefficients $\Pi_\phi^{(i)} $ arise from a series expansion of the self energies at $p^2= -\mu^2$, namely 
\begin{align}\label{eq:finiteSE}
\Pi_\phi (p^2= -\mu^2) = \Pi_\phi^{(1)} a + \Pi_\phi^{(2)} a^2 + \ldots\,.
\end{align}
We have included an overall normalisation factor $\mathcal{N}_\phi$, which originates from the freedom to rescale wave-function renormalisation $Z_\phi$ in \cref{eq:dmu-bare-G}. The coefficients $\Pi_\phi^{(i)}$ have been computed in \cite{Chetyrkin:2000dq, Ruijl:2017eht} up to four-loop order. The final result for the CS resummed two-loop propagator is
\begin{align}
	G_\phi 
	&= \frac{1}{p^2} \frac{\mathcal{N}_\phi}{1 + \Pi_\phi^{(1)} \bar{a} + \Pi_\phi^{(2)} \bar{a}^2} \left( \frac{a}{\bar{a}} \right)^{\!\gamma_\phi^{(1)}/\beta_1} \left( \frac{a - a_*}{\bar{a} - a_*} \right)^{\!\gamma_\phi (a_*)/\theta}\! .
	\label{eq:propresummed}
\end{align}
From this expression, we find the UV behaviour at the Gaussian fixed point and the IR behaviour at the BZ fixed point. Close to the Gaussian fixed point, the term $(a/\bar{a})^{\gamma_\phi^{(1)}/\beta_1} $ in \cref{eq:propresummed} dominates, while in the IR, the last term in \cref{eq:propresummed} corresponding to the BZ fixed point produces an IR behaviour characterised by the anomalous dimension evaluated at the fixed point and its critical exponent. Using the asymptotics of the running coupling in \cref{eq:runcoupasym}, the explicit large and small momentum asymptotics of the propagator are
\begin{align}	\label{eq:proplimitIR}
	G_\phi =
	\left\{
	\begin{array}{ll}
		\frac{N^{\text{UV}}_{\phi}}{p^2} \left[ \log( - \frac{p^2}{\mu^2}) \right]^{\gamma^{(1)}_\phi/\beta_1} 
		&\text{for } \left|p^2\right| \to \infty  \,,  \\[2ex]
		\frac{N^{\text{IR}}_{\phi}}{p^2} \left(-\frac{p^2}{\mu^2}\right)^{-\gamma_\phi(a_*)} 	
		&\text{for } \left|p^2\right| \to 0  \,.
	\end{array}\right.
\end{align}
The large momentum asymptote is well known and relates to the Oehme-Zimmermann superconvergence relation \cite{Oehme:1979ai,Oehme:1979bj,Oehme:1990kd}.
Note that the critical exponent in \cref{eq:propresummed} cancels against the critical exponent from the power law of the running coupling in the IR such that the leading power in the IR limit of \cref{eq:proplimitIR} is independent of the critical exponent. 
The normalisation factors $N^{\text{IR}}_\phi$ and $N^{\text{UV}}_\phi$ in \cref{eq:proplimitIR} read 
\begin{align}
	N^{\text{IR}}_{\phi} &= \mathcal{N}_\phi \frac{ e^{\left(1 - a_*/a\right)\gamma_\phi (a_*)/\theta}}{1 + \Pi_\phi^{(1)} a_* + \Pi_\phi^{(2)} a_*^2} \left( \frac{a}{a_*} \right)^{\gamma_{\phi}^{(1)}/\beta_1 + \gamma_\phi (a_*)/\theta} , \notag \\
	N^{\text{UV}}_{\phi} &= \mathcal{N}_\phi \left(- \beta_1 a\right)^{\gamma_{\phi}^{(1)}/\beta_1} \left( \frac{a_* - a}{a_*} \right)^{\gamma_\phi (a_*)/\theta}  .
	\label{eq:NUV-and-NIR}
\end{align}
and we choose the normalisation factor $\mathcal{N}_\phi$ such that $N^{\text{UV}}_\phi=1$. This choice will later ensure that the spectral function is properly normalised. 

\begin{figure}[t]
	\includegraphics[width=\linewidth]{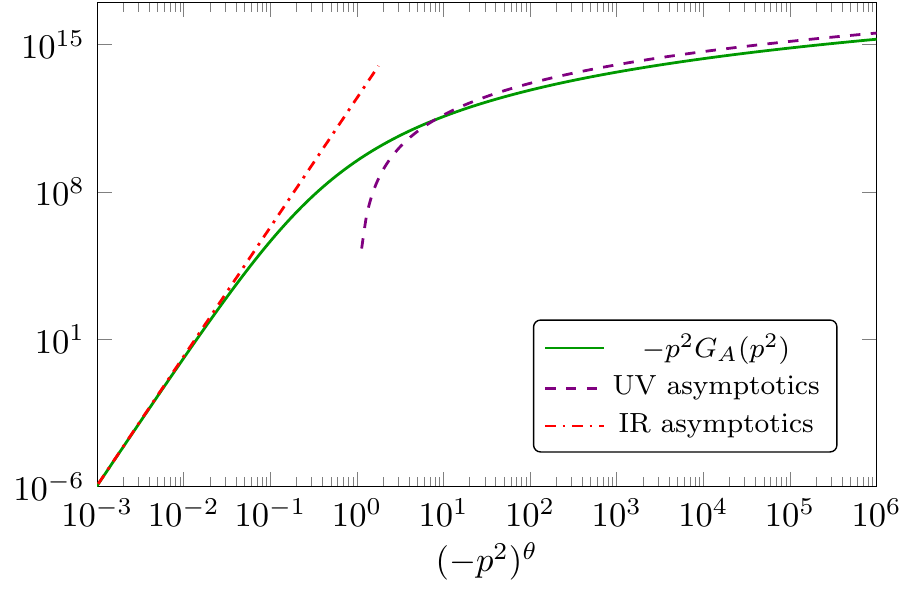}
	\caption{The resummed gluon propagator \cref{eq:propresummed} and its UV/IR asymptotics  \cref{eq:proplimitIR} are shown as a function of $(-p^2)^\theta$ in the Veneziano limit ($\varepsilon = \frac12$,  $\xi = 1$).\footref{footnote_plot_arg}}
	\label{fig:glu_prop}
\end{figure}

The CS-resummed gluon propagator $G_A(p^2)$  at two-loop is shown in \cref{fig:glu_prop} in the Veneziano limit ($\varepsilon = \frac12$, $\xi = 1$). We observe that the propagator quickly approaches its asymptotic limits given by \cref{eq:proplimitIR}, with a smooth crossover regime in between.\footnote{Here and below, we display propagators and spectral functions as functions of $(p^2)^\theta$, which arises naturally in the argument of the $W$-Lambert function, \cref{eq:def-abar,eq:zbardef}. This choice naturally accounts for the parametrically slow running of the gauge coupling. The range shown in \cref{fig:glu_prop} corresponds to scales between $p^2 = \mathcal{O} (10^{-300})$ and $p^2 = \mathcal{O} (10^{300})$. For smaller $\varepsilon$, the range quickly becomes larger. \label{footnote_plot_arg}}

%%%%%%%%
%%%%%%%%

%%%%%%%%
\begin{figure*}[t]
	\includegraphics[width=0.42\linewidth]{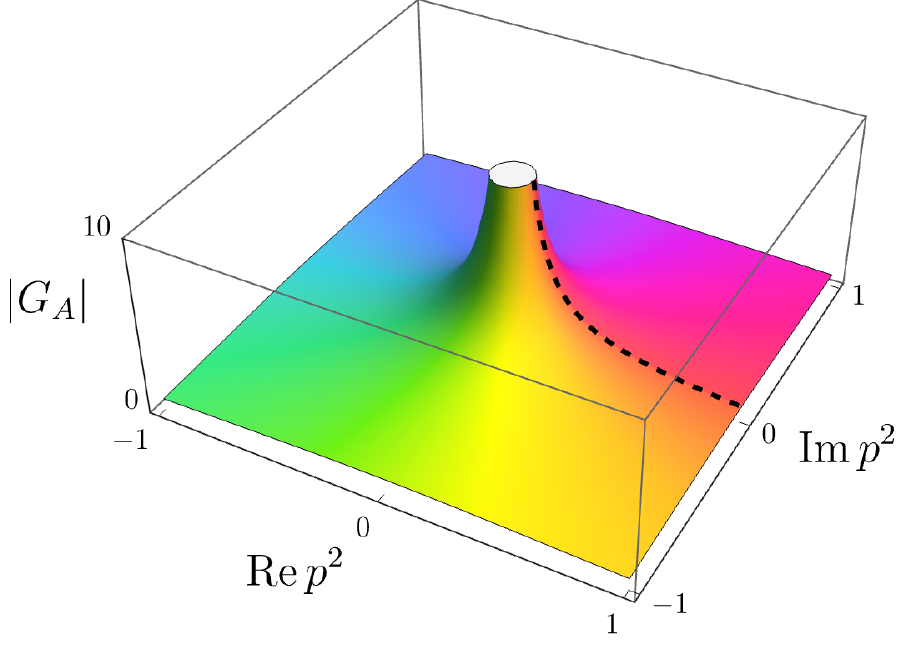}
	\hspace{.3cm}\includegraphics[width=0.42\linewidth]{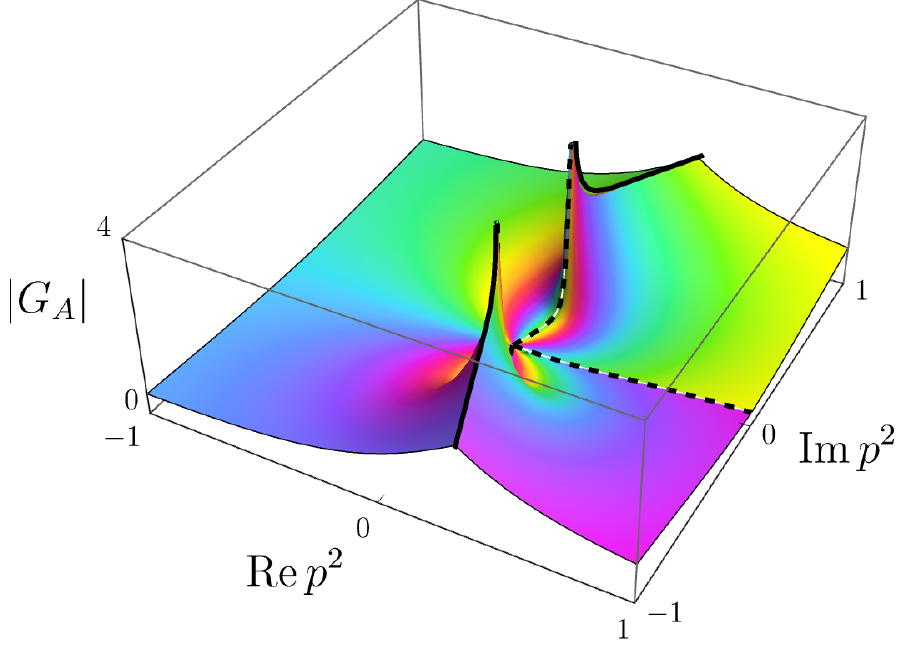}
	\includegraphics[width=0.42\linewidth]{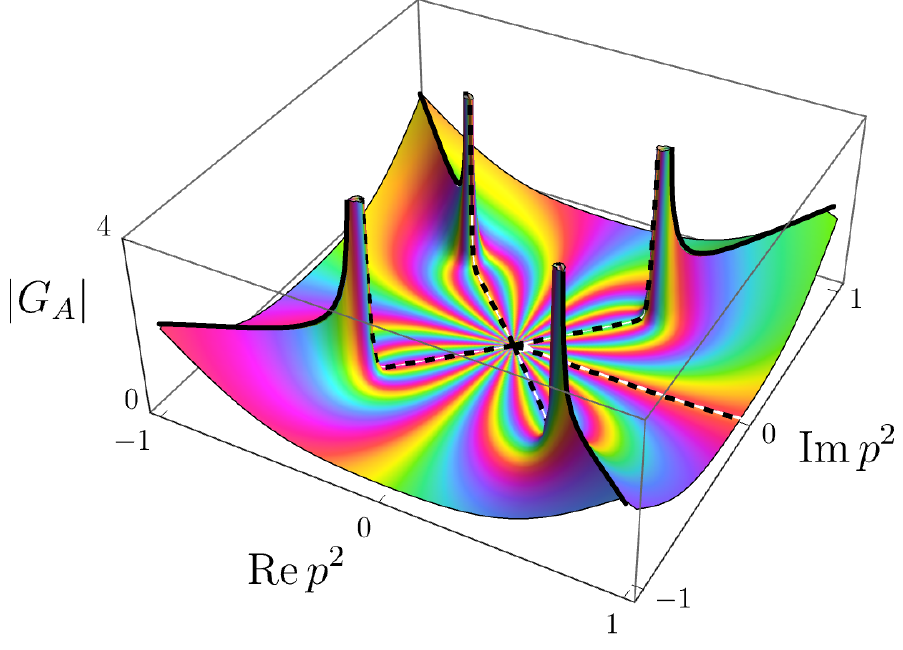}
	\hspace{2em}
	\includegraphics[width=.07\linewidth]{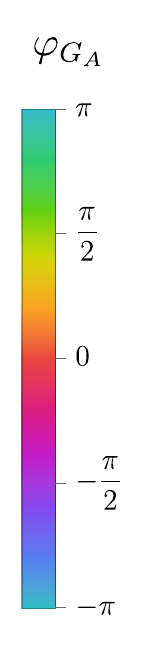}
	\hfill
	\caption{Magnitude $|G_A|$ and phase $\varphi_{G_A}$ of the gluon propagator in the complex plane at two-loop order. We use $\nc \rightarrow \infty$, $\xi = 1$, as well as $\varepsilon = 0.5$ (top left), $\varepsilon = 2.5$ (top right), and $\varepsilon = 2.7$ (bottom). Branch cuts originating from power laws (the $W$-Lambert function) are illustrated with dashed (solid) black lines. There is only one branch cut for $\varepsilon = 0.5$ while the other plots show additional branch cuts related to the properties of the running coupling and the $W$-Lambert function.}
	\label{fig:prop_complexplots}
\end{figure*}
%%%%%%%%

\subsection{Propagators in the Complex Plane}
\label{sec:complex-prop}
Next, we need to understand the analytical properties of propagators in the plane of complexified momenta $p^2$. In particular, we need to understand whether the propagator admits branch cuts or poles. There are three possible origins for the latter. These are
\begin{itemize}
	\item[(i)] branch cuts from the running coupling $\bar{a}$,
	\item[(ii)] branch cuts from the explicit power laws in \cref{eq:propresummed},
	\item[(iii)] poles from the self-energy term in \cref{eq:propresummed}.
\end{itemize}
Next, we discuss the different cases one by one.

{\bf Case (i).} Branch cuts originating from  the running coupling $\bar{a}$ have been discussed in \cref{sec:complex-plane-2-loop}. To apply these findings to the propagator, we replace $\mu^2 \rightarrow -p^2$ and $\mu_0^2 \rightarrow \mu^2$ as well as $a \rightarrow \bar{a}$ and $a \rightarrow a_0$. Then, the running coupling leads to two kinds of branch cuts in the propagator:
\begin{itemize}
	\item[a)] The branch cut at $p^2/\mu^2 < 0$ originating from the power law in the definition of $\bar{z}$.
	\item[b)] Branch cuts for $p^2/\mu^2$ such that $\bar{z} < -1/e$, which originate from the branch cuts of $W_0 (\bar{z})$.
\end{itemize}
As discussed in \cref{sec:complex-plane-2-loop}, the branch cut a) is important for the physics. In turn, branch cuts of the type b) are “dangerous” in that they spoil the existence of a standard KL spectral representation.

{\bf Case (ii).} The explicit power laws in \cref{eq:propresummed} lead to branch cuts when
\begin{align}
	\frac{a}{\bar{a}} &< 0 \, ,
	&&\text{or} &
	\frac{a - a_*}{\bar{a} - a_*} &< 0 \, .
	\label{eq:propaddbranch}
\end{align}
Since $a > 0$, the former is reached only if $\bar{a} < 0$, i.e., $W_0(\bar{z}) < - 1$. There are no values $\bar{z}$ where this equation is fulfilled if the principal branch of the W-Lambert function is used. Instead, the W-Lambert function only takes these values in the branch $W_{-1} (\bar{z})$ for $-1/e < \bar{z} < 0$. Hence, if we stick to the definition \cref{eq:afullsol} in the complex plane, this does not play a role.

The second condition in \cref{eq:propaddbranch} can only be fulfilled if
\begin{align}
	\bar{a} &> a_* \, , 
	&& \text{i.e.}
	&	 -1 &< W_0 (\bar{z}) < 0 \, .
\end{align}
These values are reached in the principal branch if
\begin{align}
	- \frac{1}{e} < \bar{z} < 0 \, .
\end{align}
Note that the phase for $\bar{z}$ where these branch cuts are reached is the same as for the branch cuts originating from the $W$-Lambert function, only the required absolute value of $\bar{z}$ is different, see \cref{eq:cond-BP}. Thus, the branch cuts originating from the explicit power laws in \cref{eq:propresummed} extend the branch cuts originating from the $W$-Lambert function. Combining these two sources of the branch cuts, we obtain branch cuts starting at $p^2 = 0$ and reaching up to $p^2 = \infty$, with the complex phase of these cuts given by \cref{eq:astdbranchstart}.

As discussed in \cref{sec:complex-plane-2-loop} this means that for small values for $\varepsilon$ only the standard branch cut at $p^2 > 0$ remains. On the other hand, for larger values of $\varepsilon$ we have inevitably additional branch cuts in the complex plane, leading to a propagator which is only analytic in separated and disconnected regions of the complex plane.

{\bf Case (iii).} A last source of non-analyticities are potential poles from self-energy corrections.  Poles arise whenever the denominator of \cref{eq:chat-2loop} vanishes in the complex plane. If they exist, and depending on their location in the complex plane, they correspond to either stable or unstable  physical (bound) states, or to stable or unstable (unphysical) tachyonic states.

For sufficiently small $\varepsilon \rightarrow 0$,  we can strictly rule out the existence of self-energy poles based on the following observation: A zero in the self energies can only arise if the coupling in the complex plane becomes of order unity. On the real axis, the coupling is bounded by the BZ fixed point $a_*$, which becomes infinitesimally small in this limit. Hence, in the complex plane, the coupling can only become of order unity if the denominator in \cref{eq:def-abar} becomes sufficiently small, which only happens at the branching point of the coupling. However, since this branching point is only reached for sufficiently large $\varepsilon$ (e.g.~\cref{fig:branching}), we conclude that there are no poles from the self energies for $\varepsilon \rightarrow 0$. For larger $\varepsilon$, the reasoning does not apply and we  investigate the poles from \cref{eq:chat-2loop} numerically below, and separately for gluons, fermions, and ghosts.

Our findings are illustrated in \cref{fig:prop_complexplots} where we show the gluon propagator in the complex plane for different values of $\varepsilon$. Depending on the value of $\varepsilon$ we observe one or several branch cuts. Additional branch cuts beyond the standard one come in pairs symmetric about the real axis. The power-law branch cuts (dashed) and the $W$-Lamber branch cuts (solid) align and are connected at the branching point where the coupling, and therefore the propagator, diverges. It follows from our previous discussion of branch cuts of running couplings in the complexified $\mu^2$ plane that branch cuts of propagators in the complexified $p^2$ plane also imply ambiguities. Similarly, choosing the branch cuts appropriately may lead to the disappearance of some of them.

\section{Spectral Functions}
\label{sec:spectral-function}
In this section, we investigate the availability of K\"all\'en-Lehmann (KL) spectral representations for gauge field, quark, and ghost propagators. The KL representation \cite{Kallen:1952zz, Lehmann:1954xi} is defined via
\begin{align} \label{eq:KS-L}
	G_\phi(p^2) = \int_{0}^\infty \frac{\mathrm d\lambda^2}{\pi}\frac{\rho_\phi(\lambda^2)}{p^2-\lambda^2} \,,
\end{align}
with
\begin{align}
	\label{eq:rho-G}
	\rho_\phi(\lambda^2) = -\lim_{\eta \to 0}\text{Im}\, G_\phi(\lambda^2 + i\eta)\,.
\end{align}
In our conventions, the timelike momenta of the propagator with the usual branch cut are on the positive real half axis and the spacelike momenta of the propagator are on the negative real half axis. On the latter, we find the standard Euclidean propagator, which is real. The propagator fulfils the relation $G_\phi(z^*) = G_\phi(z)^*$.

In unitary theories with physical particles as asymptotic states, spectral functions are positive. In contrast, for fields that are not asymptotic states even the existence of a KL representation is not guaranteed and if the spectral function exists then it may be gauge-dependent, e.g., \cite{Cyrol:2018xeq,Dudal:2019pyg, Li:2019hyv,Dudal:2020uwb, Bonanno:2021squ, Fehre:2021eob}. An example for a gauge-invariant and positive-definite spectral function is the Higgs-Higgs bound state spectral function \cite{Maas:2020kda}. In situations with complex conjugated poles or branch cuts in the complex plane of the propagator, the KL spectral representation needs to be generalised, e.g., \cite{Binosi:2019ecz, Dudal:2019aew, Hayashi:2021nnj, Hayashi:2021jju}.
Recently, a lot of progress has been achieved in the direct non-perturbative computation of spectral function, see \cite{Horak:2020eng, Fehre:2021eob, Roth:2021nrd, Horak:2021pfr, Horak:2022myj, Braun:2022mgx}. For a recent discussion of unitarity and causality criteria  for propagators, see, e.g.,~\cite{Platania:2022gtt}.

\subsection{Existence}
\label{sec:existence-2loop}
In our setting, the existence of a KL representation cannot be taken for granted, given that neither gluons nor quarks or ghosts qualify as physical asymptotic states. Still, we are interested in conditions under which their propagators can, nevertheless, be represented by a KL spectral function. A first such condition is that propagators only have a single branch cut on the positive real axis, and are analytic otherwise. At two-loop, we know that this holds for $\varepsilon < \varepsilon_{\rm branch\,cut}$, see \cref{eq:epscrit}. For $\varepsilon > \varepsilon_{\rm branch\,cut}$ this is spoiled by the existence of additional branch cuts arising from the running coupling \cref{eq:BP-two-loop}. The standard KL representation is violated, and a spectral representation would need to be modified, see for example \cite{Horak:2022myj} for a respective generalisation.

%%%%%%
\begin{figure}[t]
	\includegraphics[width=0.45\textwidth]{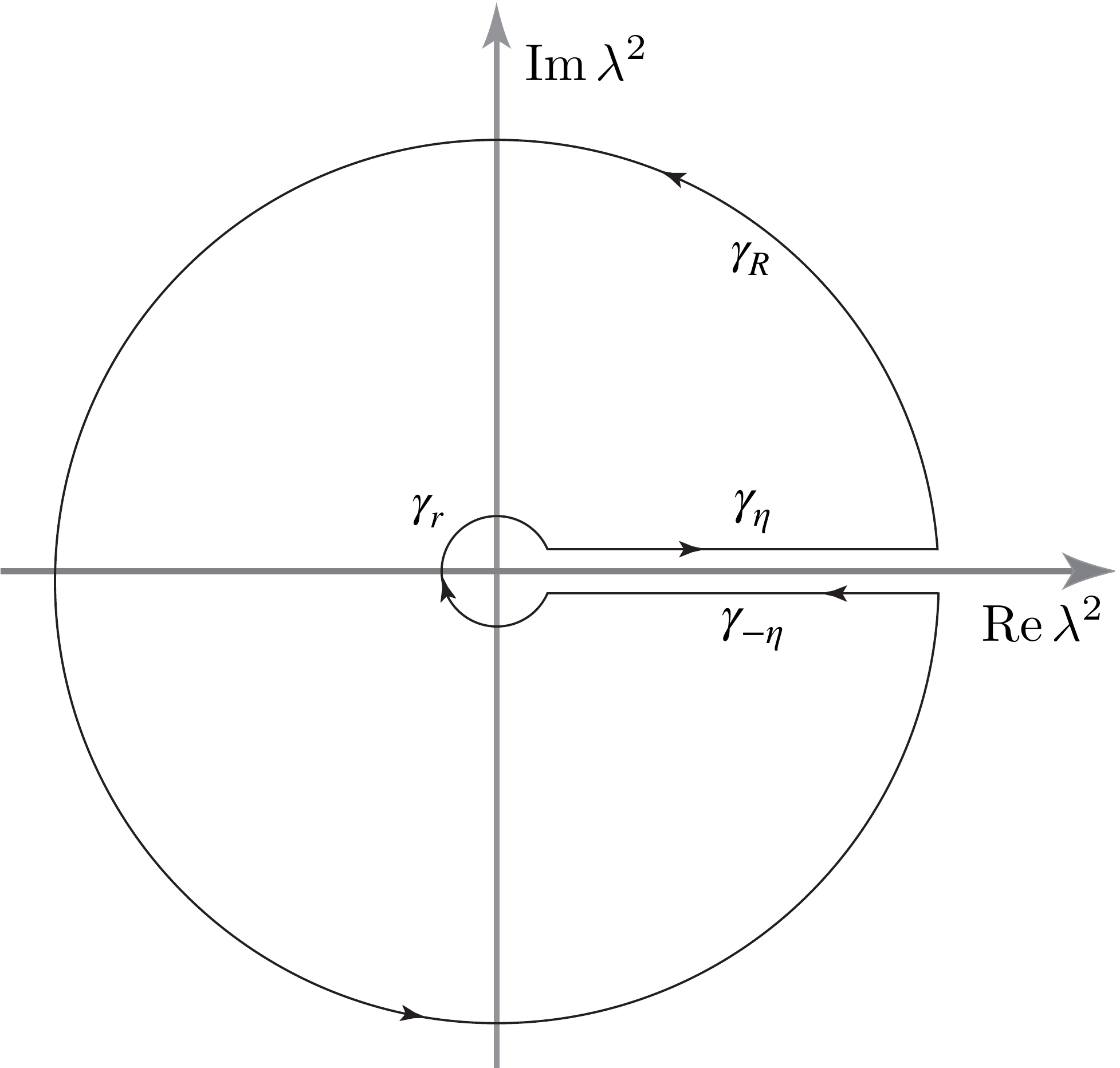}
	\caption{The contour integration of the propagator in the complex plane.}
	\label{fig:keyhole}
\end{figure}
%%%%%%

For the single branch cut, we need to check that the spectral integral is convergent. For this we integrate the propagator divided by $p^2-\lambda^2$ along the keyhole integration contour shown in \cref{fig:keyhole}. Assuming $p^2$ to have a non-trivial imaginary part or being negative, the propagator is analytic in the entire area enclosed by the integration contour except at a pole at $\lambda^2 = p^2$.\footnote{For $p^2 > 0$ we can add a small imaginary part, i.e. $p^2 \rightarrow p^2 + i \delta$ such that the pole is within the integration contour.} Using Cauchy's residue theorem, we find
\begin{align}
	\int_\Gamma \text{d} \lambda^2 \, \frac{G_\phi (\lambda^2)}{p^2 - \lambda^2} = - 2 \pi i\, G_\phi (p^2) \, .
	\label{eq:contourres}
\end{align}
with $\Gamma = \gamma_R \circ \gamma_{-\eta} \circ \gamma_r \circ \gamma_\eta$. Due to the asymptotic behaviour of the propagator in the UV, the integrand vanishes fast enough on the outer contour integration $\gamma_R$ and the integral over this part vanishes. For the integration along $\gamma_{-\eta}$ and $\gamma_\eta$, we have
\begin{align}
	\int_{\gamma_\eta \circ \gamma_{- \eta}} \! \text{d} \lambda^2 &\, \frac{G_\phi (\lambda^2)}{p^2 - \lambda^2} =  2 i \int_{0}^{\infty} \!\text{d} \lambda^2 \, \frac{\Im G_\phi (\lambda^2 + i \eta)}{p^2 - \lambda^2} 
	\label{eq:cauchyintreal}
\end{align}
in the limit $\eta\to 0$, since $G_\phi (z)^* = G_\phi (z^*)$. Lastly, we consider the small contour $\gamma_r$. Using the asymptotic properties of the propagator in the IR, see \cref{eq:proplimitIR}, we derive
\begin{align}
	\int_{\gamma_r} \text{d} \lambda^2 \, \frac{G_\phi (\lambda^2)}{p^2 - \lambda^2} = i \frac{\Omega_\phi}{p^2} \, ,
\end{align}
with
\begin{align}
	\Omega_\phi =N^{\text{IR}}_\phi\, \frac{1 - e^{- 2 \pi i \gamma_\phi (a_*)}}{1 + \gamma_\phi (a_*)}\, \lim_{\lambda^2 \rightarrow 0} \left( \frac{1}{\lambda^2} \right)^{\gamma_{\phi} (a_*)}  .
	\label{eq:Omegadef}
\end{align}
From this expression, we see that the integral on $\gamma_r$ only vanishes if either
\begin{align}
	\gamma_\phi (a_*)& \leq 0\,,
	&&\text{or}&
	\gamma_\phi (a_*) & \in  \mathbb{N} \, .
	\label{eq:cond-existence-KL}
\end{align}
Putting these results into \cref{eq:contourres} gives
\begin{align}
	G_\phi (p^2) =	- \lim_{\eta \to 0}\int_0^\infty \frac{\text{d} \lambda^2}{\pi} \, \frac{\Im \, G_\phi (\lambda^2 + i \eta)}{p^2 - \lambda^2} - \frac{1}{2 \pi} \frac{\Omega_\phi}{p^2}\, .
\end{align}
We observe that a KL spectral representation for the propagator is only fulfilled if $\Omega_\phi = 0$, which corresponds to the requirement that the spectral integral is convergent in the IR. For an asymptotically safe theory, a similar condition would appear for the contour $\gamma_R$ corresponding to a convergent spectral integral in the UV. With $\Omega_\phi = 0$, the spectral density is given in \cref{eq:rho-G}. The case $\Omega_\phi \neq 0$ would require a generalisation of the KL spectral representation, which we do not consider here.

Remarkably, the different types of KL spectral functions that emerge from the conditions in \cref{eq:cond-existence-KL} have very different properties. For $\gamma_\phi (a_*)=0$,  spectral functions have a single-particle delta peak at vanishing spectral values, which for $\gamma_\phi (a_*)=n$ with $n\in \mathbb{N}^+$ becomes the $n$-th derivative of a delta function at vanishing frequencies. For $-1 < \gamma_\phi (a_*)<0$ on the other hand, the spectral functions still diverge for $\lambda \rightarrow 0$ but with vanishing integration measure such that this pole does not contribute to the KL spectral representation. For $\gamma_\phi (a_*) \leq -1$, the spectral functions become constant or vanish as $\lambda \rightarrow 0$.

We conclude that the existence of KL spectral representations centrally depends on the anomalous dimensions of fields at the BZ fixed point, see \cref{eq:cond-existence-KL}. To leading order in $\varepsilon$ and in the Veneziano limit, they are given by 
\begin{align}
	\gamma_A (a_*) =& \, -\frac{2 \varepsilon}{25} - \frac{2 \varepsilon}{75} \xi + \order{\varepsilon^2} \,, \notag \\
	\gamma_\psi (a_*) =& \, - \frac{2 \varepsilon}{75} \xi 	 + \order{\varepsilon^2} \,, \notag\\
	\gamma_c (a_*) =& \ \ \ \, \frac{\varepsilon}{25} - \frac{\varepsilon}{75} \xi + \order{\varepsilon^2}\,.
	\label{eq:gammaexp}
\end{align}
Higher-order contributions in $\varepsilon$ also involve higher orders in $\xi$, which may become relevant quantitatively for large gauge-fixing parameters. Here, we consider small $\xi$ where these effects play no role.

Focusing on the first conditions in \cref{eq:cond-existence-KL}, $	\gamma_\phi (a_*) \leq 0$, the ranges of $\xi$ that lead to well-defined spectral function are
\begin{align}
	\text{Gluons:}&&  
	\xi &\geq  -3 + \order{\varepsilon} \, ,\notag  \\[1ex]
	\text{Quarks:}&&  
	\xi &\geq  \ \ \, 0 + \order{\varepsilon} \, ,\notag \\[1ex]
	\text{Ghosts:}&&  \xi &\geq  \ \ \, 3 + \order{\varepsilon} \, .
	\label{eq:KL-existence-2loop}
\end{align}
For these choices of gauge parameters (except for the boundary value), the spectral function consists only of a multi-particle continuum and not a single-particle delta-peak. Only at the boundary value, the spectral function contains a delta peak at vanishing frequencies corresponding to a massless particle.

The second condition in \cref{eq:cond-existence-KL}, $\gamma_\phi (a_*) \in  \mathbb{N}$, singles out $n$ fine-tuned gauge parameters for which the spectral function exists. For $\gamma_\phi (a_*) = n$, the spectral function contains a delta function with $n$ derivatives at vanishing frequencies. In the remainder of the paper, we focus on the first condition in \cref{eq:cond-existence-KL} since it allows for a range of the gauge-fixing parameter.

The conditions for existence \cref{eq:KL-existence-2loop} depend mildly on $\varepsilon$ and the order of the loop expansion. We do not display expressions at higher loop order as these are rather lengthy. However, we note that the bounds  \cref{eq:KL-existence-2loop} only receive minor corrections even at large $\varepsilon$.\footnote{For example, at $\varepsilon =1$ and $\nc=3$, the bounds become $\xi\geq -2.7$ for the gluon, $\xi\geq -0.08$ for the quark, and $\xi\geq 3.1$ for the ghosts.}

In summary, we find that sufficiently large gauge-fixing parameters excluding Landau gauge are a necessity for the existence of spectral functions for all fields, \cref{eq:KL-existence-2loop}. On the other hand, smaller gauge-fixing parameters including the Landau gauge can be chosen as long as we demand the existence of spectral functions only for the gluons and quarks. These conditions must additionally be met with the single branch-cut constraint. The latter imposes a condition on the Veneziano parameter constant being limited to small or moderate $\varepsilon <  \varepsilon_\text{branch cut}$. Importantly, the latter bound is independent of the gauge parameter, while the other constraints discussed here are gauge dependent.

%%%%%%%%
\begin{figure*}[t]
	\includegraphics[width=\linewidth]{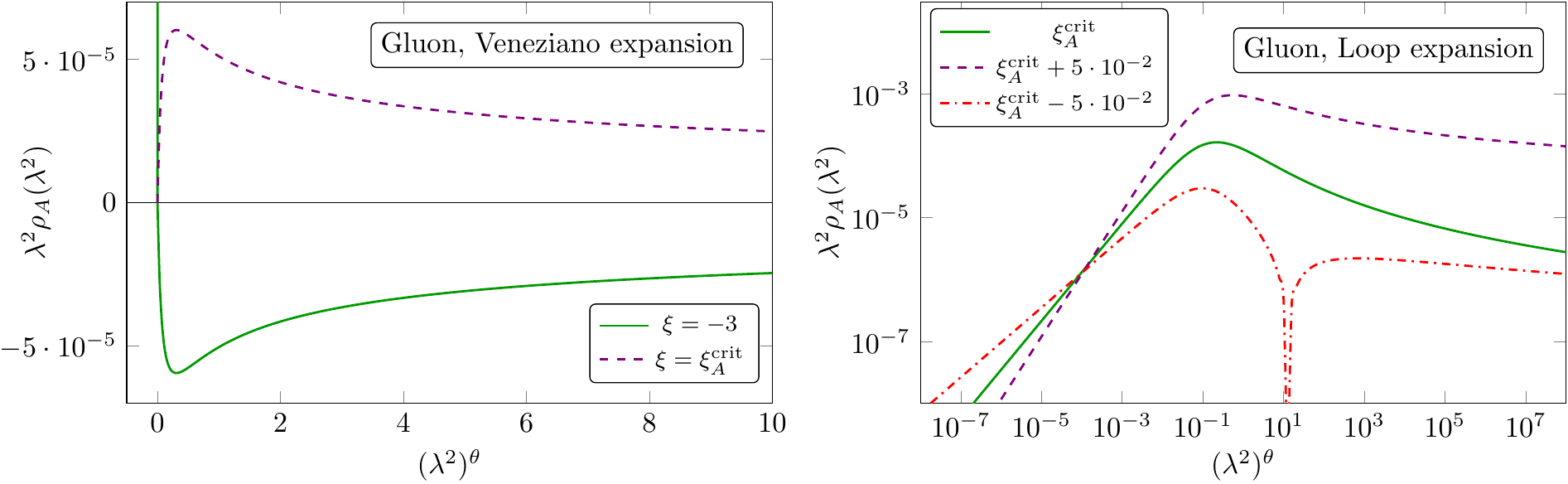}
	\caption{Gluon spectral function in the leading order Veneziano expansion (left) and in the loop expansion at two-loop order (right). In the left panel, we are using $\varepsilon = \frac{1}{10}$ and $\nc \rightarrow \infty$, while in the right panel we are using $\nc = 3$ and $\nf = 16$ ($\varepsilon = \frac{1}{6}$). We display the spectral functions as a function of $(\lambda^2)^{\theta}$.\footref{footnote_plot_arg}}
	\label{fig:spec-gluon}
\end{figure*}
%%%%%%%%

\subsection{Normalisation}
\label{sec:sumrules}
We now compute the sum rules for the spectral functions under consideration from the known UV asymptotic behaviour of the propagators \cref{eq:proplimitIR}. Evaluating the KL representation \cref{eq:KS-L} for $p^2 \rightarrow \infty$ we obtain
\begin{align}
	\lim_{p^2 \rightarrow \infty} G_\phi (p^2) = \frac{1}{p^2} \int_0^\infty \frac{\text{d} \lambda^2}{\pi} \, \rho_\phi (\lambda^2) \,.
\end{align}
With  \cref{eq:proplimitIR}, the sum rule for the spectral function reads
\begin{align}
	\int_{0}^\infty \frac{\text{d} \lambda^2}{\pi} \, \rho_\phi (\lambda^2) =  \lim_{\lambda^2 \rightarrow \infty} \left(\log \lambda^2\right)^{\gamma_\phi^{(1)}/\beta_1} \, .
	\label{eq:sumrule}
\end{align}
Here we have used that the UV asymptotic behaviour of the propagator is normalised with $N_{\text{UV},\phi} = 1$, see the discussion below \cref{eq:proplimitIR}. From this expression, we see that the spectral function only has a proper normalisation if the one-loop coefficient of the anomalous dimension vanishes, $\gamma_\phi^{(1)} = 0$. For gluons, ghosts, and quarks, they are given by
\begin{align}
	\frac{\gamma_A^{(1)}}{\beta_1} &= \frac{3 \xi + 9}{4 \varepsilon} - 1 \, , \notag \\
	\frac{\gamma_\psi^{(1)}}{\beta_1} &= \frac{3}{4 \varepsilon } \left(1 - \frac{1}{\nc^2}\right) \xi\, , \notag\\
	\frac{\gamma_c^{(1)}}{\beta_1} &= \frac{3}{8 \varepsilon } (\xi -3) \, ,
\end{align}
For each type of particle, there is one critical value $\xi^\text{crit}$ where a normalisation of its spectral function can be achieved. This value is different for all species,
\begin{align}
	\label{eq:xi-crit}
	\xi_A^\text{crit} &=-3 +\frac{4}{3} \varepsilon  \, , \notag\\
	\xi_\psi^\text{crit} &= \ \ \, 0  \, , \notag\\
	\xi_c^\text{crit} &= \ \ \, 3\, . \
\end{align}
These values are exact to all orders in $\varepsilon$ and $1/\nc$ since they only depend on the one-loop coefficient of the field anomalous dimension. For other values of the gauge-fixing parameter, the norm of the spectral function either vanishes or diverges,
\begin{align}
	\int_{0}^\infty \frac{\text{d} \lambda^2}{\pi} \, \rho_\phi (\lambda^2) = \left\{ 
	\begin{matrix}
		0 &&&& \xi < \xi_\phi^\text{crit}\,, \\[1ex]
		1&& \text{if} && \xi = \xi_\phi^\text{crit} \,, \\[1ex]
		\infty &&&& \xi > \xi_\phi^\text{crit} \,.
	\end{matrix} \right.
	\label{eq:xi-crit-norm}
\end{align}
If the spectral function exists and has a vanishing norm, it must contain positive and negative parts. 

We remark that $\xi_\phi^\text{crit}$ coincides with the gauge parameter choice for which the anomalous dimension vanishes to linear order in $\varepsilon$. This is not a coincidence and originates from the fact that the two-loop coefficient of the anomalous dimension is only relevant at the next-to-leading order in the Veneziano expansion. Hence, the zeroth-order coefficients of \cref{eq:xi-crit} and \cref{eq:KL-existence-2loop} in the Veneziano expansion must agree. This has consequences for the term $((a - a_*)/(\bar{a} - a_*))^{\gamma_\phi (a_*)/\theta}$ in the CS resummed propagator, see \cref{eq:propresummed}. In the Veneziano expansion, the eigenvalue of the BZ fixed point $\theta$ is quadratic in $\varepsilon$ while the fixed-point anomalous dimension $\gamma_\phi (a_*)$ is linear in $\varepsilon$ for general gauge parameters $\xi$. Thus, in general the exponent $\gamma_\phi (a_*)/\theta$ diverges for $\varepsilon\to 0$. The only exception is given by \cref{eq:KL-existence-2loop} up to subleading terms in $\varepsilon$, which ensures that $\gamma_\phi (a_*) = \mathcal{O}(\varepsilon^2)$. This explains why the spectral function is normalisable for $\xi_\phi^\text{crit}$. 

\subsection{Gluons}	
\label{sec:glue-veneziano}
We first discuss the gluon spectral function at the leading order in the Veneziano expansion and at the two-loop order in the loop expansion. For the Veneziano expansion, \cref{eq:afullsol} is used together with \cref{eq:BZVeneziano,eq:BZVenezianoeigv}. Furthermore, the gluon anomalous dimension and self-energy are expanded to the leading order in $\varepsilon$. In contrast, for the loop expansion, \cref{eq:afullsol} is used together with \cref{eq:BZ2loop,eq:BZ2loopeigv} and all quantities include all two-loop contributions.

The gluon propagator and its spectral function depend on the gauge parameter $\xi$ and they show qualitative differences depending on the chosen gauge. As discussed in the previous section, the gluon spectral function only exists for $\gamma_A (a_*) \leq 0$ or $\gamma_A (a_*) \in \mathbb{N}$. At leading order in the Veneziano parameter, we have $\gamma_A (a_*) \leq 0$ for $\xi \geq -3$. For $\xi = -3$, the anomalous dimension vanishes at the BZ fixed point and the spectral function contains a $\delta$-peak in the IR. This $\delta$-peak is not present for other choices of the gauge parameter. However, peaks related to derivatives of $\delta$-distributions can be found by tuning $\xi$ such that $\gamma_A (a_*) \in \mathbb{N}$. 

In the left panel of \cref{fig:spec-gluon}, we show $\lambda^2 \rho_A(\lambda^2)$ in the leading order Veneziano expansion at $\varepsilon = 1/10$ for $\xi = -3$ as well as $\xi = \xi_A^\text{crit} = -3 +\tfrac{4}{3} \varepsilon$.\footnote{We display $\lambda^2 \rho_\phi(\lambda^2)$ instead of $\rho_\phi(\lambda^2)$, since $\rho_\phi(\lambda^2)$ contains a prefactor $1/\lambda^2$ which leads to very large numerical values of $\approx 10^{5000}$ for $(\lambda^2)^\theta \ll 1$ and very small values of $\approx 10^{-1500}$ for $(\lambda^2)^\theta \gg 1$. This is avoided by multiplication with $\lambda^2$. Since we display $\lambda^2 \rho_\phi (\lambda^2)$, technically speaking we have $\lambda^2 \delta (\lambda^2) \rightarrow 0$. Nevertheless, for illustrative purposes we show the $\delta$-function as a divergence at $\lambda^2 = 0$.} For $\xi = -3$, the spectral function contains a $\delta$-peak at vanishing spectral values $\lambda^2 = 0$, while the continuum part is negative. Since $\xi = -3 <\xi_A^\text{crit}$, the spectral function must have a vanishing norm, see \cref{eq:xi-crit,eq:xi-crit-norm}, and therefore it must contain positive as well as negative contributions. In this case, the negative continuum part cancels the positive contribution from the $\delta$-peak. For $\xi = \xi_A^\text{crit}$, the spectral function is very different as it is positive but does not contain a $\delta$-peak in the IR. Since $-1 < \gamma_A (a_*) < 0$, the spectral function $\rho_A (\lambda^2)$ vanishes as $\lambda^2 \to 0$ in \cref{fig:spec-gluon}.

%%%%%%%%%
\begin{figure*}[t]
	\includegraphics[width=\linewidth]{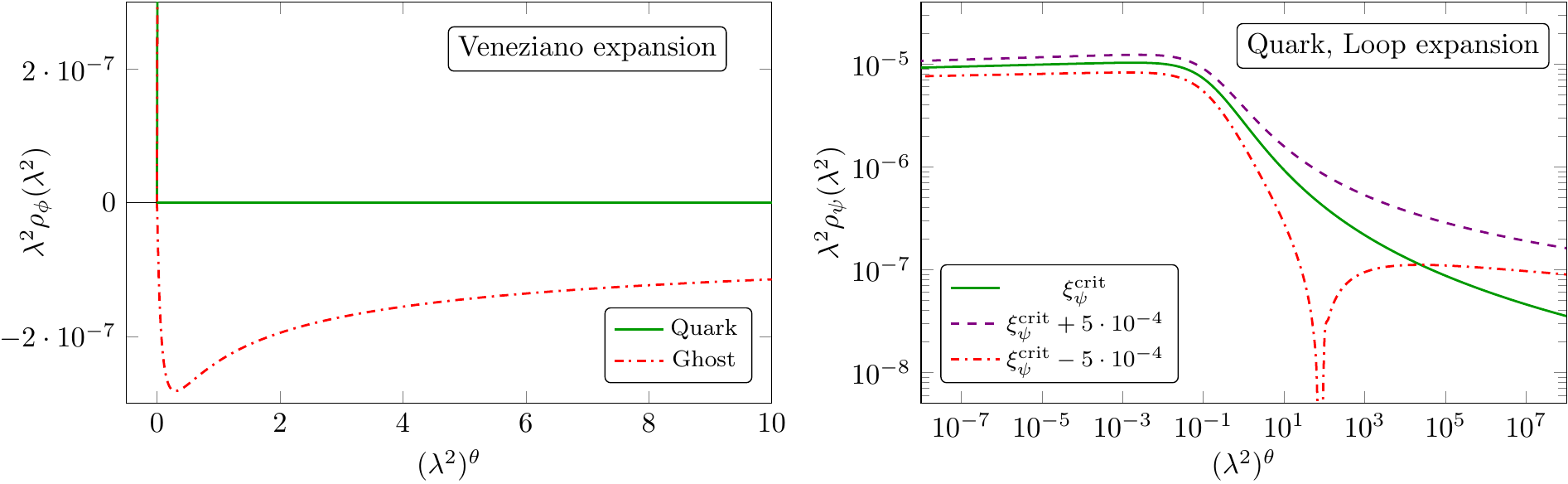}
	\caption{Spectral function of the quark and ghost in the Veneziano expansion (left) and the spectral function of the quark in the loop expansion (right). In the left panel, we use $\varepsilon = \frac{1}{10}$ and $\nc \to \infty$, while in the right panel we use $\nc = 3$ and $\nf = 16$ ($\varepsilon = \frac{1}{6}$). We display the spectral functions as a function of $(\lambda^2)^\theta$.\footref{footnote_plot_arg}}
	\label{fig:spec-quark-ghost}
\end{figure*}	
%%%%%%%%%

In the right panel of \cref{fig:spec-gluon}, we show the gluon spectral function at two-loop order. We use the values $\nc = 3$ and $\nf = 16$, which corresponds to $\varepsilon = 1/6$. The gauge parameter is chosen at its critical value $\xi = \xi_A^\text{crit} =-3 +\tfrac{4}{3} \varepsilon$ where the spectral function is normalised, as well as  slightly above and below. For $\xi = \xi_A^\text{crit}$, the gluon spectral function is positive definite and normalisable. While the latter is guaranteed by the gauge choice, the former is non-trivial. Even for very small changes below this critical gauge parameter, $\xi < \xi_A^\text{crit}$, the integral over the spectral function vanishes and the spectral function must contain positive and negative parts. For $\xi > \xi^\text{crit}_A$, the integral over the spectral function diverges which can be seen from the slower fall-off in the UV. Also in the loop expansion, we can tune the gauge parameter such that the spectral function has a $\delta$-peak at vanishing frequencies. This happens however not at $\xi=-3$ as in the Veneziano expansion due to the subleading contributions that are taken into account in the loop expansion.

Next, we comment on the existence of poles from self-energy corrections. 
With increasing $\varepsilon$ we find no poles from self-energies for any $N_c$ 
at least up until $\varepsilon_{\rm branch\,cut}$ where additional branch cuts arise.\footnote{\label{poles}The query for poles becomes ambiguous as soon as additional branch cuts are present, the reason being  that cuts can always be chosen in such a way that poles are moved to a different sheet in the complex plane,  the sole exception being poles on the real axis.}

We briefly discuss the comparison to the gluon spectral function in the confining QCD region. There the spectral function is analytically continued from the spacelike momenta of the gluon propagator which has been obtained by functional or Lattice methods in Landau gauge $\xi=0$, see, e.g., \cite{Cyrol:2018xeq, Ilgenfritz:2017kkp, Fischer:2017kbq, Binosi:2019ecz, Horak:2021syv}. Then, the gluon spectral function typically has a vanishing norm: it is negative in the IR and UV, with a positive spectral density around the confinement scale. In this respect, it is similar to the red dash-dotted curve in the right panel of \cref{fig:spec-gluon}, where the gauge parameter is chosen below its critical value $\xi<\xi_A^\text{crit} = -3 +\tfrac{4}{3} \varepsilon$. In comparison to the Landau gauge, we have $\xi_A^\text{crit}<0$ for matter content of the BZ window but $0<\xi_A^\text{crit}$ for standard QCD and this is the main contributor to the similarities between the spectral functions.

\subsection{Quarks and Ghosts}
\label{sec:glue-quark-ghost}
We now discuss the spectral functions of the quark and ghost fields. Also, these spectral functions are gauge-dependent and in \cref{eq:KL-existence-2loop,eq:xi-crit} we show the values of the gauge parameter for which these spectral functions exist or are normalisable. These values for the gauge parameter are different for the three species under consideration.

For the quark spectral function to exist in the Veneziano expansion, we require $\xi \geq 0$ or $\xi$ tuned such that $\gamma_\psi (a_*) \in \mathbb{N}$. At this order, the gauge for which the quark anomalous dimension at the fixed point vanishes and for which the spectral function is normalisable coincide and is given by the Landau gauge, $\xi = 0$. Thus, we can obtain a well-defined normalisable quark spectral function featuring a $\delta$-peak in the IR. Inserting leading order expressions in the Veneziano limit, we obtain for the quark propagator in Landau gauge,
\begin{align}
	G_\psi = \frac{1}{p^2} \frac{\mathcal{N}_\psi}{1 + \Pi_\psi^{(1)} \bar{a} + \mathcal{O} \left( \varepsilon^2 \right)} \, .
\end{align}
The exponents present in the CS resummed propagator \cref{eq:propresummed} become trivial in this gauge. Furthermore, the two-loop self-energy only gives subleading corrections which should be neglected at leading order in the Veneziano expansion. Lastly, we observe that $\Pi_\psi^{(1)} \propto \xi$. Hence, in the Landau gauge, the one-loop self-energy becomes trivial and we are left with a free propagator for the quark field. In consequence, the spectral function $\rho_\psi (\lambda^2)$ only contains a $\delta$-peak at $\lambda^2 = 0$ and vanishes everywhere else. This is displayed in the left panel of \cref{fig:spec-quark-ghost}. It is remarkable that the quark propagator appears to be free within an interacting theory. We emphasise that this is only present in the leading-order Veneziano expansion and higher orders in $\varepsilon$ will inevitably introduce a multi-particle continuum. 

For the ghost spectral function in the Veneziano expansion, the existence condition \cref{eq:KL-existence-2loop} requires $\xi \geq 3$, unless we tune the anomalous dimension such that $\gamma_c (a_*) \in \mathbb{N}$. As for the quarks, at leading order in the Veneziano expansion, this coincides with the critical value $\xi_c^\text{crit}$ for which the ghost spectral function is normalisable. Thus, this value of the gauge parameter gives rise to a normalisable ghost spectral function featuring a $\delta$-peak in the IR. In contrast to the quarks, the spectral function is non-trivial in this gauge. This is due to the fact that $\Pi_c^{(1)} = -\nc + \mathcal{O} (1/\nc)$ for $\xi = 3$. Hence, the propagator and the ghost spectral function pick up a nontrivial contribution from the one-loop self-energy in the CS resummed propagator. The resulting spectral function in the left panel of  \cref{fig:spec-quark-ghost} shows a $\delta$-peak in the IR and gives a negative continuum part thereafter. Thus, even though the gauge parameter is chosen such that the spectral function is normalisable, it does not imply that the spectral function is positive as we observe that the continuum part is strictly negative. 

In the right panel of \cref{fig:spec-quark-ghost}, we show the quark spectral function at two-loop order. We use the values $\nc = 3$ and $\nf = 16$, which corresponds to $\varepsilon = 1/6$. The gauge parameter is chosen at its critical value $\xi = \xi_\psi^\text{crit} =0$ where the spectral function is normalised, as well as  slightly above and below. The spectral functions are very similar to the gluon case in the loop expansion, c.f., the right panel of \cref{fig:spec-gluon}. For $\xi = \xi_\psi^\text{crit}$, the quark spectral function is positive definite and normalisable. For gauge parameters just below the critical value, $\xi < \xi_\psi^\text{crit}$, the integral over the spectral function vanishes and the spectral function contains positive and negative parts. For gauge parameters above the critical value $\xi > \xi^\text{crit}_\psi$, the integral over the spectral function diverges. A remarkable difference to the gluon case is that the quark spectral function approaches zero much slower in the IR. The reason is that the quark anomalous dimension evaluated at the BZ fixed point is tiny. In the Veneziano expansion, it would be actually vanishing for $\xi=\xi_\psi^\text{crit}$ and in the loop expansion, it is only modified by subleading corrections. We do not display the ghost spectral functions in the loop expansion since they are very similar to the quark spectral function.

Once more we have looked for poles from self energies. As for the gluons, we find that there are none in the quark propagator. However, we observe a tachyonic pole in the ghost propagator as soon as  $\varepsilon > 2.19$ at $N_c=3$ and for $\varepsilon > 2.25$ in the Veneziano limit.

As a final remark, we comment on how results for gauge-variant propagators and spectral functions can be used to extract gauge-invariant, physical information. In \cite{Capri:2016aqq, Capri:2016gut, Capri:2017abz}, and in the context of the Gribov-Zwanziger action, gauge-invariant gluon and quark fields have been constructed out of gauge-variant ones by providing them with a non-local dressing. The thereby constructed propagators have been found to be gauge invariant, and, curiously, identical to the standard quark and gluon propagators in the Landau gauge. At weak coupling, we expect that the findings of \cite{Capri:2016aqq, Capri:2016gut, Capri:2017abz} apply equally in our setting. Then, the green curve displayed in the right panel of \cref{fig:spec-quark-ghost} precisely corresponds to the quark spectral function in the Landau gauge. It is positive and normalisable, and equal to the spectral function of a non-local and gauge-invariant quark field. Notice though that while the gluon spectral function in the Landau gauge is positive, unlike the quark one, it is not normalisable. As an aside, we also observe that non-normalisable spectral functions genuinely arise for fields which asymptote to strictly positive anomalous dimensions in the UV, see \cref{sec:sumrules}.

%%%%%
\begin{table*}[t!]
	\aboverulesep = 0mm
	\belowrulesep = 0mm
	\addtolength{\tabcolsep}{1pt}
	\setlength{\extrarowheight}{1pt}
	\begin{tabular}{`c?cc`}
		\toprule
		\rowcolor{Yellow}
		\bf Veneziano limit & \;\, $\bm{\varepsilon_{\rm branch\,cut}}$  \;\, & $\bm{\varepsilon_{\rm max}}$  \\
		2-loop & 2.3285 & 2.8846 \\
		\rowcolor{LightGray}
		3-loop & 2.7240 & 3.5889 \\		
		4-loop & 2.7265 & 3.4601 \\		
		\rowcolor{LightGray}
		5-loop & --  & \;\, 1.1774$^*$ \;\,\\		
		\;\, 5-loop Pad\'e [1,3] \;\, & --  & 2.0646  \\		
		\rowcolor{LightGray}
		5-loop Pad\'e [2,2] & --  &  1.4609$^*$ \\		
		5-loop Pad\'e [3,1] & --  & 0.7203$^*$ \\		
		\bottomrule
	\end{tabular}
	\hspace{2cm}
	\begin{tabular}{`c?cc`}
		\toprule
		\rowcolor{Yellow}
		$\bm{\nc =3}$ & \;\, $\bm{\varepsilon_{\rm branch\,cut}}$ \;\, & $\bm{\varepsilon_{\rm max}}$  \\
		2-loop & 2.2723 & 2.8158 \\
		\rowcolor{LightGray}
		3-loop & 2.6798 & 3.5520 \\		
		4-loop & 2.6817 & 3.0538$^*$ \\		
		\rowcolor{LightGray}
		5-loop & -- & \;\, 1.2019$^*$ \;\, \\		
		\;\, 5-loop Pad\'e [1,3] \;\, & --  & 2.2183  \\		
		\rowcolor{LightGray}
		5-loop Pad\'e [2,2]  & --  & 1.6993$^*$ \\
		5-loop Pad\'e [3,1] & --  & 0.7304$^*$  \\		
		\bottomrule
	\end{tabular}
	\caption{\label{tab:xi-max}
		Values of $\varepsilon$ where branch cuts occur ($\varepsilon_\text{branch\,cut}$) and where the BZ fixed point disappears ($\varepsilon_\text{max}$) at different loop orders for $\nc\to\infty$ (left) and for $\nc=3$ (right). The values of $\varepsilon_\text{max}$ marked with an asterisk occur due to a FP merger.}
\end{table*}
%%%%%

%%%%%%%%%%%%%%%%%%%%%%%%%%%%
\section{Higher Loops}
\label{sec:higher-orders}
Thus far we have studied the propagators and spectral functions in the two-loop limit where we have full analytic control. In this section, we address the effects of higher-loop orders. We show how explicit analytical solutions for the running coupling can be found in the Veneziano expansion, and also report results from numerical and implicit solutions.

Further, $\beta$-function and field anomalous dimensions are known up to five-loop order \cite{Herzog:2017ohr, Chetyrkin:2017bjc} while the finite parts of the propagators have been computed up to four-loop order \cite{Ruijl:2017eht}. These contributions become important for finite values of the Veneziano parameter $\varepsilon$, and when exploring the size of the conformal window. Therefore, we derive expressions for propagators and spectral functions to higher orders and study the convergence at finite  $\varepsilon$.

\subsection{Running  Coupling from Higher Loops}
\label{sec:Veneziano-nloop}
Here, we show that explicit analytic solutions for the running gauge coupling can be found at any order in the Veneziano expansion, generalising the two-loop result \cref{eq:afullsol}. We explain the underlying systematics and illustrate the construction in the Veneziano limit. 

We begin by noting that the left- and right-hand sides of \cref{eq:runningcoupling1} start out at order $\varepsilon$ and $\varepsilon^3$, respectively, indicating that the RG running is at least as slow as $\varepsilon^2$. We can make \cref{eq:runningcoupling1} more amenable to a systematic solution as a power series in $\varepsilon$ by performing a change of variables from $a(\mu^2)$ to a rescaled version $\hat a(z)$, with
\begin{align}\label{eq:aahat}
	a(\mu^2) = \frac{\varepsilon}{\nc} \hat{a}(z)  \,,
\end{align}
and $z=z(\mu^2)$ as given in  \cref{eq:zdef}. The prefactor accounts for the fact that $a\sim \varepsilon$ at a fixed point, while the substitution $\mu^2\to z(\mu^2)$ accounts for the parametrically slow running  $\lesssim \theta(\varepsilon)$. In combination, the original $\mu^2 {\dd a}/{\dd \mu^2}$ beta function \cref{eq:runningcoupling1}   turns into
\begin{align}
	z \dderiv{\hat{a}(z)}{z} =  \sum_{n = 1}^\infty \hat\beta_n \,  \hat{a}^{n + 1}\,,
	\label{eq:a_ven_exp}
\end{align}
where the rescaled loop coefficients 
\begin{align}
	\label{eq:beta_hat}
	\hat\beta_n=  \frac{\beta_n}{\nc^n} \, \frac{{\varepsilon}^{\!n}}{\theta(\varepsilon)}
\end{align}
now involve the universal scaling exponent $\theta(\varepsilon)$ as defined in \cref{eq:theta}, and whose $\varepsilon$-expansion is given in \cref{eq:thetaexp,eq:thetacoeffs}. In the Veneziano limit, any $\nc$-dependence drops out and the rescaled loop coefficients $\hat\beta_n$  are polynomials in $\varepsilon$,
\begin{align}
	\label{eq:beta_hat_nm}
	\hat\beta_n(\varepsilon)=\sum_m\hat\beta_{n,m}\varepsilon^m\,,
\end{align}
and whose leading order terms scale  as $\hat\beta_{1},\hat\beta_{2}\sim {\cal O}(1)$ and $ \hat\beta_{n\ge 3}\sim {\cal O}(\varepsilon^{n-2})$ with $\varepsilon$.

The virtue of the rescaled $\beta$-function \cref{eq:a_ven_exp} is that its left- and right-hand sides both start out at order unity. Hence, expanding the running gauge coupling as a series in the Veneziano parameter,
\begin{align}
	\hat{a}(z) = \sum_{n = 1}^\infty \hat{a}_{n} (z)\,\varepsilon^{n-1} \,,
	\label{eq:e_ven_exp_ansatz}
\end{align}
leads to a hierarchy of differential equations for the coefficient functions $\hat{a}_n$ which can be solved recursively. Using $t=\ln z$ we find
\begin{align} \label{eq:dan}
	\partial_t \hat a_1 &=   \hat\beta_{1,0} \,  \hat{a}_1^2+ \hat\beta_{2,0} \,  \hat{a}_1^3\,,\nonumber \\
	\partial_t \hat a_{n\ge 2} &=   ( 2\hat\beta_{1,0}  \hat{a}_1 + 3\hat\beta_{2,0} \,  \hat{a}_1^2)\hat a_n+I_n[\{\hat a_{i<n}\}] \,,
\end{align}
where the inhomogeneous terms $I_n$ are independent of $ \hat a_n$ and only depend on the functions $\hat a_{ i<n}(z)$.  The leading order differential equation for $\hat a_1$ is equivalent to the two-loop $\beta$-function and  integrated in terms of the $W$-Lambert function
\begin{align}\label{eq:a0}
	\hat{a}_1(z) = \frac{4}{75} \frac{1}{1 + W_0 (z)} \, .
\end{align}
All higher order corrections $\hat a_n$ can be found by solving first-order linear differential equations whose inhomogeneous term depends on the lower order solutions $\hat a_{i<n}$, see \cref{eq:dan}. For example, $I_2=\hat\beta_{1,1} \,  \hat{a}_1^2+\hat\beta_{2,1} \,  \hat{a}_1^3+\hat\beta_{3,1} \,  \hat{a}_1^4$, and similarly to higher order. Using \cref{eq:dan} with \cref{eq:a0}, the next-to-leading order correction is found to be
\begin{align}\label{eq:a1}
	\hat a_2(z) &= \frac{2192}{421875} \frac{1 + \frac{975}{274} W_0(z)\left[C + W_0(z) \right]}{[1+W_0(z)]^3} \, ,
\end{align}
where $C$ is an integration constant determined by the initial condition $a (\mu_0^2) = a_0$. At $z=0$, we observe that the solutions  \cref{eq:a0,eq:a1} match the exact fixed-point coefficients at the corresponding order in $\varepsilon$, see  \cref{eq:acoeffs}. We also have computed the next-to-next-to-leading order correction analytically, though the result is not given explicitly as it is rather lengthy without offering additional insights.

We emphasize that the explicit solution for the running gauge coupling can naturally be extended beyond the Veneziano limit. The key is to stick to $\varepsilon$ as the central expansion parameter but to retain the parametric dependence on $\nc$. In the above, this turns the polynomials $\hat\beta_n(\varepsilon)$  and the scaling exponent $\theta(\varepsilon)$ in \cref{eq:beta_hat}, and the coefficients $\hat\beta_{n,m}$ in \cref{eq:beta_hat_nm}  into $\nc$-dependent quantities, which then feed into the solutions $\hat a_n$, but without structurally changing the hierarchy \cref{eq:dan}. Thus, the solutions at finite $\nc$  smoothly approach the solution in the Veneziano limit for sufficiently small $\varepsilon$, as they must.

\subsection{Gauge Coupling in the Complex Plane}
\label{sec:complex-plane-coupling-n-loop}
In the analytic $W$-Lambert solution at two-loop order, we observed branching points in the complex plane for $\varepsilon> \varepsilon_\text{branch\,cut} =  2.3285$ at $\nc\to\infty$, see \cref{fig:branching}. We will now extend this result to higher-loop orders. We can do this either analytically in the Veneziano expansion, or numerically in the loop expansion. We choose the latter since we expect $\varepsilon_\text{branch\,cut}$ to remain at rather large values at higher orders. The existence of additional branch cuts in the former is solely tied to the argument of the $W$-Lambert function and the dependence of $z$ on $\mu^2$. This suggests a generalisation of \cref{eq:BP-two-loop} to higher loop orders by replacing the two-loop eigenvalue with the appropriate eigenvalue at higher loops. 

We check the existence of branch points with numerical integration curves in the complex plane of $a(\mu)$. We choose a closed half-circle integration contour with the straight line slightly above the real axis and the half-circle closing in the upper half. In the case of a branch cut, this closed integration contour returns a non-vanishing imaginary part and in the case of a pole in the complex plane, the integrated solution gives the residuum of the complex pole. We evaluate this complex integral as a function of the Veneziano parameter $\varepsilon$ for each loop order. The value $\varepsilon_\text{branch\,cut}$ is defined as the lowest $\varepsilon$ at which branch cuts or complex conjugated poles appear in the complex plane. 

We show the results of our numerical investigation in \cref{tab:xi-max} for $\nc\to\infty$ and $\nc=3$ together with values of the Veneziano parameter $\varepsilon_\text{max}$ where the BZ fixed point vanishes. From two- to three-loop order, both $\varepsilon_\text{max}$ and $\varepsilon_\text{branch\,cut}$ increase, while from three- to four-loop order, both values barely change at all. The reason is that the four-loop coefficient of the $\beta$ function is strongly suppressed in this regime of $\nf$ and $\nc$. A remarkable difference at four-loop order is that for $\nc=3$ the BZ fixed point disappears with a fixed-point merger, instead of a divergence of the fixed-point value. The suggested generalisation of \cref{eq:BP-two-loop} to higher loops by adapting the eigenvalue is in good agreement with \cref{tab:xi-max} provided we use the full eigenvalue and do not expand in $\varepsilon$ or $\nc$.

The inclusion of the five-loop order has the biggest impact on $\varepsilon_\text{max}$ and $\varepsilon_\text{branch\,cut}$. The BZ fixed point vanishes already at very small values of $\varepsilon$  via a fixed-point merger. The existence of branch cuts in the complex plane does not provide a stronger bound on $\varepsilon$. At five-loop order, we also include the Pad\'e approximants $[n,m]$  defined by
\begin{align}
	\beta_{[n,m]}	 =   a^2 \frac{\sum_{i=0}^n \gamma_i a^i }{1 + \sum_{i=1}^m \delta_i a^i}\,,
\end{align}
where the coefficients $\gamma_i$ and $\delta_i$ are determined such that the perturbative expansion agrees with the original $\beta$-function up to order $a^{n+m+1}$. In \cref{tab:xi-max}, we show our results for the five-loop Pad\'e approximants $[1,3]$,  $[2,2]$, and $[3,1]$. In all cases, the fixed point vanishes before additional branch-cuts show up in the complex plane of the coupling. The average value $\varepsilon_\text{max}$ of the five-loop Pad\'e approximants is larger than that of the standard five-loop $\beta$-function, which might hint that the standard five-loop gives a too small value for $\varepsilon_\text{max}$. Our results are well compatible with the Pad\'e estimations from \cite{DiPietro:2020jne} where it was estimated that $\varepsilon_\text{max}>1.3$ for $\nc \to\infty$ and $\varepsilon_\text{max}>1.5$ for $\nc =3$.

\subsection{Callan-Symanzik with Higher Loops}
The resummation of the propagator via the CS equation at $n$-loop order is performed in full analogy to \cref{sec:CS-prop}. The first hurdle is to solve the integral in \cref{eq:callimpprop}, for which we use the partial fraction decomposition of the integrand,\footnote{Note that this result in general only holds if the anomalous dimension and the $\beta$-function are used at the same loop order.}
\begin{align}
	\frac{\gamma_\phi (a)}{\beta (a)} = \frac{\gamma_\phi^{(1)}}{\beta_1 a} + \sum_i \frac{\gamma_\phi (a_{i,*})}{\theta_i} \frac{1}{a - a_{i,*}} \, .
\end{align}
In this expression, the sum goes over all non-trivial fixed points $a_{i,*}$ of the $\beta$-function and $\theta_i$ denotes their eigenvalues 
\begin{align}\label{eq:thetai}
	\theta_i = \left. \frac{\partial\beta_a}{\partial a} \right|_{a = a_{i,*}}\,.
\end{align}
The general result of the integration boils down to
\begin{align}
	\int_{a}^{\bar{a}}\! \frac{\text{d} a'}{\beta (a')} \, \gamma_\phi (a')
	=& \, \frac{\gamma_\phi^{(1)}}{\beta_1} \log( \frac{\bar{a}}{a} ) \notag\\
	& + \sum_i \frac{\gamma_\phi (a_{i,*})}{\theta_i} \log( \frac{\bar{a} - a_{i,*}}{a - a_{i,*}}).
	\label{eq:CZint-gen}
\end{align}
The asymptotic UV behaviour at the Gaussian fixed point and the IR behaviour at the BZ fixed point are extracted as follows. Close to the Gaussian fixed point, the first term in \cref{eq:CZint-gen} dominates and reproduces the known UV behaviour of the propagator. Close to any of the non-trivial fixed points, the term in the product of \cref{eq:CZint-gen} corresponding to the fixed point dominates. In particular, in the IR close to the BZ fixed point we obtain the correct IR behaviour including corrections to the two-loop behaviour found previously.

The integration constant is again obtained via comparison to the perturbative propagator. Our final result for the CS resummed propagator is
\begin{align}
	G_\phi 
	&= \frac{1}{p^2} \frac{\mathcal{N}_\phi}{1 + \sum_{n = 1}^\infty \Pi^{(n)}_\phi \bar{a}^n} \left( \frac{a}{\bar{a}} \right)^{\!\frac{\gamma_\phi^{(1)}}{\beta_1}} \prod_i \left( \frac{\bar{a} - a_{i,*}}{a - a_{i,*}} \right)^{\!\frac{\gamma_\phi (a_{i,*})}{\theta_i}}\!  .
	\label{eq:propresummedgen}
\end{align}
The large and small momentum asymptotics of the propagator agree structurally with  \cref{eq:proplimitIR}, amended by improved values for the fixed point, eigenvalue, anomalous dimensions, and self-energies. The normalisation factor $\mathcal{N}_\phi$ is  again chosen such that $N_{\text{UV},\phi}=1$, see \cref{eq:proplimitIR}.

%%%%%%%
\begin{table*}[t]
	\aboverulesep = 0mm
	\belowrulesep = 0mm
	\addtolength{\tabcolsep}{1pt}
	\setlength{\extrarowheight}{1pt}
	\begin{tabular}{`c?ccc`}
		\toprule
		\rowcolor{Yellow}
		$\bm{\varepsilon^{\rm conv}|_{\nc =3}}$& \;\, \bf Gluons  \;\, & \bf  Quarks  & \bf  Ghosts \\
		{2-loop} & $0.20$ & \;\, $6.7 \cdot 10^{-3}$ \;\, & \;\, $1.0 \cdot 10^{-2}$\;\,  \\
		\rowcolor{LightGray}
		{3-loop} & $1.5$ & $0.12$ & $1.73$ \\		
		{4-loop} & $0.95$ & $0.19$ & $0.65$ \\
		\bottomrule
	\end{tabular}
	\hspace{1cm}
	\begin{tabular}{`c?ccc`}
		\toprule
		\rowcolor{Yellow}
		$\bm{\varepsilon^{\rm conv} |_{1/\nc =0}}$	& \;\, \bf  Gluons  \;\, & \bf Quarks  & \bf Ghosts \\
		{2-loop}	& $0.22$ & \;\, $-$ \;\, & \;\, $1.1 \cdot 10^{-2}$\;\,  \\
		\rowcolor{LightGray}
		{3-loop}	& $1.6$ & $8.2 \cdot 10^{-2}$ & $0.81$ \\
		{4-loop}	& $1.0$ & $0.30$ & $0.83$ \\
		\bottomrule
	\end{tabular}
	\caption{\label{tab:criteps}
		Values for $\varepsilon$ at which the next loop order leads to a change of the anomalous dimension by more than $10\%$ at $\xi=\xi_\phi^\text{crit}$. The left table shows the results for $\nc = 3$ and the right table for the Veneziano limit $\nc \rightarrow \infty$.
	}
\end{table*}
%%%%%%%

\subsection{Existence of Spectral Functions}
\label{sec:existence-nloop}
The existence of the spectral function is related to the value of $\gamma_\phi(a_*)$, see \cref{eq:cond-existence-KL} in \cref{sec:existence-2loop}. We display the values of $\xi$ for which the spectral function stops to exist, corresponding to  $\gamma_\phi(a_*) =0$. These values are written in the Veneziano expansion with subleading $\nc$ corrections, $\xi^\text{no-spec}_\phi = \xi^\text{no-spec}_{\phi,0}  + \xi^\text{no-spec}_{\phi,1}/\nc^2 + \dots$, and for the leading contribution we find
\begin{align}
	\xi^\text{no-spec}_{A,0} (\varepsilon)&\approx  -3+0.267 \varepsilon +0.241 \varepsilon^2  -0.0436 \varepsilon^3 \,, \notag \\[1ex]
	\xi^\text{no-spec}_{\psi,0} (\varepsilon)&\approx  \ \  0.0145 \varepsilon^2  - 0.00518 \varepsilon^3 \,, \notag \\[1ex]
	\xi^\text{no-spec}_{c,0} (\varepsilon)&\approx \ \ 3 -0.0267 \varepsilon -0.168 \varepsilon^2+ 0.0540 \varepsilon^3 \,.
	\label{eq:xinospec}
\end{align}
This expression includes contributions up to $\varepsilon^3$ or equivalently up to five-loop order. For $\xi < \xi^\text{no-spec}_\phi$, the spectral function of the field $\phi$ does suffer from a divergence in the IR and does not exist, with the exception of certain fine-tuned values where $\gamma_\phi(a_*)\in \mathbb{N}$, see \cref{eq:cond-existence-KL}. We compare \cref{eq:xinospec} to $\xi^\text{crit}_\phi$ where the spectral function is normalisable as given in \cref{eq:xi-crit}. For gluons and ghosts, we observe that $\xi^\text{no-spec}_\phi < \xi^\text{crit}_\phi$ and thus a normalisable spectral function exists for these species in the Veneziano limit. This is not the case for the quarks since for $\varepsilon \rightarrow 0$, we have $\xi_\psi^\text{no-spec} > \xi_\psi^\text{crit}$. A normalisable spectral function for the quarks field, therefore, does not exist in the Veneziano limit as it requires infinite $\nc$.

For a better understanding of the behaviour of the quark spectral function, we consider finite $\nc$ corrections. The next-to-leading order contributions are given by
\begin{align}
	\xi^\text{no-spec}_{A,1} (\varepsilon)&\approx \ \ 0.117 \varepsilon +0.262 \varepsilon^2  - 0.137 \varepsilon^3 \,, \notag \\[1ex]
	\xi^\text{no-spec}_{\psi,1} (\varepsilon)&\approx  - 0.04 \varepsilon +   0.0201 \varepsilon^2  +0.00999 \varepsilon^3 \,, \notag \\[1ex]
	\xi^\text{no-spec}_{c,1} (\varepsilon)&\approx -0.0117 \varepsilon -0.232 \varepsilon^2- 0.0117 \varepsilon^3 \,.
	\label{eq:xinospecsub}
\end{align}
Most importantly, the quark term $\xi^\text{no-spec}_{\psi,1}$ has a negative linear $\varepsilon$ contribution at $1/\nc^2$, while it starts with a positive $\varepsilon^2$ contribution in $\xi^\text{no-spec}_{\psi,0}$. Hence, for finite $\nc$, it might be possible to tune $\varepsilon$ such that $\xi^\text{no-spec}_\psi < \xi^\text{crit}_\psi$ and a normalisable quark spectral function exists. However, for a proper analysis we should take into account that field multiplicities are integers. The smallest possible Veneziano parameter $\varepsilon_\text{min}$ for each $\nc$ is given by
\begin{align}
	\varepsilon_\text{min} = \left\{ \begin{matrix}
		\frac{1}{\nc} & \text{if} & \nc = \text{even}\,, \\[1ex]
		\frac{1}{2 \nc} & \text{if} & \nc = \text{odd}\,. \\
	\end{matrix} \right.
	\label{eq:mineps}
\end{align}
Using this in \cref{eq:xinospec,eq:xinospecsub} for the quark, we find
\begin{align}
	\xi_{\psi}^\text{no-spec} (\varepsilon_\text{min})&\approx \left\{ \begin{matrix}
		\frac{0.0145}{\nc^2} - \frac{0.0452}{\nc^3} & \text{if} & \nc = \text{even}\,, \\[1ex]
		\frac{0.00363}{\nc^2} - \frac{0.0206}{\nc^3}  & \text{if} & \nc = \text{odd}\,. \\
	\end{matrix} \right.
\end{align}
The leading contribution to $\xi_{\psi,\varepsilon_\text{min}}^\text{no-spec}$ is always positive and thus we still find $\xi_\psi^\text{no-spec} > \xi_{\psi,\varepsilon_\text{min}}^\text{crit}$, meaning that a normalisable quark spectral function is not well-defined. For small $\nc$, $\xi_{\psi,\varepsilon_\text{min}}^\text{no-spec}$ becomes negative but that is outside of the validity of the expansion.

As a further check, we performed a scan for integer $\nc$ up to including $\nc = 15$ and all integer $\nf$ such that $\varepsilon > 0$. There are only two combinations which lead to a negative quark anomalous dimension consistently at different loop orders. These are $(\nc,\,\nf) = (3,\,16)$ and $(\nc,\,\nf) = (5,\,27)$. At five-loop order, the quark anomalous dimension is also negative for $(\nc,\,\nf) = (2,\,10)$, however, it is positive at four-loop order. We conclude that all other integer choices and in particular the Veneziano limit lead to a positive quark anomalous dimension. As such, a normalisable quark spectral function is absent for these settings. Instead, the quark spectral function may only be defined for gauge parameters not equal to the critical value $\xi_\psi^\text{crit}$. The spectral function then has either a diverging or a vanishing norm.

%%%%%%%%
\begin{figure*}[t]
	\includegraphics[width=\textwidth]{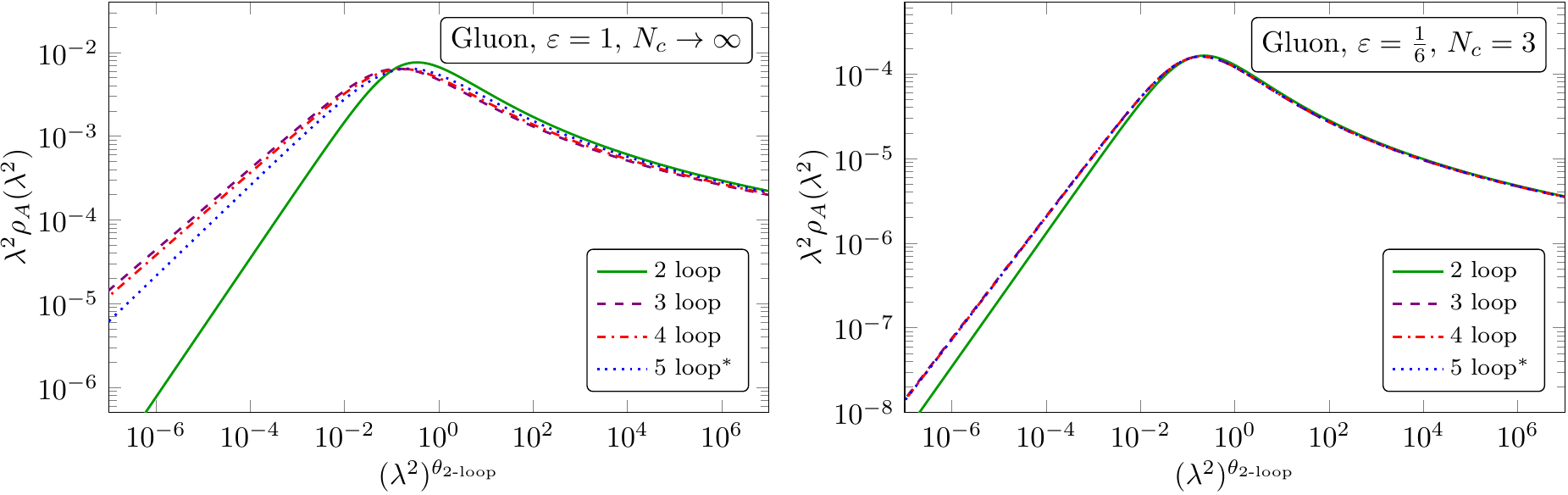}
	\caption{Gluon spectral functions in the loop expansion at $\xi =\xi_A^\text{crit}$ for $\nc \rightarrow \infty$ with $\varepsilon =1$ (left) and at $\nc = 3$ with $\nf = 16$ ($\varepsilon = 1/6$) (right). The asterisk at the highest expansion order indicates that the highest order contribution the finite parts is missing.}
	\label{fig:spectral-nloop-glue}
\end{figure*}
%%%%%%%

\subsection{Convergence}
\label{sec:conv}
Next, we assess the radius of convergence of the loop expansion in terms of the Veneziano parameter $\varepsilon$. Remarkably, the convergence depends strongly on the field species. A central quantity that enters the computation of the spectral function is the field anomalous dimension evaluated at the BZ fixed point, $\gamma_{\phi}(a_*)$. We evaluate $\gamma_{\phi}(a_*)$ for each field at each loop order and determine the maximal $\varepsilon$ for which the change due to the next loop order remains below $10\%$. The results are shown in \cref{tab:criteps} for $\nc=3$ and $\nc\to\infty$, and we discuss our findings for the different particle types one by one.

A relatively stable picture is obtained for gluons. Allowing a change of $10\%$, we are able to consider Veneziano parameters of the order of $\varepsilon \approx 1$ suggesting a radius of convergence roughly in line with the expected size of the conformal BZ window \cref{tab:xi-max}. This result is qualitatively independent of sending $\nc \rightarrow \infty$ or $\nc$ being finite.

For quarks, this observation is rather different. Going from two-loop to three-loop, the $10\%$ change of the anomalous dimension is ill-defined for $\nc \rightarrow \infty$ since the three-loop result does not match the two-loop anomalous dimension at $\xi=\xi_\psi^\text{crit}$ for $\varepsilon \rightarrow 0$. This is because
\begin{align}
	\left. \gamma_\psi (a_*) \right|_{\xi = \xi_\psi^\text{crit}} = \mathcal{O}\! \left(\varepsilon^3\right) + \mathcal{O}\! \left(\frac{1}{\nc}\right) .
\end{align}
The universal $\varepsilon^3$ contribution is only found from the three-loop perturbation theory onwards and has not converged yet at two-loop.\footnote{For $\xi = \xi_\psi^\text{crit}$, the $\varepsilon^3$ is already obtained at three-loop since the one-loop coefficient of the anomalous dimension vanishes.} Since the $\varepsilon^2$ contribution vanishes, the two-loop result of the loop expansion does not contain the correct leading-order contribution of the Veneziano expansion. This has already been discovered in \cref{sec:glue-quark-ghost} where we noted that the Veneziano expansion of the quark propagator at two-loop does not give rise to any non-trivial corrections and we are left with a one particle pole.

At finite $\nc$, the quark anomalous dimension at the BZ fixed point carries a universal and non-vanishing $\varepsilon^2$ piece. Thus, the leading order of the Veneziano expansion is contained in the two-loop result for finite $\nc$ and we obtain a well-defined $\varepsilon_\text{2-loop}^\text{conv}$ for quarks, see \cref{tab:criteps}. For $\nc = 3$, the numerical value for $\varepsilon^\text{conv}_\text{2-loop}$ is rather small and not reachable if $\nf$ is an integer.

Higher loop corrections enhance the observed convergence for the quarks at $\nc \rightarrow \infty$ as well as finite $\nc$. Including up to five-loop corrections, the anomalous dimension does not become as stable as for the gluons, however. Asking for a $10\%$ change of the anomalous dimension, we find $\varepsilon^\text{conv}_\text{4-loop} \approx 0.3$.

A possible explanation for the reduced convergence of the quark anomalous dimension at the BZ fixed point might be related to the non-existence of a normalisable quark spectral function in the Veneziano limit. For its existence we require $\gamma_\psi (a_*) < 0$. While this is true at two-loop for $\nc \rightarrow \infty$, it can only be achieved at higher loop orders if $ \varepsilon \ll 1/\nc$ and $\nf$ not an integer except for a few cases given at the end of \cref{sec:existence-nloop}. Thus, the two-loop approximation shows a qualitative difference compared to higher loop orders possibly leading to the observed convergence properties.

Finally, the convergence of the ghosts lies somewhat in between the quarks and ghosts. Even though the two-loop approximation converges substantially worse than higher-order results, it still gives a good convergence up until $\varepsilon \approx 0.01$. Including higher-order corrections, the convergence becomes similar to the observed convergence of the gluons. As for the quarks, we can interpret the bad convergence of the two-loop approximation by the fact that the ghost spectral function in the Veneziano expansion at two-loop does not include all non-trivial contributions which are present at higher loop orders. In \cref{sec:glue-quark-ghost} we have seen that the only non-trivial correction comes from the self-energy, while other corrections, in particular corrections due to $\gamma_c (a_*) \neq 0$ at $\xi = \xi_c^\text{crit}$ at higher loop orders are absent. Starting from three-loop onwards, these corrections are taken into account.

In summary, we have established that the convergence of field anomalous dimensions at the BZ fixed point and of spectral functions with the conformal parameter $\varepsilon$ depend strongly on the type of field. A reliable estimate for the radius of convergence or the non-perturbative size of the BZ conformal window cannot  be obtained in this manner. To put this outcome into  perspective, we compare results with a similar type of analysis that has been performed recently in 4d supersymmetric gauge-matter theories with an interacting conformal fixed point \cite{Bond:2022xvr}. In supersymmetry, the chiral superfield anomalous dimensions are known exactly, and the quality of perturbative approximations can be checked. Good convergence is observed at weak coupling while at strong coupling,  convergence depends more substantially on the type of matter field. In particular, examples are found where three loop is a good approximation for some superfield anomalous dimensions but not for others, and many cases exist where three loops  fail miserably in  estimating the non-perturbative radius of convergence  \cite{Bond:2022xvr}. We conclude that the disparities in the convergence for different fields observed in this study appear to be a genuine feature of 4d QFTs with conformal fixed points \cite{Bond:2016dvk, Bond:2018oco}, supersymmetric or otherwise, rather than a specific feature of this theory.

%%%%%%%%
\begin{figure*}[t]
	\includegraphics[width=\textwidth]{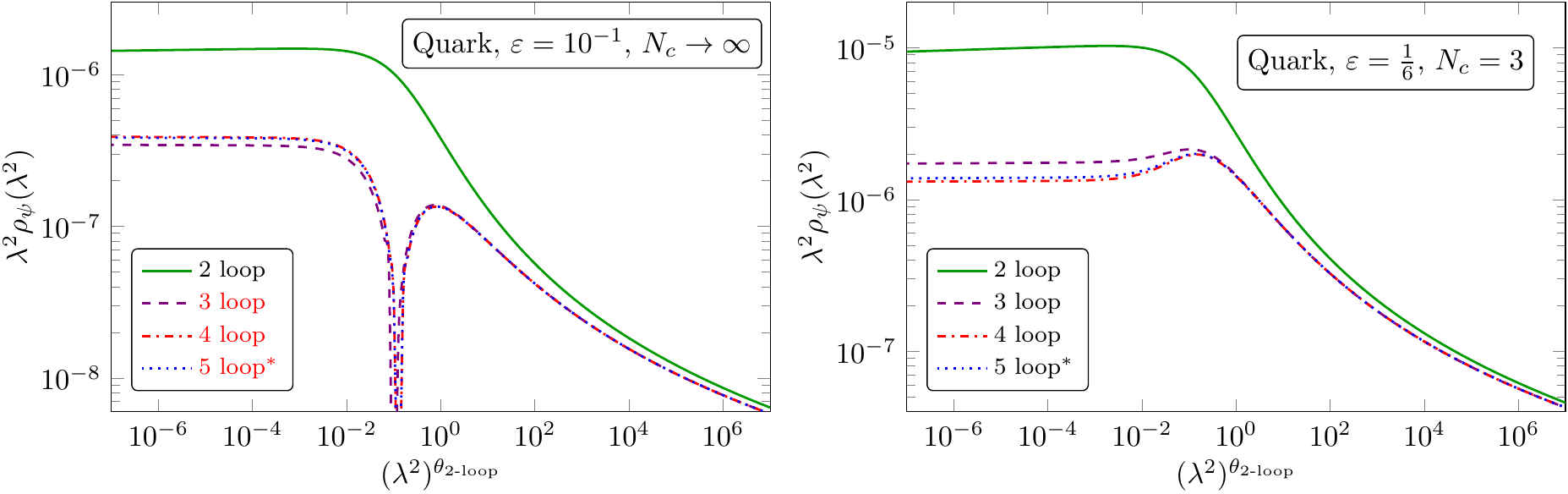}
	\caption{Quark spectral functions in the loop expansion at $\xi =\xi_\psi^\text{crit}$ for $\nc \rightarrow \infty$ with $\varepsilon =1/10$ (left) and at $\nc = 3$ with $\nf = 16$ ($\varepsilon = 1/6$) (right). The asterisk at the highest expansion order indicates that the highest order contribution the finite parts is missing. The red colour that the spectral function is not integrable in the IR.}
	\label{fig:spectral-nloop-quark}
\end{figure*}
%%%%%%%

%%%%%%%%
\begin{figure*}[t]
	\includegraphics[width=\textwidth]{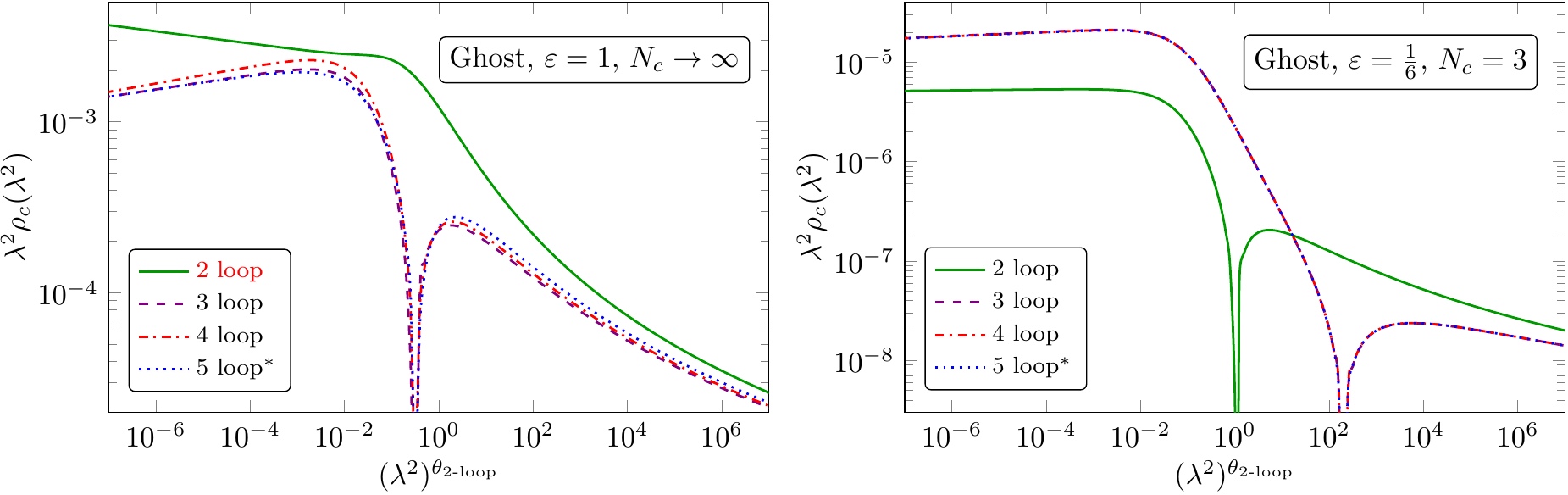}
	\caption{Ghost spectral functions in the loop expansion at $\xi =\xi_c^\text{crit}$ for $\nc \rightarrow \infty$ with $\varepsilon =1$ (left) and in the loop expansion at $\nc = 3$ and $\nf = 16$ ($\varepsilon = 1/6$) (right). The asterisk at the highest expansion order indicates that the highest order contribution the finite parts is missing. The red colour that the spectral function is not integrable in the IR.}
	\label{fig:spectral-nloop-ghost}
\end{figure*}
%%%%%%%

\subsection{Spectral Functions at Higher Loops}
\label{sec:spec-nloop}
The spectral functions of the fields are obtained from \cref{eq:propresummedgen} via numerical integration of the gauge coupling in the complex plane. From our analysis in \cref{sec:complex-plane-coupling-n-loop}, we know that there are no additional branch cuts and poles in the complex plane for small $\varepsilon$ and therefore we know that this numerical integration is justified. We compute the spectral functions of the gluon, quark, and ghost at two-, three-, four-, and five-loop order and consider finite $\nc$ as well as $\nc \rightarrow \infty$. The five-loop order only includes the five-loop contribution in the beta function and anomalous dimensions, while the five-loop self-energy corrections are missing. We expect that these are subleading since also the lower-order finite parts only play a subleading role.

In \cref{fig:spectral-nloop-glue}, we show the gluon spectral functions as a function of $(\lambda^2)^{\theta_\text{2-loop}}$. The left panel displays the loop expansion in the $\nc \rightarrow \infty$ limit and we have chosen $\varepsilon=1$ around the convergence value $\varepsilon_\text{conv}$, see \cref{tab:criteps}, which is well within the conformal BZ window, see \cref{tab:xi-max}. In the right panel of \cref{fig:spectral-nloop-glue}, we show the loop expansion at $\nc = 3$ and $\nf = 16$ corresponding to $\varepsilon = 1/6$. The gauge parameter is always chosen on the critical value according to \cref{eq:xi-crit} leading to normalisable spectral functions. The gluon spectral function is positive definite and converges well. Only the leading-order is significantly different from the higher orders at small spectral values.

In \cref{fig:spectral-nloop-quark}, we show the loop expansion of the quark spectral functions in the $\nc \rightarrow \infty$ limit at  $\varepsilon=1/10$ and at  $\nc = 3$ and $\nf = 16$. We have chosen a smaller $\varepsilon$ in the $\nc\to\infty$ limit compared to the gluon due to the slower convergence of the quark spectral function. We observe that the leading-order behaviour differs qualitatively from higher orders, in particular for $\nc \rightarrow \infty$. There, the higher-order quark spectral functions do not exist due to the wrong sign of the fixed-point anomalous dimension, see \cref{eq:cond-existence-KL}. In consequence, the spectral functions shown in the left panel of \cref{fig:spectral-nloop-quark} have been obtained from the discontinuity of the propagator via \cref{eq:rho-G} but integrating over them with \cref{eq:KS-L} does not give back the propagator. This strong difference between the leading and higher orders is explained by the fact that the quark spectral function does not obtain any non-trivial corrections at the leading order in the Veneziano expansion.

In \cref{fig:spectral-nloop-ghost}, we show the loop expansion of the ghost spectral functions for $\nc \rightarrow \infty$ at  $\varepsilon=1$ in the left panel and at  $\nc = 3$ and $\nf = 16$ in the right panel.  The ghost spectral function shares a similar property in that it also is not positive definite beyond two-loop. However, the ghost anomalous dimension stays negative at loop orders higher than two, only the two-loop result becomes positive. Thus, we find the opposite of the quark case and in the left panel of  \cref{fig:spectral-nloop-ghost} the ghost spectral function is well-defined for all but the two-loop result. Furthermore, this means that we obtain a normalisable, but not positive definite ghost spectral function for $\varepsilon = 1$. The sign change in the ghost spectral function can be avoided by using a smaller $\varepsilon$.

We furthermore checked the existence of poles from the self energies. We restricted our search to values of $\varepsilon < \varepsilon_{\rm branch\,cut}$ given in \cref{tab:xi-max}, for reasons detailed in footnote \ref{poles}. In this regime, we find that the quark and gluon propagators never have a pole from the self energies at any loop order and at any $N_c$. For the ghost propagator, we already observed a tachyonic pole from  self energies at two-loop for $\varepsilon > 2.2$, see \cref{sec:glue-quark-ghost}. At three- and four-loop order,  instead, we find a pair of complex conjugated poles. These poles show up for $\varepsilon> 2.66$ (three-loop, $N_c=3$), $\varepsilon> 2.71$ (three-loop, $N_c=\infty$), $\varepsilon> 2.62$ (four-loop, $N_c=3$), and $\varepsilon> 2.68$ (four-loop, $N_c=\infty$), see \cref{tab:self-energy-poles}. Overall, and for any $N_c$ and loop order,  we find 
\begin{equation}\label{eq:poles}
1-\frac{\varepsilon_{\rm poles}}{\varepsilon_{\rm branch\,cut}}\lesssim\mathcal{O}(10^{-2} )\,,
\end{equation}
stating that poles in the ghost sector only arise very close to the boundary $\varepsilon_{\rm branch\,cut}$ where additional branch cuts arise. At this point perturbation theory has become unreliable, and the conformal window ceases to exist.

In summary, up to including four-loop orders, we have established that neither the gluon nor the quark self-energy corrections  lead to bound state (or tachyonic) poles in their propagators. Our result is valid for any $\varepsilon<\varepsilon_{\rm branch\,cut}$ (\cref{tab:xi-max}) covering  the entire BZ conformal window. Further, poles in the self energies of the ghost propagator (\cref{tab:self-energy-poles}), only arise at strong coupling and very close to the onset of branch cuts (\cref{tab:xi-max}) where perturbation theory becomes unreliable. We therefore conclude that the theory does not offer hints for bound states for any UV-free trajectory connecting with the BZ fixed point in the IR.

%%%%%%%
\begin{table}[t]
	\aboverulesep = 0mm
	\belowrulesep = 0mm
	\addtolength{\tabcolsep}{1pt}
	\setlength{\extrarowheight}{1pt}
	\begin{tabular}{`c?cc`}
		\toprule
		\rowcolor{Yellow}
		$\bm{\varepsilon_{\rm \bf poles}}$& \;\, \bf  $\bm{N_c = 3}$ \;\, &  \;\, \bf $\bm{N_c=\infty}$   \;\, \\
		\;\,  2-loop  \;\, 	& $2.19$ & $2.25$\\
		\rowcolor{LightGray}
		3-loop	& $2.66$ & $2.71$  \\
		4-loop	& $2.62$ & $2.68$  \\
		\bottomrule
	\end{tabular}
	\caption{
		As soon as $\varepsilon\ge \varepsilon_{\rm poles}$, the ghost propagator develops poles through self-energy corrections. At two-loop, the pole is tachyonic, while at three- and four-loop order, we find a pair of complex conjugated poles. Self-energy corrections do not induce  poles in the quark and gluon propagator.
	}
	\label{tab:self-energy-poles}
\end{table}
%%%%%%%

\subsection{Functional Renormalisation Group}
\label{sec:non-perturbative}
We use the functional renormalisation group (fRG) as a non-perturbative method to compare to the perturbative analysis. The main goal is to understand how the CS resummation that was necessary for perturbation theory to get rid of large logarithms is reflected in another method. Therefore we only compute here the flow of the gluon two-point function in a simple approximation. 

The fRG is based on a flow equation for the scale-dependent effective action $\Gamma_k$, the Wetterich equation \cite{Wetterich:1992yh, Ellwanger:1993mw, Morris:1993qb}
\begin{align}\label{eq:Wetterich}
	\partial_t\Gamma_k =  \frac12 \text{Tr}\, \mathcal{G}_k\,\partial_t R_k \,.
\end{align}
The flow equation interpolates between the classical action $S$ at the initial scale $k_\text{in}$ and the full quantum effective action $\Gamma$ in the limit $k\to 0$. The RG time $t$ is defined as $t = \ln k/k_\text{in}$. The regulator $R_k$ implements the Wilsonian integration-out of momentum shells and $\mathcal{G}_k$ is the full field dependent propagator, $\mathcal{G}_k =(\Gamma_k^{(2)}+ R_k)^{-1}$, where $\Gamma_k^{(n)} = \delta^n \Gamma_k /\delta \phi^n$.

The fRG equation is a one-loop equation but takes into account higher-loop orders via resummation. For example, at the initial scale $k_\text{in}$ the two-point function $\Gamma_{k}^{(2)}$, which enters on the right-hand side of \cref{eq:Wetterich}, is given by the classical two-point function $S^{(2)}$. However, already after one RG step at $k_\text{in} -\delta k$, the two-point function is modified and includes the quantum corrections from the integrated out RG step. This emphasises that the propagator at vanishing RG scale $\mathcal{G}_{k=0}$ is already the resummed object that we want to compare to and it does not contain large logarithms.

The flow equation for the gluon two-point function is obtained from \cref{eq:Wetterich} via two field derivatives. Since Euclidean signature is used in the flow equation, we only compute and compare the gluon propagator for space-like momenta at $k=0$. The flow of the gluon two-point function depends on all propagators $\mathcal{G}_\phi$, the three- and four-point vertices, $\Gamma_{k}^{(3)}$ and $\Gamma_{k}^{(4)}$, as well as on the regulator function $R_k$. To simplify our computation as much as possible, we approximate the vertices with the classical vertices, $\Gamma^{(n)} = S^{(n)}$, and use the perturbative two-loop trajectory \cref{eq:afullsol} as an input. Furthermore, we choose the regulator proportional to the two-point function 
\begin{align}
	\label{eq:Regulator} 
	R_k(p) &= \Gamma^{(2)}_k(p)\, r_k(x)\,,
	&
	\text{with} \quad x&=\frac{p^2}{k^2}\,, 
\end{align}
and use a Litim-type cutoff  \cite{Litim:2000ci,Litim:2001up} for the shape function 
\begin{align}
	\label{eq:Litim-cutoff}
	r_k (x) =  (1/x -1) \, \Theta(1-x)\,.
\end{align}
With this setup, we can evaluate the diagrams numerically for all space-like momenta. We parameterise the transversal part of the gluon two-point function with 
\begin{align}
	\Gamma^{(AA)}_{T,k} = Z_{A,k}(p\,) p^2\,,
\end{align}
where $Z_{A,k}(p)$ is the momentum dependent gluon wave-function renormalisation, whose flow follows straightforwardly
\begin{align}
	\label{eq:flow-Zk}
	\partial_t Z_{A,k} (p) = \frac{\partial_t\Gamma^{(AA)}_{T,k} }{p^2}  \,,
\end{align}
where the right-hand side is given by one-loop diagrams. We integrate \cref{eq:flow-Zk} on the perturbative two-loop trajectory from \cref{eq:afullsol} to $k=0$ where $Z_A(p) = Z_{A,k=0}(p)$ is the full wave-function renormalisation and the full propagator function is given by $G_A = 1/(Z_A(p)p^2)$, which we can compare to the CS resummed propagator given in \cref{eq:propresummed}. The overall normalisation of the wave function renormalisation is arbitrary and we choose the same normalisation as in the perturbative computation, see \cref{eq:NUV-and-NIR}.

%%%%%%
\begin{figure}[t]
	\includegraphics[width=\linewidth]{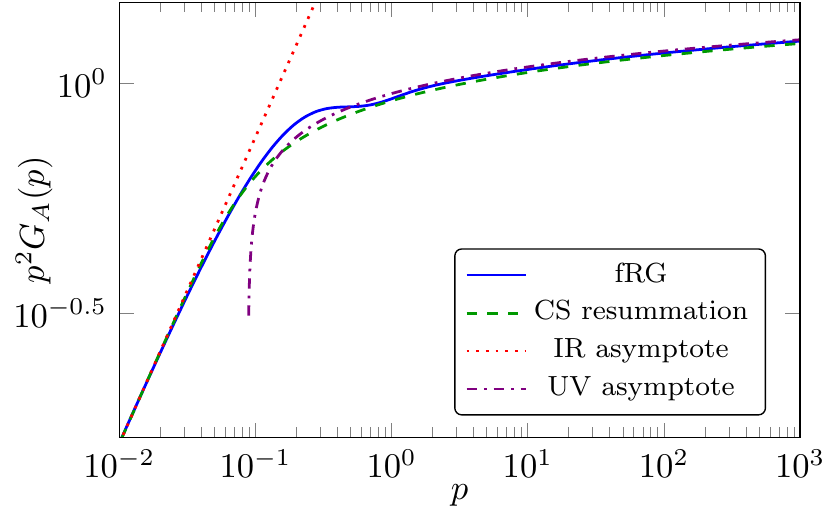} 
	\caption{Gluon propagator computed from the fRG in comparison with the perturbative result for $\varepsilon = \frac52 $, $\nc = 100$, and $\xi = 1$.}
	\label{fig:ZA-fRG}
\end{figure}
%%%%%

The resulting gluon propagator is displayed in \cref{fig:ZA-fRG} and compared to the perturbative resummation at one-loop, i.e., setting $\Pi_A^{(2)}=0$ and $\gamma_A(a_*) = \gamma_A^{(1)} a_*$ in \cref{eq:propresummed}. The approaches agree remarkably well and we only find small differences due to the different regularisation schemes. It has to be remarked that the UV and IR asymptotic behaviour has to be identical as it is fully determined by the universal one-loop anomalous dimension and $\beta$-function in the UV, and by the fixed-point anomalous dimension in the IR, see \cref{eq:proplimitIR}. Only the normalisations $N_\text{IR}$ and $N_\text{UV}$ are non-universal and depend on the unphysical choice $a(\mu)$, see \cref{eq:NUV-and-NIR}. In \cref{fig:ZA-fRG}, we have chosen $a(\mu)\approx  0.138 \, a_*$ in the perturbative computation such that the UV and IR normalisations match. Consequently, only in the region around $p \approx 1$ do we encounter differences due to the regularisation scheme. Most notably the fRG computation displays a bump in the propagator while the CS resummation is a monotonic interpolation between the IR and UV asymptotes. In the fRG, the individual contributions from the quark and the gauge loops are monotonic in their momentum dependence but the overlay creates the bump at the transition scale. In the CS, a non-trivial structure can arise from the self energy-contribution or additional terms coming from the integral of the anomalous dimension \cref{eq:CZint-gen} at higher loop orders. While these non-trivial structures are present at higher loop orders, the resulting modifications around $p \approx 1$ are very small and would not be visible in \cref{fig:ZA-fRG}. In particular, we do find a visible bump using CS at higher loop orders in contrast to the fRG result.

%%%%%%%%%%%%%%%%%%%%%%%%%%%%
\section{Extensions}
\label{sec:general}
We have seen that the running of couplings in the plane of complexified RG momentum  $\mu^2$  offers important insights into the existence of a spectral representation for field propagators because branch cuts of the former translate directly into branch cuts of the latter. In previous sections, we have exploited this link to find branch cut conditions (\cref{tab:xi-max}) both analytically (at two-loop) and numerically (for higher loops), and to link these to the size of universal scaling exponents \cref{eq:BP-two-loop}.

In this section, we extend the scope and derive criteria for branch cuts in more general theories, expressed again in terms of universal scaling exponents. This will be done for general perturbative $n$-loop $\beta$-functions as well as for suitably resummed expressions,  and their solutions.

\subsection{Branch Cut Conditions from Higher Loops}
\label{sec:CS-higher-loops}
We consider a general quantum field theory with a single coupling of canonical  mass dimension $d_a$ whose perturbative  $n$-loop  $\beta$-function is given by
\begin{align}
	\mu^2 \frac{\dd a}{\dd \mu^2} \equiv \beta (a) = d_a a + \beta_1 a^2 + \dots +\beta_n a^{n+1}\,.
	\label{eq:beta-gen}
\end{align}
If $d_a$ can be taken as a small parameter (such as $\epsilon=4-d$ in the conventional $\epsilon$-expansion \cite{Wilson:1971dc}), the flow \cref{eq:beta-gen} can be integrated analytically, and order by order in $d_a$, using the method described in \cref{sec:Veneziano-nloop}. Here, we solve this differential equation implicitly using the partial fraction decomposition for the inverse of the $\beta$-function,
\begin{align}
	\label{eq:partial-fraction}
	\frac{1}{\beta (a)} = \frac{1}{d_a  a}  + \sum_i \frac{1/\theta_i}{a - a_{i,*}} \, .
\end{align}
Once more, $a_{i,*}$ denote all non-trivial zeros of the $\beta$-function, i.e.~all non-trivial fixed points, and $\theta_i$ the corresponding eigenvalues \cref{eq:theta}. For \cref{eq:partial-fraction}, we assumed that there are no degenerate fixed points. The differential equation can now be integrated,
\begin{align}
	\frac1{d_a} \log \frac{a}{a_0} + \sum_i \frac{1}{\theta_i} \log \frac{a - a_{i,*}}{a_0 - a_{i,*}} = \log \frac{\mu^2}{\mu_0^2} \, .
\end{align}
Using the eigenvalue sum rule
\begin{align}
	\label{eq:identity}
	\sum_i 1/\theta_i = - 1/d_a \,,
\end{align}
which holds true at any finite loop order (see \cref{app:betaidentity}), the result simplifies into
\begin{align}
	\sum_i \frac{1}{\theta_i} \log \frac{1 - a_{i,*}/a}{1 - a_{i,*}/a_0} = \log \frac{\mu^2}{\mu_0^2} \, .
	\label{eq:a-implicit}
\end{align}
For dimensionless couplings $d_a=0$, the sum rule \cref{eq:identity} reads instead 
\begin{align}
	\sum_i 1/\theta_i = \beta_2/\beta_1^2 \,,
\end{align}
and \cref{eq:a-implicit} has an additional term $1/\beta_1 (1/a_0 - 1/a )$ on the left-hand side to account for the logarithmic running at the Gaussian fixed point.

Next, we apply the implicit function theorem to \cref{eq:a-implicit}. Assuming a function $F$ depending on two variables, $F (a, \mu^2)$, it states that for any point $(\hat{a}, \hat{\mu}^2)$ where $F(a, \mu^2) |_{\hat{a}, \hat{\mu}^2}$ is analytic with
\begin{align}
	\label{eq:impl-function-conditions}
	F (\hat{a}, \hat{\mu}^2) &= 0 \, ,
	&&\text{and}&
	\left. \pderiv{F}{a} \right|_{\hat{a}, \hat{\mu}^2} &\neq 0 \, ,
\end{align}
there is an analytic function $a(\mu^2)$ in the neighbourhood of $(\hat{a}, \hat{\mu}^2)$ fulfilling $F (a, \mu^2) = 0$. Thus, everywhere where the given requirements of the implicit function theorem are fulfilled, we can solve for an analytic function $a(\mu^2)$. Possible singularities and branching points can only occur at points where one of the requirements in \cref{eq:impl-function-conditions} is violated. However, we emphasise that there are not necessarily non-analyticities if \cref{eq:impl-function-conditions} is violated. We can only exclude non-analyticities if  \cref{eq:impl-function-conditions} is fulfilled, and find candidate non-analyticities if it is violated.

In our case, the function $F(a, \mu^2)$ is given by 
\begin{align}
	F(a, \mu^2) =\log \frac{\mu^2}{\mu_0^2} - \sum_i \frac{1}{\theta_i} \log \frac{1 - \frac{a_{i,*}}{a}}{1 - \frac{a_{i,*}}{a_0}} \, .
\end{align}
As expected, $F(a, \mu^2)$ is not analytic at $a = 0$. The two-loop running coupling can be applied locally around $a = 0$ at general loop orders. Using the properties of the two-loop running coupling, we conclude that the non-analyticity at $a = 0$ corresponds to a branching point. This is in agreement with common expectations of a vanishing radius of convergence of the perturbative series and also the occurrence of a branch cut in the propagator for timelike momenta.

The point $a = 0$ is not the only non-analyticity appearing in the running coupling. There are additional points due to
\begin{align}
	\label{eq:F-der}
	\pderiv{F}{a} = \frac{1}{\beta(a)} = 0 \, .
\end{align}
This can only be fulfilled if $a \rightarrow \infty$ since we are assuming $\beta(a)$ to be a polynomial, see \cref{eq:beta-gen}. To see where $a \rightarrow \infty$ can be fulfilled, we solve $F(a, \mu^2) = 0$ for $\mu^2$ in the limit of $a \rightarrow \infty$. This leads to the equation
\begin{align}
	\prod_i \left( 1 - \frac{a_{i,*}}{a_0} \right)^{\!-1/\theta_i} = \frac{\mu^2}{\mu_0^2} \, .
	\label{eq:branches}
\end{align}
Each complex $\mu^2$ that solves this equation is a candidate for a branch cut or singularity. The determination of which kind of singularities can be found at these points is more difficult than at $a = 0$. This is because all higher-order derivatives of $F$ by $a$ vanish at $a \rightarrow \infty$, thus, the Jacobian vanishes to all orders. In the two-loop case, the points with $a \rightarrow \infty$ corresponds to $W(z) = -1$, where the $W$-Lambert function has a branching point. This behaviour might generalise to higher orders and the points fulfilling \cref{eq:branches} might lead to additional branching points in the complex plane. 

We now translate the condition \cref{eq:branches}, into a strict relation for the eigenvalues. In a first step, we assume that all the fixed points are real and we write the factors $(1 - a_{i,*}/a_0)^{-1/\theta_i}$ as $r_i e^{-i \varphi(a_{i,*})/\theta_i}$ where $\varphi(a_{i,*})$ is the complex phase of the factor $(1 - a_{i,*}/a_0)$ and $r_i$ is the absolute value of the total factor. If $a_{i,*}<a_0$ then the factor is positive and complex phase is zero, $\varphi(a_{i,*}) = 0$, while if  $a_{i,*}>a_0$ then the factor is negative and $\varphi(a_{i,*}) = \pi$. Note, that we are using the principal branch of the roots. Furthermore, the factors $\mu_0$ and $e^{1/(\beta_1 a_0)}$ are real and positive. Thus we end up with the equation
\begin{align}
	\label{eq:total-phases}
	\mu^2  = C \exp(-i \pi \sum_i^{a_{i,*}>a_0} \frac1{\theta_i}) \,,
\end{align}
where the sum runs \emph{only} over the eigenvalues belonging to fixed points with $a_{i,*}>a_0$. The complex phase of $\mu^2$ is between $-\pi$ and $\pi$, and therefore the equation has no solutions in the principal branch if 
\begin{align}
	\label{eq:master-equation}
	\left| \sum_i^{a_{i,*}>a_0} \frac1{\theta_i} \right| >1 \,.
\end{align}
In this case, we have no branching points in the complex plane.  Conversely, branching points \emph{might} appear if the sum over the inverse eigenvalues is smaller than unity. For beta functions at two- and three-loop order, this relation was observed in \cite{Gardi:1998ch}.

Let us now extend this argument to also include fixed points in the complex plane. Since the $\beta$-function is real, complex fixed points and their eigenvalues always appear as complex conjugate pairs. In \cref{eq:branches}, they show up as
\begin{align} \label{eq:complex-theta}
	\left( 1 - \frac{a_{*,\text{cc}}}{a_0} \right)^{\!-1/\theta_\text{cc}}&\left( 1 - \frac{(a_{*,\text{cc}})^\dagger}{a_0} \right)^{\!-1/\theta_\text{cc}^\dagger}  \notag \\
	&= r_\text{cc}^2\exp(i \frac{\varphi(a_{*,\text{cc}})}{\theta_\text{cc}^\dagger} - i \frac{\varphi(a_{*,\text{cc}})}{\theta_\text{cc}}) \notag \\
	&= r_\text{cc}^2 \exp(\frac{2 \varphi(a_{*,\text{cc}}) \Im(\theta_\text{cc})}{|\theta_\text{cc}|^2}) \,,
\end{align}
which establishes that their combined complex phase is vanishing and thus they do not contribute to the complex phase on the right-hand side of \cref{eq:total-phases}. In summary, \cref{eq:master-equation} also holds in the presence of complex conjugated fixed points and the sum only runs over the eigenvalues $\theta_i$ belonging to real fixed points with $a_{i,*}>a_0$. 

We emphasise that a solution to \cref{eq:branches} only give candidates for branch cuts and it is not clear if these candidates are realised in the explicit solution of the $\beta$-function equation. Intuitively, one can imagine that the branch point is located in a different branch of the solution.

We evaluated \cref{eq:master-equation} for the gauge $\beta$-function in the BZ phase and computed the critical Veneziano parameter for which branch cuts appear in the complex plane at each loop order. The values for $\varepsilon_\text{branch\,cut}$ agree exactly with the ones found numerically in \cref{tab:xi-max}. We also numerically explored $\beta$-functions with arbitrary coefficients and observed that the branch point candidates that we obtain from \cref{eq:branches} indeed always give rise to a branch cut. While it seems intriguing that a violation of \cref{eq:master-equation} always gives rise to a branch cut in the complex plane, this stems from a numerical search and should not be taken as a general statement. 

In summary, we conclude that if \cref{eq:master-equation} is fulfilled then there are no additional non-analyticities in the complex plane. Conversely, if \cref{eq:master-equation} is violated, we cannot make a definite statement but our numerical analysis suggests that additional non-analyticities are very likely. We also emphasise that \cref{eq:master-equation} is independent of the mass dimension of the coupling $a$.

\subsection{Veneziano Limit and Fixed Point Merger}
In the Veneziano limit, and for small $\varepsilon \ll 1$, it is worth pointing out that the condition  \cref{eq:master-equation} is always fulfilled and the BZ fixed point is guaranteed to be free from additional non-analyticities in the complex plane. This can be appreciated as follows. For small $\varepsilon \ll 1$, the leading eigenvalue reads $\theta_\text{BZ}= \frac{8}{225} \varepsilon ^2+{\cal O}(\varepsilon ^3)$. We choose $0 < a_0 < a_{*,\text{BZ}}$ to ensure that the coupling runs from close to the free fixed point into the BZ fixed point. In consequence, $\theta_\text{BZ}$ provides a parametrically large contribution $\sim 1/\varepsilon ^{2}$ to the sum in \cref{eq:master-equation}. This contribution could be tamed by another eigenvalue of the order $1/\varepsilon ^{2}$. Within perturbation theory, this would require the existence of a second fixed point, parametrically close to the BZ fixed point. In particular, this necessitates that the two-loop coefficient $\beta_2$  becomes parametrically small with the Veneziano parameter. However, this is impossible:  the two-loop coefficient $\beta_2$ is of order unity and positive, for any $4d$ quantum gauge theory coupled to any type of matter as long as the one-loop coefficient $\beta_1$ is parametrically small or vanishing \cite{Bond:2016dvk}. We conclude that the BZ fixed point is guaranteed to be free from additional non-analyticities in the complex plane in the Veneziano limit, and for small $\varepsilon $.

For large $\varepsilon$, on the other hand, the propagators contain branch cuts as shown in \cref{tab:xi-max}. An exception arises at five-loop order, where the BZ fixed point disappears into the complex plane at moderately small $\varepsilon$ due to a fixed point merger before the branch cut bound appears at the previous loop order. With this motivation in mind, let us briefly discuss how \cref{eq:master-equation}  behaves in the vicinity of a general fixed-point merger. In fact, \cref{eq:master-equation} is not required to be continuous function of $\varepsilon$ close to a merger.  Consider two fixed points $a_1$ and $a_2$ merging at the point $a_m$. For any such system, the $\beta$-function can be expanded around the fixed point merger $a = a_m$,
\begin{align}
	\label{eq:beta-merger}
	\beta (a) &= \delta^2 + \delta \,c_0 (a - a_m) + c_1 (a - a_m)^2 \notag\\
	& \quad \,+ c_2 (a - a_m)^3 + \dots \,,
\end{align}
where $\delta$ is a small parameter. It describes how two fixed points $a_1$ and $a_2$  merge in dependence of an external parameter and become complex afterwards. The coefficients $\delta$ and $c_i$ are functions of this external parameter. At the fixed point merger, we must have $\delta = 0$ and we can expand this problem in powers of $\delta$, in particular $a_{1,2} = a_m \pm  b \,\delta +{\cal O}(\delta^2)$, and $b$ of order unity. Note that all the terms in the first line of \cref{eq:beta-merger} are of order $\delta^2$, while the term in the second line is subleading of order $\delta^3$. Computing the eigenvalues of the fixed point $a_{1,2}$ as an expansion in $\delta$ gives
\begin{align}
	\frac{1}{\theta_1} + \frac{1}{\theta_2} = - \frac{c_2}{c_1^2} + \mathcal{O} (\delta) \, .
	\label{eq:fp_merge}
\end{align}
In general, $c_1$ and $c_2$ are not going to vanish at the fixed point merger. Thus,  the inverse eigenvalues do not cancel each other in \cref{eq:master-equation} and give a finite contribution. However, after the fixed point merger, they do not contribute to \cref{eq:master-equation} any more, see the discussion around \cref{eq:complex-theta}, and \cref{eq:master-equation} has a discontinuity. Remarkably, it is the first subleading contribution parameterised by the coefficient $c_2$ that is responsible for the discontinuity since the leading contributions cancel.

We can apply this to the fixed point merger of the BZ fixed point at the five-loop order of perturbation theory. There we have the situation that we have exactly two positive fixed points that merge at $\varepsilon = 1.2019$ at $\nc = 3$, see \cref{tab:xi-max}. After the fixed point merger, there is not a single critical exponent contributing to \cref{eq:master-equation}, and nonetheless no proliferation of branch cuts before the merger. In fact we can numerically compute $\sum_i 1/\theta_i \approx 12.5$ just before the fixed point merger. This agrees with the analytic considerations of \cref{eq:fp_merge} when we expand the five-loop $\beta$-function around a fixed point merger. This discontinuity prevents the existence of a branch cut regime before the BZ fixed point disappears. In comparison in four-loop case at $\nc=3$, the BZ fixed point also vanishes with a fixed point merger at $\varepsilon_\text{max} = 3.0538$ but in contrast shows a proliferation of branch cuts already at $\varepsilon_\text{branch cut} = 2.6817$, see \cref{tab:xi-max}. Matching the four-loop case to the merger template in \cref{eq:fp_merge}, we obtain $\sum_i 1/\theta_i \approx 0.46$ right at the merger and hence the discontinuity is small enough in order to allow for the existence of a branch cut regime before the merger.

\subsection{Branch Cut Conditions from Resummations}
\label{sec:resummed-beta}
We consider a general quantum field theory with a single coupling of vanishing canonical mass dimension whose $[n,k]$~Pad\'e-resummed $(n+k)$-loop $\beta$-function is given by
\begin{align}
	\label{eq:pade-beta}
	\mu^2 \frac{\mathrm d a}{\mathrm d \mu^2} \equiv \beta (a) =   a^2 \frac{\gamma_1 + \gamma_2 a  +\dots+ \gamma_n a^{n-1}}{1 + \delta_1 a +   \dots+\delta_k a^k}\,.
\end{align}
The coefficients $\gamma_i$ and $\delta_i$ are uniquely linked to the original $\beta$-function coefficients.\footnote{A canonical mass dimension $d_a\neq0$ will alter some intermediate expressions, such as \cref{eq:pade_frac,eq:implicit-resummed-beta,eq:identity-resummed-beta}, without affecting the final results and conclusions.} If $k\leq n+1$, the partial fraction decomposition for $1/\beta$ takes the form
\begin{align}
	\frac{1}{\beta (a)} = \frac{1}{\gamma_1 a^2} - \frac{\gamma_2 - \gamma_1 \delta_1}{\gamma_1^2 a} + \sum_i \frac{1/\theta_i}{a - a_{i,*}} \, .
	\label{eq:pade_frac}
\end{align}
Using the sum rule
\begin{align}
	\sum_i \frac{1}{\theta_i} =	- \frac{\gamma_1 \delta_1-\gamma_2}{\gamma_1^2} \,,
	\label{eq:identity-resummed-beta}
\end{align}
which is valid for  $k\leq n$, this equation is readily  integrated, and we write the result as $F(a, \mu^2) =0$ with
\begin{align}
	\label{eq:implicit-resummed-beta}
	F(a, \mu^2) =\log \frac{\mu^2}{\mu_0^2} - \frac1{\gamma_1}\left(\frac1{a_0}-\frac1a \right) - \sum_i \frac{1}{\theta_i} \log \frac{1 - \frac{a_i^*}{a}}{1 - \frac{a_i^*}{a_0}} \,.
\end{align}
Applying the implicit function theorem to $F$ indicates that a first non-analyticity arises at $a =0$, as expected, corresponding to a branch cut on the negative half axis. To find further non-analyticities, we have to solve $\partial_a F = 1/\beta(a) = 0$, which identifies two potential sources for non-analyticities:
\begin{itemize}
	\item[i)] The limit  $a\to\infty$ is a solution. In consequence, \cref{eq:implicit-resummed-beta} falls back onto the non-resummed setting \cref{eq:branches}, and we find that that the condition \cref{eq:master-equation} entails  the same non-analyticities  as in ordinary perturbation theory.
	\item[ii)] New types of solutions are given by the roots of the denominator, $1+ \sum \delta_i a^i = 0$. 
\end{itemize}
A notable difference between these is that case i) leads to \emph{one} condition for the absence of non-analyticities in terms of the critical exponents, while case ii) leads to $n$ conditions where $n$ is the number of roots of the Pad\'e denominator.

To simplify the analysis of ii), we assume that the denominator only has real roots which we denote by $\tilde a_j$, i.e., $1+ \sum \delta_i (\tilde a_j)^i = 0$. This leads for each root $\tilde a_j$ to
\begin{align}
	e^{\frac1{\gamma_1}\left(\frac1{a_0} - \frac{1}{\tilde a_j}\right)}\prod_i \left( \frac{1 - \frac{a_i^*}{a_0}}{1- \frac{a_i^*}{\tilde a_j}} \right)^{\!-1/\theta_i} = \frac{\mu^2}{\mu_0^2} \,.
	\label{eq:branches2}
\end{align}
The real parts as well as the combined contributions from the complex conjugated fixed points can be absorbed by an appropriate choice of $\mu_0^2$. Compared to \cref{eq:branches}, there is a new contribution if $1-\frac{a_i^*}{\tilde a_j} <0$. Since $\tilde a_j$ can be negative, this implies that now also negative fixed points can contribute. Eventually, we find for each root $\tilde a_j$ a condition to avoid additional non-analyticities 
\begin{align}
	\label{eq:master-equation-resummed}
	\left| \sum_i^{a_i^*>a_0} \frac1{\theta_i}  - \sum_i^{a_i^*\gtrless \tilde a_j } \frac1{\theta_i} \right| &>1 \,,
	&&\text{if}&  \tilde a_j  &\gtrless 0 \,.
\end{align}
Notably, in the case where $\tilde a_j<0$ also negative fixed points with $a_i^*<\tilde a_j$ contribute. This condition has to be fulfilled for each root $\tilde a_j$ as well as \cref{eq:master-equation} to guarantee the absence of additional non-analyticities. Conversely, there are not necessarily additional analyticities if one of these conditions is violated, we only find a candidate for a non-analyticity. Indeed, we did some numerical studies with arbitrary $\beta$-function coefficients and found cases where one condition was violated but no additional non-analyticities showed up in the complex plane.

In summary, we found conditions for the absence of additional non-analyticities in the complex plane of the coupling for perturbative $\beta$-functions \cref{eq:master-equation} as well as resummed $\beta$-functions \cref{eq:master-equation,eq:master-equation-resummed}. If one condition is violated, we can find a candidate for a non-analyticity with \cref{eq:branches} for perturbative $\beta$-functions and with \cref{eq:branches,eq:branches2} in the case of resummed $\beta$-functions. If the non-analyticity is realised has to be checked on a case by case basis. Within the numerical search of perturbative $\beta$-functions, we found that all candidates were indeed realised by the solution. We remind the reader, that the non-analyticities in the complex plane of the coupling which propagate to the field propagators via the CS resummation, see \cref{sec:CS-prop}, and the absence of non-analyticities in the field propagators is a necessary but not sufficient criterion for the existence of a KL spectral representation, see \cref{sec:existence-2loop}.

%%%%%%%%%%%%%%%%%%%%%%%%%%%%
\section{Discussion and Conclusions}
\label{sec:conluctions}
We have put forward a comprehensive study of unitary and asymptotically free quantum gauge theories with  Banks-Zaks fixed points in the IR. By construction, these theories are conformal both in the asymptotic UV and the asymptotic IR, connected by a perturbatively-controlled separatrix inbetween (\cref{fig:BZ}). As such, weakly interacting quarks and gluons remain good degrees of freedom to describe the system at all scales. A central question is whether the field propagators allow for a K\"all\'en-Lehmann spectral representation. To that end, we performed a detailed investigation of  the propagators in the complex momentum plane, both analytically and numerically, using results from perturbation theory for the running gauge coupling (up to five loop) and self-energy corrections (up to four loop). Performing a Callan-Symanzik resummation of large logarithms from self-energy corrections also proved important.

At weak coupling, and in the Veneziano limit, the smallness of the conformal expansion parameter $\varepsilon$ ensures strict perturbative control. In this regime, we find that the running of the gauge coupling along the entire UV-IR connecting separatrix can be determined analytically as a systematic power series in $\varepsilon$. Propagators in the complex plane are  dominated by the running gauge coupling, while the finite self-energy corrections turn out to be parametrically subleading after the large logarithms have been resummed. Most notably, this does not introduce new poles or cuts in the complex plane as long as interactions remain weak.
If, additionally, field anomalous dimensions in the deep IR are either negative or integers, the availability of a K\"all\'en-Lehmann spectral representation is guaranteed. 

On the other hand, the complex structure of propagators changes significantly as soon as interactions become strong. This qualitative change shows up through new branch cuts and poles (\cref{fig:prop_complexplots}) once interactions exceed a characteristic strength $(\varepsilon\ge \varepsilon_{\rm branch\,cut})$. Invariably, a standard K\"all\'en-Lehmann spectral representation is lost even before the fixed point disappears. These new effects originate from the running coupling, whose non-analyticities in the complex plane transfer directly to the field propagators. In addition, new poles  arise in  ghost propagators from self-energy corrections at strong coupling $\varepsilon\ge \varepsilon_{\rm poles}$ (see \cref{tab:self-energy-poles}) where perturbation theory becomes unreliable. Despite of their different origins, we notice that $\varepsilon_{\rm branch\,cut}$ and $\varepsilon_{\rm poles}$  only deviate at the percent level or below, \cref{eq:poles}. We take the interaction-induced proliferation of cuts and poles (\cref{fig:branching}) and the qualitative change in the propagators to indicate that the BZ conformal window is smaller than expected from beta functions (\cref{tab:xi-max}). Further, and even though all theories are characterised by the scale $\Lambda_c$, \cref{eq:Lambdac}, the spectral representation of  propagators does not offer  indication for  bound  states  with masses  $\propto\Lambda_c$. This result holds true for any setting with $\varepsilon<\varepsilon_{\rm poles}$, and covers all theories whose RG flow terminates at the BZ fixed point in the IR.

Even though weakly coupled quarks and gluons  are ideally suited to describe the system   at all scales,  it is important to remember that their propagators  are not gauge invariant variables.  Interestingly, however, the complex structure of poles and cuts induced by the running coupling is insensitive to gauge dependences. Still, field anomalous dimensions which control the scaling of propagators in the deep IR,  are manifestly gauge dependent and influence the existence, normalisability,  and positive definiteness of spectral functions. This is illustrated in \cref{fig:spec-gluon,fig:spec-quark-ghost,fig:spectral-nloop-glue,fig:spectral-nloop-quark,fig:spectral-nloop-ghost}, where we have  computed quark, gluon, and ghost spectral functions for a range of loop orders, Veneziano parameters, and gauge-fixing parameters. As a consistency check, we have  also used the fRG to  compute the mid-momentum regime of propagators and found that results (\cref{fig:ZA-fRG}) compare very well with those from the Callan-Symanzik flow (\cref{fig:glu_prop}), as they must. We further noticed that the apparent convergence of the loop expansion is different for different fields (\cref{tab:criteps}),  in qualitative agreement with findings elsewhere \cite{Bond:2022xvr}. Most importantly, we find a wide range of parameters where all field propagators simultaneously have a well-defined KL spectral representation. This key result indicates that spectral representations -- ordinarily adopted for the study of scalar theories or for correlation functions of gauge-invariant quantities -- may very well be of use for the study of gauge theories with matter and ghosts. 

While our results have been achieved specifically for $SU(\nc)$ gauge theories with $\nf$ Dirac fermions in the fundamental representation, they equally hold true for any other 4d gauge-matter theory with a weakly interacting Banks-Zaks fixed point. The reason for this is that the running of the gauge coupling is invariably dictated by the $W$-Lambert function to any loop order, see \cref{eq:aahat,eq:a0,eq:a1}, leading to identical complex structures for propagators. This is particularly true in the Veneziano limit where large-$N$ equivalences amongst $SU$, $SO$, and $Sp$ gauge theories coupled to matter also ensure identical field anomalous dimensions in the deep IR \cite{Lovelace:1982hz, Bond:2019npq}. We therefore conclude that any 4d QFT with a  weakly interacting Banks-Zaks fixed point universally admits a KL spectral representation for their propagators.

Since the loss of a KL spectral representation and the emergence of complex conjugated propagator poles and branch cuts are intimately linked to the running coupling in the complex plane, we have extended investigations towards theories with more general beta functions. As a first step, we looked into the complex structures of theories with finite order perturbative \cref{eq:beta-gen} or resummed $\beta$-functions \cref{eq:pade-beta} and identified  conditions for the absence of non-analyticities in terms of universal scaling exponents characterising the fixed points of the theory, see \cref{eq:master-equation,eq:master-equation-resummed}. We also studied fingerprints for fixed-point mergers \cref{eq:fp_merge} which should prove useful for investigating endpoints of conformal windows \cite{Bond:2017tbw, Benini:2019dfy, Bond:2021tgu}.

It will be interesting to expand our investigations towards theories with several running couplings such as 4d gauge-Yukawa theories \cite{Litim:2014uca, Bond:2017lnq, Bond:2019npq}, and more strongly coupled theories including Einstein Hilbert \cite{Fehre:2021eob} or higher curvature gravities. On a different tack, it will also be important to extend our studies towards gauge-invariant objects such as bound state correlation functions and scattering amplitudes, or with the help of dressing functions \cite{Capri:2016aqq, Capri:2016gut, Capri:2017abz}. The resummed field propagators presented here are basic building blocks for this, and it will be intriguing to observe how gauge dependences ultimately cancel out. We hope to come back to these topics in the future.
\bigskip

\centerline{\bf Acknowledgements}
This work is supported  by the Science and Technology Research Council (STFC) under the Consolidated Grant ST/T00102X/1, and by the STFC Studentship Grant ST/S505766/1 (YK).

\appendix

\section{Running Gauge Coupling}
\label{app:analyticsol}
At two-loop, the $\beta$-function for the running gauge coupling \cref{eq:runningcoupling1} can be solved explicitly. Starting from
\begin{align}
	\mu^2 \dderiv{a}{\mu^2} = \beta_1 a^2 + \beta_2 a^3  \, ,
	\label{eq:runcoup2loop}
\end{align}
we first display the implicit solution of \cref{eq:runcoup2loop} which is obtained by integrating the differential equation,
\begin{align}
	\frac{1}{\theta} \log(\frac{a_* - a}{a_* - a_0} \frac{a_0}{a})-\frac{1}{\beta_1} \left( \frac{1}{a} - \frac{1}{a_0} \right) = \log (\frac{\mu ^2}{\mu_0^2}) \, ,
	\label{eq:betasolexact}
\end{align}
where $a_* = -\beta_1/\beta_2$ is the BZ fixed point and $\theta = \beta_1^2/\beta_2$ the corresponding eigenvalue. An explicit solution to this equation can be found by usage of the $W$-Lambert function $W(z)$, which is defined as the solution to the equation
\begin{align}
	W(z) e^{W (z)} = z \, ,
	\label{eq:WLambert}
\end{align}
To bring \cref{eq:betasolexact} into this form, we exponentiate it and introduce the new variables 
\begin{align}
	\omega &= \frac{a_* - a}{a} \, , 
	&
	\omega_0 &= \frac{a_* - a_0}{a_0} \, .
\end{align}
The solution can then be read-off,
\begin{align}
	a &= \frac{a_*}{1 + W_i(z)} \, ,
	&
	z &= \omega_0 e^{\omega_0} \left(\frac{\mu ^2}{\mu_0^2}\right)^{\!\theta} \,,
	\label{eq:afullsolpreapp}
\end{align}
see also \cref{eq:afullsol}. Eq.~\cref{eq:WLambert} has a countable infinite number of complex solutions leading to infinitely many branches $W_i (z)$. Using that the coupling $a$ is real for all $\mu^2/\mu_0^2 > 0$ and that $a_0 < a_*$, it follows that the principal branch $W_0 (z)$ is the unique solution for $0 < z < \infty$. This explicit solution can be extended to three-loop order with the help of a Pad\'e approximant \cite{Gardi:1998qr}. In this way, we can write
\begin{align}
	\beta (a) = a^2 \frac{\beta_1 \beta_2 + (\beta_2^2 - \beta_1 \beta_3)a}{\beta_2 - \beta_3 a} + \order{a^5} \,,
\end{align}
which allows for an explicit solution of the type \cref{eq:afullsolpreapp} with $a^*=(\beta_1 \beta_2)/(\beta_1 \beta_3-\beta_2^2)$ and $\theta =  \beta_1^2/\beta_2$.

At two-loop, the asymptotic properties of the gauge coupling can be obtained directly from the asymptotic behaviour of the $W$-Lambert function,
\begin{align}
	W_0 (z) = \begin{cases}
		\log z - \log \log z + \order{\frac1z} & \text{for} \quad z \rightarrow \infty\,, \\[1ex]
		z + \order{z^2} & \text{for} \quad z \rightarrow 0\,.
	\end{cases}
\end{align}
For the running coupling, this gives us the asymptotic behaviour
\begin{align}
	a(\mu^2) =
	\begin{cases}
		- \frac1{\beta_1 \log(\mu^2/\mu_0^2)} &\text{for }\quad \left|\mu^2\right| \to \infty  \\[2ex]
		a_* - a_* \omega_0 e^{\omega_0} \left(\frac{\mu^2}{\mu_0^2}\right)^{\!\theta}	
		&\text{for }\quad \left|\mu^2\right| \to 0\,.
	\end{cases}
	\label{eq:runcoupasym}
\end{align}
Both asymptotes can also be directly extracted from the $\beta$-function by linearising and solving the $\beta$-function around the given fixed point.

At higher loop orders, the algorithm of \cref{sec:Veneziano-nloop} can be used to find explicit solutions for the running gauge coupling in a systematic expansion in $\varepsilon\ll 1$.

\section{Five Loop Fixed Point and Exponent}
\label{sec:AppBZ}
In the Veneziano limit, the BZ fixed point expands as
\begin{align}
	\nc a_*&=\sum_{n=1} \hat a^*_n\,\varepsilon^{n} \,,
\end{align}
for small $\varepsilon$. Using the five-loop beta function \cite{Herzog:2017ohr, Chetyrkin:2017bjc}, the first four coefficients of the BZ fixed point are found to be
\begin{align}\label{eq:acoeffs}
	\hat a^*_1&= \frac{4}{75}\,, \nonumber\\
	\hat a^*_2&= \frac{2192}{421875}\,, \nonumber\\
	\hat a^*_3&= \frac{5844232}{2373046875} +\frac{1408}{421875}\zeta_3\,, \nonumber\\
	\hat a^*_4&= \frac{2226607268}{2669677734375} +\frac{935296}{791015625} \zeta_3 -\frac{45056}{31640625}\zeta_5\,.
\end{align}
Similarly, the universal scaling exponent expands as
\begin{align}\label{eq:thetaexp}
	\theta=\sum_{n=2} \hat\theta_n \, \varepsilon^{n}\,,
\end{align}
and its first four coefficients are found to be
\begin{align}\label{eq:thetacoeffs}
	\hat \theta_2&=  \frac{8}{225} \,,\nonumber\\
	\hat \theta_3&= \frac{208}{16875} \,,\nonumber\\
	\hat \theta_4&=   \frac{2934256}{2373046875} -\frac{2816}{1265625}\, \zeta_3\,, \nonumber\\
	\hat \theta_5&= \frac{4771112816}{4449462890625} -\frac{722432}{474609375}\,\zeta_3 +\frac{180224}{94921875}\,\zeta_5 \,.
\end{align}

\section{Sum Rules for  Scaling Exponents}
\label{app:betaidentity}
We are going to establish a series of sum rules for universal critical exponents related to fixed points of perturbative $n$-loop $\beta$-functions. We assume that the $\beta$-function takes the shape
\begin{align}
	\mu^2 \frac{\dd a}{\dd \mu^2} \equiv \beta (a) = \beta_1 a^2 + \beta_2 a^3  + \dots + \beta_n a^{n+1}\,.
	\label{eq:beta-gen-app}
\end{align}
Besides the double-zero at $a=0$, the beta function has $n-1$ non-trivial zeros $a_{i,*}$ in the complexified $a$-plane, with scaling exponents 
\begin{align}
	\theta_i=\left.\frac{\partial\beta(a)}{\partial a}\right|_{a=a_i^*}\,.
\end{align}
Of physical interest are the real zeros, and in particular the one closest to the origin. 

Sum rules for scaling exponents are found by  starting from the partial fraction decomposition for the inverse $\beta$-function,
\begin{align}
	\frac{1}{\beta (a)} = \frac{1}{\beta_1 a^2} - \frac{\beta_2}{\beta_1^2 a} + \sum_i \frac{1/\theta_i}{a - a_{i,*}} \, ,
	\label{eq:betapfdecomp}
\end{align}
which is then integrated in the complex plane along a circle including all poles of \cref{eq:betapfdecomp} with its radius going to infinity. Since the finite order $\beta$-function is a polynomial of at least quadratic order, its inverse goes to zero fast enough at infinity such that its integral along the curve vanishes,
\begin{align}
	\oint \text{d} a \, \frac{1}{\beta (a)} = 0 \, .
\end{align}
On the other hand, for each term on the right-hand side of \cref{eq:betapfdecomp} we can use the residue theorem. This leads to our first sum rule for critical exponents
\begin{align}
	\sum_i \frac{1}{\theta_i} = \frac{\beta_2}{\beta_1^2} \,.
\end{align}
If the coupling $a$ has a non-trivial  mass dimension $d_a$ and the $\beta$-function is given by \cref{eq:beta-gen}, then the corresponding sum rule reads 
\begin{align}
	\sum_i \frac1{\theta_i} =  -\frac1{d_a} \,,
\end{align}
instead, see \cref{eq:identity}. Also for resummed $\beta$-functions of the type \cref{eq:pade-beta}, an analogous sum rule can be derived straightforwardly, see \cref{eq:identity-resummed-beta}.

We can derive further identities by exploiting that
\begin{align}
	\oint \text{d} a \, \frac{a}{\beta (a)} = 0 \, ,
\end{align}
provided the $\beta$-function is two-loop or higher. Multiplying \cref{eq:betapfdecomp} by $a$ and using the residue theorem once more, we obtain the sum rule
\begin{align}
	\sum_i \frac{a_i^*}{\theta_i} = \frac{1}{\beta_1} \,.
\end{align}
It states that the sum of inverse scaling exponents, weighted by the corresponding fixed point coupling, is given by the inverse one-loop coefficient. 

Generalising to multiplications with higher powers of $a$ and assuming that the $\beta$-function is evaluated at high enough loop orders, we also obtain the sum rules
\begin{align}
	\sum_i \frac{(a_i^*)^m}{\theta_i} = 0 \, .
\end{align}
It states that the sum of inverse scaling exponents, weighted by any integer power $1<m<n$ of the corresponding fixed point coupling, vanishes. These sum rules and variants have been tested numerically, and are used throughout the main text.

\section{Anomalous Dimensions and Self-Energies}
\label{app:explicit-loops}
In this appendix, we summarise expressions for the anomalous dimensions and the self energies up to two-loop order. Five loop results for  the $\beta$-function and anomalous dimensions, and four loop results in the self energies can be found in \cite{Herzog:2017ohr, Chetyrkin:2017bjc} and  \cite{Chetyrkin:2000dq, Ruijl:2017eht}, respectively. 

The one and two-loop coefficients of the field anomalous dimensions
$$\gamma_\phi (a) = \gamma_\phi^{(1)} a + \gamma_\phi^{(2)} a^2 + \dots $$ are given by 
\begin{align}
	\gamma_A^{(1)} &= - \frac{\nc}{2} \left(\xi + 3 \right)+\frac{2  \nc}{3} \varepsilon \,, \notag \\
	\gamma_A^{(2)} &= \frac{44-\left(2 \xi ^2+11 \xi +95\right) \nc^2}{8}  +   \left(\frac{7 \nc^2}{2}-1\right) \varepsilon\,, 
\end{align}
for the gluons, by
\begin{align}
	\gamma_\psi^{(1)} &= \frac{\xi  \left(1 - \nc^2\right)}{2 \nc} \,, \notag \\
	\gamma_\psi^{(2)} &= \frac{1 - \nc^2}{8 \nc^2} \left[3 + \xi (\xi +8) \nc^2\right] + \frac{1-\nc^2}{2} \varepsilon \,, 
\end{align}
for the quarks, and by
\begin{align}
	\gamma_c^{(1)} &= \frac{3 - \xi}{4} \nc \,, \notag \\
	\gamma_c^{(2)} &= \frac{\xi - 5}{16} \nc^2+\frac{5 \nc^2}{12} \varepsilon 
\end{align}
for the ghosts. Similarly, the one and two-loop self energy corrections
$$\Pi_\phi (p^2= -\mu^2) = \Pi_\phi^{(1)} a + \Pi_\phi^{(2)} a^2 + \dots$$ read 
\begin{align}
	\Pi_A^{(1)} &= -\frac{\nc}{12} \left(3 \xi ^2+6 \xi -41\right) - \frac{10 \nc}{9} \varepsilon \,, \notag \\
	\Pi_A^{(2)} &= \left(3 \zeta_3+\frac{1311}{32}\right)\! \nc^2+22 \zeta_3-\frac{605}{24} - \frac{115}{48} \xi ^2 \nc^2 \notag \\
	& \quad\, - \frac{\nc^2}{96} \xi  (192 \zeta_3+139)+ \frac{\nc^2}{16} \xi ^3 (\xi - 1)  	\notag\\
	& \quad\,+ \varepsilon  \left(\frac{55}{12} - 4 \zeta_3 - \frac{287 \nc^2}{24} + \frac{5 \nc^2}{9} \xi(\xi + 1) \right) , 
\end{align}
for the gluons,
\begin{align}
	\Pi_\psi^{(1)} &= \frac{\xi  \left(\nc^2-1\right)}{2 \nc} \,, \notag \\
	\Pi_\psi^{(2)} &= \frac{9}{16} \xi ^2 \left(\nc^2-1\right) + \frac{1}{4} \xi  (13-6 \zeta_3) \left(\nc^2-1\right) \notag\\
	& \quad\,+\frac{\nc^2 - 1}{32 \nc^2} \left(5 (\nc^2 + 1) - 48 \zeta_3\right)+\frac{7}{8} \varepsilon  \left(\nc^2-1\right) ,
\end{align}
for the quarks, and 
\begin{align}
	\Pi_c^{(1)} &= -\nc \,, \notag \\
	\Pi_c^{(2)} &= \frac{3 \nc^2}{16} \xi ^2 (\zeta_3-2) + \frac{\nc^2}{64} \xi  (7-24 \zeta_3)\notag \\
	&\quad\, + \frac{\nc^2}{64} (60 \zeta_3 +113) - \frac{95 \nc^2}{48} \varepsilon \,,
\end{align}
for the ghosts.

%%%%%% References %%%%%%%%%%%%%%%%%%%%
\bibliographystyle{mystyle}
\bibliography{Gravity}

\providecommand{\href}[2]{#2}\begingroup\raggedright\begin{thebibliography}{10}

\bibitem{Kallen:1952zz}
G.~Kallen, \emph{{On the definition of the Renormalization Constants in Quantum
  Electrodynamics}},
  \href{https://doi.org/10.1007/978-3-319-00627-7_90}{\emph{Helv. Phys. Acta}
  {\bfseries 25} (1952) 417}.

\bibitem{Lehmann:1954xi}
H.~Lehmann, \emph{{On the Properties of propagation functions and
  renormalization contants of quantized fields}},
  \href{https://doi.org/10.1007/BF02783624}{\emph{Nuovo Cim.} {\bfseries 11}
  (1954) 342}.

\bibitem{Gross:1973id}
D.~J. Gross and F.~Wilczek, \emph{{Ultraviolet Behavior of Nonabelian Gauge
  Theories}}, \href{https://doi.org/10.1103/PhysRevLett.30.1343}{\emph{Phys.
  Rev. Lett.} {\bfseries 30} (1973) 1343}.

\bibitem{Politzer:1973fx}
H.~D. Politzer, \emph{{Reliable Perturbative Results for Strong
  Interactions?}},
  \href{https://doi.org/10.1103/PhysRevLett.30.1346}{\emph{Phys.Rev.Lett.}
  {\bfseries 30} (1973) 1346}.

\bibitem{Litim:2014uca}
D.~F. Litim and F.~Sannino, \emph{{Asymptotic safety guaranteed}},
  \href{https://doi.org/10.1007/JHEP12(2014)178}{\emph{JHEP} {\bfseries 12}
  (2014) 178} [\href{https://arxiv.org/abs/1406.2337}{{\ttfamily 1406.2337}}].

\bibitem{Bond:2016dvk}
A.~D. Bond and D.~F. Litim, \emph{{Theorems for Asymptotic Safety of Gauge
  Theories}}, \href{https://doi.org/10.1140/epjc/s10052-017-4976-5}{\emph{Eur.
  Phys. J.} {\bfseries C77} (2017) 429}
  [\href{https://arxiv.org/abs/1608.00519}{{\ttfamily 1608.00519}}].

\bibitem{Bond:2022xvr}
A.~D. Bond and D.~F. Litim, \emph{{Asymptotic safety guaranteed for strongly
  coupled gauge theories}},
  \href{https://doi.org/10.1103/PhysRevD.105.105005}{\emph{Phys. Rev. D}
  {\bfseries 105} (2022) 105005}
  [\href{https://arxiv.org/abs/2202.08223}{{\ttfamily 2202.08223}}].

\bibitem{Coleman:1973sx}
S.~R. Coleman and D.~J. Gross, \emph{{Price of asymptotic freedom}},
  \href{https://doi.org/10.1103/PhysRevLett.31.851}{\emph{Phys. Rev. Lett.}
  {\bfseries 31} (1973) 851}.

\bibitem{Bond:2018oco}
A.~D. Bond and D.~F. Litim, \emph{{Price of Asymptotic Safety}},
  \href{https://doi.org/10.1103/PhysRevLett.122.211601}{\emph{Phys. Rev. Lett.}
  {\bfseries 122} (2019) 211601}
  [\href{https://arxiv.org/abs/1801.08527}{{\ttfamily 1801.08527}}].

\bibitem{Weinberg:1980gg}
S.~Weinberg, \emph{{Ultraviolet Divergences in Quantum Theories of
  Gravitation}}, pp.~790--831.
\newblock 1, 1980.

\bibitem{Gardi:1998ch}
E.~Gardi and G.~Grunberg, \emph{{The Conformal window in QCD and supersymmetric
  QCD}}, \href{https://doi.org/10.1088/1126-6708/1999/03/024}{\emph{JHEP}
  {\bfseries 03} (1999) 024}
  [\href{https://arxiv.org/abs/hep-th/9810192}{{\ttfamily hep-th/9810192}}].

\bibitem{Fehre:2021eob}
J.~Fehre, D.~F. Litim, J.~M. Pawlowski and M.~Reichert, \emph{{Lorentzian
  Quantum Gravity and the Graviton Spectral Function}},
  \href{https://arxiv.org/abs/2111.13232}{{\ttfamily 2111.13232}}.

\bibitem{Belavin:1974gu}
A.~A. Belavin and A.~A. Migdal, \emph{{Calculation of Anomalous Dimensions in
  Non-Abelian Gauge Field Theories}}, {\emph{Pisma Zh. Eksp. Teor. Fiz.}
  {\bfseries 19} (1974) 317}.

\bibitem{Caswell:1974gg}
W.~E. Caswell, \emph{{Asymptotic Behavior of Nonabelian Gauge Theories to Two
  Loop Order}}, \href{https://doi.org/10.1103/PhysRevLett.33.244}{\emph{Phys.
  Rev. Lett.} {\bfseries 33} (1974) 244}.

\bibitem{Banks:1981nn}
T.~Banks and A.~Zaks, \emph{{On the Phase Structure of Vector-Like Gauge
  Theories with Massless Fermions}},
  \href{https://doi.org/10.1016/0550-3213(82)90035-9}{\emph{Nucl. Phys. B}
  {\bfseries 196} (1982) 189}.

\bibitem{Chetyrkin:2000dq}
K.~G. Chetyrkin and A.~Retey, \emph{{Three Loop Three Linear Vertices and Four
  Loop Similar to Mom Beta Functions in Massless QCD}},
  \href{https://arxiv.org/abs/hep-ph/0007088}{{\ttfamily hep-ph/0007088}}.

\bibitem{Baikov:2016tgj}
P.~A. Baikov, K.~G. Chetyrkin and J.~H. K\"uhn, \emph{{Five-Loop Running of the
  QCD coupling constant}},
  \href{https://doi.org/10.1103/PhysRevLett.118.082002}{\emph{Phys. Rev. Lett.}
  {\bfseries 118} (2017) 082002}
  [\href{https://arxiv.org/abs/1606.08659}{{\ttfamily 1606.08659}}].

\bibitem{Herzog:2017ohr}
F.~Herzog, B.~Ruijl, T.~Ueda, J.~A.~M. Vermaseren and A.~Vogt, \emph{{The
  five-loop beta function of Yang-Mills theory with fermions}},
  \href{https://doi.org/10.1007/JHEP02(2017)090}{\emph{JHEP} {\bfseries 02}
  (2017) 090} [\href{https://arxiv.org/abs/1701.01404}{{\ttfamily
  1701.01404}}].

\bibitem{Luthe:2017ttg}
T.~Luthe, A.~Maier, P.~Marquard and Y.~Schroder, \emph{{The five-loop Beta
  function for a general gauge group and anomalous dimensions beyond Feynman
  gauge}}, \href{https://doi.org/10.1007/JHEP10(2017)166}{\emph{JHEP}
  {\bfseries 10} (2017) 166}
  [\href{https://arxiv.org/abs/1709.07718}{{\ttfamily 1709.07718}}].

\bibitem{Chetyrkin:2017bjc}
K.~G. Chetyrkin, G.~Falcioni, F.~Herzog and J.~A.~M. Vermaseren,
  \emph{{Five-loop renormalisation of QCD in covariant gauges}},
  \href{https://doi.org/10.1007/JHEP10(2017)179}{\emph{JHEP} {\bfseries 10}
  (2017) 179} [\href{https://arxiv.org/abs/1709.08541}{{\ttfamily
  1709.08541}}].

\bibitem{Polchinski:1987dy}
J.~Polchinski, \emph{{Scale and Conformal Invariance in Quantum Field Theory}},
  \href{https://doi.org/10.1016/0550-3213(88)90179-4}{\emph{Nucl. Phys.}
  {\bfseries B303} (1988) 226}.

\bibitem{Komargodski:2011vj}
Z.~Komargodski and A.~Schwimmer, \emph{{On Renormalization Group Flows in Four
  Dimensions}}, \href{https://doi.org/10.1007/JHEP12(2011)099}{\emph{JHEP}
  {\bfseries 12} (2011) 099} [\href{https://arxiv.org/abs/1107.3987}{{\ttfamily
  1107.3987}}].

\bibitem{Komargodski:2011xv}
Z.~Komargodski, \emph{{The Constraints of Conformal Symmetry on RG Flows}},
  \href{https://doi.org/10.1007/JHEP07(2012)069}{\emph{JHEP} {\bfseries 07}
  (2012) 069} [\href{https://arxiv.org/abs/1112.4538}{{\ttfamily 1112.4538}}].

\bibitem{Luty:2012ww}
M.~A. Luty, J.~Polchinski and R.~Rattazzi, \emph{{The $a$-theorem and the
  Asymptotics of 4D Quantum Field Theory}},
  \href{https://doi.org/10.1007/JHEP01(2013)152}{\emph{JHEP} {\bfseries 01}
  (2013) 152} [\href{https://arxiv.org/abs/1204.5221}{{\ttfamily 1204.5221}}].

\bibitem{Litim:2015iea}
D.~F. Litim, M.~Mojaza and F.~Sannino, \emph{{Vacuum stability of
  asymptotically safe gauge-Yukawa theories}},
  \href{https://doi.org/10.1007/JHEP01(2016)081}{\emph{JHEP} {\bfseries 01}
  (2016) 081} [\href{https://arxiv.org/abs/1501.03061}{{\ttfamily
  1501.03061}}].

\bibitem{Bond:2017wut}
A.~D. Bond, G.~Hiller, K.~Kowalska and D.~F. Litim, \emph{{Directions for model
  building from asymptotic safety}},
  \href{https://doi.org/10.1007/JHEP08(2017)004}{\emph{JHEP} {\bfseries 08}
  (2017) 004} [\href{https://arxiv.org/abs/1702.01727}{{\ttfamily
  1702.01727}}].

\bibitem{Bond:2017tbw}
A.~D. Bond, D.~F. Litim, G.~Medina~Vazquez and T.~Steudtner, \emph{{UV
  conformal window for asymptotic safety}},
  \href{https://doi.org/10.1103/PhysRevD.97.036019}{\emph{Phys. Rev. D}
  {\bfseries 97} (2018) 036019}
  [\href{https://arxiv.org/abs/1710.07615}{{\ttfamily 1710.07615}}].

\bibitem{Bond:2017lnq}
A.~D. Bond and D.~F. Litim, \emph{{More Asymptotic Safety Guaranteed}},
  \href{https://doi.org/10.1103/PhysRevD.97.085008}{\emph{Phys. Rev.}
  {\bfseries D97} (2018) 085008}
  [\href{https://arxiv.org/abs/1707.04217}{{\ttfamily 1707.04217}}].

\bibitem{Bond:2017suy}
A.~D. Bond and D.~F. Litim, \emph{{Asymptotic safety guaranteed in
  supersymmetry}},
  \href{https://doi.org/10.1103/PhysRevLett.119.211601}{\emph{Phys. Rev. Lett.}
  {\bfseries 119} (2017) 211601}
  [\href{https://arxiv.org/abs/1709.06953}{{\ttfamily 1709.06953}}].

\bibitem{Bond:2019npq}
A.~D. Bond, D.~F. Litim and T.~Steudtner, \emph{{Asymptotic safety with
  Majorana fermions and new large $N$ equivalences}},
  \href{https://doi.org/10.1103/PhysRevD.101.045006}{\emph{Phys. Rev. D}
  {\bfseries 101} (2020) 045006}
  [\href{https://arxiv.org/abs/1911.11168}{{\ttfamily 1911.11168}}].

\bibitem{Gies:2005as}
H.~Gies and J.~Jaeckel, \emph{{Chiral phase structure of QCD with many
  flavors}}, \href{https://doi.org/10.1140/epjc/s2006-02475-0}{\emph{Eur. Phys.
  J. C} {\bfseries 46} (2006) 433}
  [\href{https://arxiv.org/abs/hep-ph/0507171}{{\ttfamily hep-ph/0507171}}].

\bibitem{Dietrich:2006cm}
D.~D. Dietrich and F.~Sannino, \emph{{Conformal Window of $SU(N)$ Gauge
  Theories with Fermions in Higher Dimensional Representations}},
  \href{https://doi.org/10.1103/PhysRevD.75.085018}{\emph{Phys. Rev. D}
  {\bfseries 75} (2007) 085018}
  [\href{https://arxiv.org/abs/hep-ph/0611341}{{\ttfamily hep-ph/0611341}}].

\bibitem{Jarvinen:2011qe}
M.~Jarvinen and E.~Kiritsis, \emph{{Holographic Models for QCD in the Veneziano
  Limit}}, \href{https://doi.org/10.1007/JHEP03(2012)002}{\emph{JHEP}
  {\bfseries 03} (2012) 002} [\href{https://arxiv.org/abs/1112.1261}{{\ttfamily
  1112.1261}}].

\bibitem{Kusafuka:2011fd}
Y.~Kusafuka and H.~Terao, \emph{{Fixed point merger in the SU(N) gauge beta
  functions}}, \href{https://doi.org/10.1103/PhysRevD.84.125006}{\emph{Phys.
  Rev. D} {\bfseries 84} (2011) 125006}
  [\href{https://arxiv.org/abs/1104.3606}{{\ttfamily 1104.3606}}].

\bibitem{Alvares:2012kr}
R.~Alvares, N.~Evans and K.-Y. Kim, \emph{{Holography of the Conformal
  Window}}, \href{https://doi.org/10.1103/PhysRevD.86.026008}{\emph{Phys. Rev.
  D} {\bfseries 86} (2012) 026008}
  [\href{https://arxiv.org/abs/1204.2474}{{\ttfamily 1204.2474}}].

\bibitem{DeGrand:2015zxa}
T.~DeGrand, \emph{{Lattice tests of beyond Standard Model dynamics}},
  \href{https://doi.org/10.1103/RevModPhys.88.015001}{\emph{Rev. Mod. Phys.}
  {\bfseries 88} (2016) 015001}
  [\href{https://arxiv.org/abs/1510.05018}{{\ttfamily 1510.05018}}].

\bibitem{Gukov:2016tnp}
S.~Gukov, \emph{{RG Flows and Bifurcations}},
  \href{https://doi.org/10.1016/j.nuclphysb.2017.03.025}{\emph{Nucl. Phys. B}
  {\bfseries 919} (2017) 583}
  [\href{https://arxiv.org/abs/1608.06638}{{\ttfamily 1608.06638}}].

\bibitem{Simmons-Duffin:2016gjk}
D.~Simmons-Duffin, \emph{{The Conformal Bootstrap}},  in \emph{{Theoretical
  Advanced Study Institute in Elementary Particle Physics}: {New Frontiers in
  Fields and Strings}}, pp.~1--74, 2017,
  \href{https://arxiv.org/abs/1602.07982}{{\ttfamily 1602.07982}},
  \href{https://doi.org/10.1142/9789813149441_0001}{DOI}.

\bibitem{Poland:2018epd}
D.~Poland, S.~Rychkov and A.~Vichi, \emph{{The Conformal Bootstrap: Theory,
  Numerical Techniques, and Applications}},
  \href{https://doi.org/10.1103/RevModPhys.91.015002}{\emph{Rev. Mod. Phys.}
  {\bfseries 91} (2019) 015002}
  [\href{https://arxiv.org/abs/1805.04405}{{\ttfamily 1805.04405}}].

\bibitem{Ryttov:2016ner}
T.~A. Ryttov and R.~Shrock, \emph{{Infrared Zero of $\beta$ and Value of
  $\gamma_m$ for an SU(3) Gauge Theory at the Five-Loop Level}},
  \href{https://doi.org/10.1103/PhysRevD.94.105015}{\emph{Phys. Rev. D}
  {\bfseries 94} (2016) 105015}
  [\href{https://arxiv.org/abs/1607.06866}{{\ttfamily 1607.06866}}].

\bibitem{Ryttov:2017lkz}
T.~A. Ryttov and R.~Shrock, \emph{{Physics of the non-Abelian Coulomb phase:
  Insights from Pad\'e approximants}},
  \href{https://doi.org/10.1103/PhysRevD.97.025004}{\emph{Phys. Rev. D}
  {\bfseries 97} (2018) 025004}
  [\href{https://arxiv.org/abs/1710.06944}{{\ttfamily 1710.06944}}].

\bibitem{Kuipers:2018lux}
F.~Kuipers, U.~G\"ursoy and Y.~Kuznetsov, \emph{{Bifurcations in the RG-flow of
  QCD}}, \href{https://doi.org/10.1007/JHEP07(2019)075}{\emph{JHEP} {\bfseries
  07} (2019) 075} [\href{https://arxiv.org/abs/1812.05179}{{\ttfamily
  1812.05179}}].

\bibitem{Hasenfratz:2018wpq}
A.~Hasenfratz, C.~Rebbi and O.~Witzel, \emph{{Determination of the N$_f$=12
  step scaling function using M\"obius domain wall fermions}},
  \href{https://doi.org/10.22323/1.334.0306}{\emph{PoS} {\bfseries LATTICE2018}
  (2019) 306} [\href{https://arxiv.org/abs/1810.05176}{{\ttfamily
  1810.05176}}].

\bibitem{Fodor:2018uih}
Z.~Fodor, K.~Holland, J.~Kuti, D.~Nogradi and C.~H. Wong, \emph{{Is SU(3) gauge
  theory with 13 massless flavors conformal?}},
  \href{https://doi.org/10.22323/1.334.0198}{\emph{PoS} {\bfseries LATTICE2018}
  (2018) 198} [\href{https://arxiv.org/abs/1811.05024}{{\ttfamily
  1811.05024}}].

\bibitem{Antipin:2018asc}
O.~Antipin, A.~Maiezza and J.~C. Vasquez, \emph{{Resummation in QFT with Meijer
  G-functions}},
  \href{https://doi.org/10.1016/j.nuclphysb.2019.02.014}{\emph{Nucl. Phys. B}
  {\bfseries 941} (2019) 72}
  [\href{https://arxiv.org/abs/1807.05060}{{\ttfamily 1807.05060}}].

\bibitem{Hasenfratz:2019dpr}
A.~Hasenfratz, C.~Rebbi and O.~Witzel, \emph{{Gradient flow step-scaling
  function for SU(3) with twelve flavors}},
  \href{https://doi.org/10.1103/PhysRevD.100.114508}{\emph{Phys. Rev. D}
  {\bfseries 100} (2019) 114508}
  [\href{https://arxiv.org/abs/1909.05842}{{\ttfamily 1909.05842}}].

\bibitem{DiPietro:2020jne}
L.~Di~Pietro and M.~Serone, \emph{{Looking through the QCD Conformal Window
  with Perturbation Theory}},
  \href{https://doi.org/10.1007/JHEP07(2020)049}{\emph{JHEP} {\bfseries 07}
  (2020) 049} [\href{https://arxiv.org/abs/2003.01742}{{\ttfamily
  2003.01742}}].

\bibitem{Kim:2020yvr}
B.~S. Kim, D.~K. Hong and J.-W. Lee, \emph{{Into the conformal window:
  Multirepresentation gauge theories}},
  \href{https://doi.org/10.1103/PhysRevD.101.056008}{\emph{Phys. Rev. D}
  {\bfseries 101} (2020) 056008}
  [\href{https://arxiv.org/abs/2001.02690}{{\ttfamily 2001.02690}}].

\bibitem{Bond:2021tgu}
A.~D. Bond, D.~F. Litim and G.~M. Vazquez, \emph{{Conformal windows beyond
  asymptotic freedom}},
  \href{https://doi.org/10.1103/PhysRevD.104.105002}{\emph{Phys. Rev. D}
  {\bfseries 104} (2021) 105002}
  [\href{https://arxiv.org/abs/2107.13020}{{\ttfamily 2107.13020}}].

\bibitem{DelDebbio:2021xwu}
L.~Del~Debbio and R.~Zwicky, \emph{{Dilaton and massive hadrons in a conformal
  phase}}, \href{https://doi.org/10.1007/JHEP08(2022)007}{\emph{JHEP}
  {\bfseries 08} (2022) 007}
  [\href{https://arxiv.org/abs/2112.11363}{{\ttfamily 2112.11363}}].

\bibitem{Aguilar:2002tc}
A.~C. Aguilar, A.~A. Natale and P.~S. Rodrigues~da Silva, \emph{{Relating a
  Gluon Mass Scale to an Infrared Fixed Point in Pure Gauge QCD}},
  \href{https://doi.org/10.1103/PhysRevLett.90.152001}{\emph{Phys. Rev. Lett.}
  {\bfseries 90} (2003) 152001}
  [\href{https://arxiv.org/abs/hep-ph/0212105}{{\ttfamily hep-ph/0212105}}].

\bibitem{Gies:2002af}
H.~Gies, \emph{{Running coupling in Yang-Mills theory: A flow equation study}},
  \href{https://doi.org/10.1103/PhysRevD.66.025006}{\emph{Phys. Rev.}
  {\bfseries D66} (2002) 025006}
  [\href{https://arxiv.org/abs/hep-th/0202207}{{\ttfamily hep-th/0202207}}].

\bibitem{Pawlowski:2003hq}
J.~M. Pawlowski, D.~F. Litim, S.~Nedelko and L.~von Smekal, \emph{{Infrared
  behavior and fixed points in Landau gauge QCD}},
  \href{https://doi.org/10.1103/PhysRevLett.93.152002}{\emph{Phys.Rev.Lett.}
  {\bfseries 93} (2004) 152002}
  [\href{https://arxiv.org/abs/hep-th/0312324}{{\ttfamily hep-th/0312324}}].

\bibitem{PalanquesMestre:1983zy}
A.~Palanques-Mestre and P.~Pascual, \emph{{The 1/$N^-$f Expansion of the
  $\gamma$ and Beta Functions in {QED}}},
  \href{https://doi.org/10.1007/BF01212398}{\emph{Commun. Math. Phys.}
  {\bfseries 95} (1984) 277}.

\bibitem{Gracey:1996he}
J.~Gracey, \emph{{The QCD Beta Function at O(1/\hbox{$N_f$})}},
  \href{https://doi.org/10.1016/0370-2693(96)00105-0}{\emph{Phys. Lett. B}
  {\bfseries 373} (1996) 178}
  [\href{https://arxiv.org/abs/hep-ph/9602214}{{\ttfamily hep-ph/9602214}}].

\bibitem{Holdom:2010qs}
B.~Holdom, \emph{{Large $N$ Flavor Beta-Functions: a Recap}},
  \href{https://doi.org/10.1016/j.physletb.2010.09.037}{\emph{Phys. Lett. B}
  {\bfseries 694} (2010) 74} [\href{https://arxiv.org/abs/1006.2119}{{\ttfamily
  1006.2119}}].

\bibitem{Martin:2000cr}
S.~P. Martin and J.~D. Wells, \emph{{Constraints on Ultraviolet Stable Fixed
  Points in Supersymmetric Gauge Theories}},
  \href{https://doi.org/10.1103/PhysRevD.64.036010}{\emph{Phys. Rev.}
  {\bfseries D64} (2001) 036010}
  [\href{https://arxiv.org/abs/hep-ph/0011382}{{\ttfamily hep-ph/0011382}}].

\bibitem{Ryttov:2019aux}
T.~A. Ryttov and K.~Tuominen, \emph{{Safe Glueballs and Baryons}},
  \href{https://doi.org/10.1007/JHEP04(2019)173}{\emph{JHEP} {\bfseries 04}
  (2019) 173} [\href{https://arxiv.org/abs/1903.09089}{{\ttfamily
  1903.09089}}].

\bibitem{Alanne:2019vuk}
T.~Alanne, S.~Blasi and N.~A. Dondi, \emph{{Critical Look at $\beta$ -Function
  Singularities at Large $N$}},
  \href{https://doi.org/10.1103/PhysRevLett.123.131602}{\emph{Phys. Rev. Lett.}
  {\bfseries 123} (2019) 131602}
  [\href{https://arxiv.org/abs/1905.08709}{{\ttfamily 1905.08709}}].

\bibitem{Leino:2019qwk}
V.~Leino, T.~Rindlisbacher, K.~Rummukainen, F.~Sannino and K.~Tuominen,
  \emph{{Safety versus triviality on the lattice}},
  \href{https://doi.org/10.1103/PhysRevD.101.074508}{\emph{Phys. Rev. D}
  {\bfseries 101} (2020) 074508}
  [\href{https://arxiv.org/abs/1908.04605}{{\ttfamily 1908.04605}}].

\bibitem{Dondi:2019ivp}
N.~A. Dondi, G.~V. Dunne, M.~Reichert and F.~Sannino, \emph{{Analytic Coupling
  Structure of Large $N_f$ (Super) QED and QCD}},
  \href{https://doi.org/10.1103/PhysRevD.100.015013}{\emph{Phys. Rev. D}
  {\bfseries 100} (2019) 015013}
  [\href{https://arxiv.org/abs/1903.02568}{{\ttfamily 1903.02568}}].

\bibitem{Dondi:2020qfj}
N.~A. Dondi, G.~V. Dunne, M.~Reichert and F.~Sannino, \emph{{Towards the QED
  beta function and renormalons at $1/N_f^2$ and $1/N_f^3$}},
  \href{https://doi.org/10.1103/PhysRevD.102.035005}{\emph{Phys. Rev. D}
  {\bfseries 102} (2020) 035005}
  [\href{https://arxiv.org/abs/2003.08397}{{\ttfamily 2003.08397}}].

\bibitem{Corless:1996zz}
R.~M. Corless, G.~H. Gonnet, D.~E.~G. Hare, D.~J. Jeffrey and D.~E. Knuth,
  \emph{{On the LambertW function}},
  \href{https://doi.org/10.1007/BF02124750}{\emph{Adv. Comput. Math.}
  {\bfseries 5} (1996) 329}.

\bibitem{Gardi:1998qr}
E.~Gardi, G.~Grunberg and M.~Karliner, \emph{{Can the QCD running coupling have
  a causal analyticity structure?}},
  \href{https://doi.org/10.1088/1126-6708/1998/07/007}{\emph{JHEP} {\bfseries
  07} (1998) 007} [\href{https://arxiv.org/abs/hep-ph/9806462}{{\ttfamily
  hep-ph/9806462}}].

\bibitem{Ruijl:2017eht}
B.~Ruijl, T.~Ueda, J.~A.~M. Vermaseren and A.~Vogt, \emph{{Four-loop QCD
  propagators and vertices with one vanishing external momentum}},
  \href{https://doi.org/10.1007/JHEP06(2017)040}{\emph{JHEP} {\bfseries 06}
  (2017) 040} [\href{https://arxiv.org/abs/1703.08532}{{\ttfamily
  1703.08532}}].

\bibitem{Oehme:1979ai}
R.~Oehme and W.~Zimmermann, \emph{{Quark and Gluon Propagators in Quantum
  Chromodynamics}}, \href{https://doi.org/10.1103/PhysRevD.21.471}{\emph{Phys.
  Rev.} {\bfseries D21} (1980) 471}.

\bibitem{Oehme:1979bj}
R.~Oehme and W.~Zimmermann, \emph{{Gauge Field Propagator and the Number of
  Fermion Fields}}, \href{https://doi.org/10.1103/PhysRevD.21.1661}{\emph{Phys.
  Rev. D} {\bfseries 21} (1980) 1661}.

\bibitem{Oehme:1990kd}
R.~Oehme, \emph{{On superconvergence relations in quantum chromodynamics}},
  \href{https://doi.org/10.1016/0370-2693(90)90499-V}{\emph{Phys. Lett. B}
  {\bfseries 252} (1990) 641}.

\bibitem{Cyrol:2018xeq}
A.~K. Cyrol, J.~M. Pawlowski, A.~Rothkopf and N.~Wink, \emph{{Reconstructing
  the gluon}},
  \href{https://doi.org/10.21468/SciPostPhys.5.6.065}{\emph{SciPost Phys.}
  {\bfseries 5} (2018) 065} [\href{https://arxiv.org/abs/1804.00945}{{\ttfamily
  1804.00945}}].

\bibitem{Dudal:2019pyg}
D.~Dudal, D.~M. van Egmond, M.~S. Guimaraes, O.~Holanda, L.~F. Palhares,
  G.~Peruzzo et~al., \emph{{Gauge-invariant spectral description of the $U(1)$
  Higgs model from local composite operators}},
  \href{https://doi.org/10.1007/JHEP02(2020)188}{\emph{JHEP} {\bfseries 02}
  (2020) 188} [\href{https://arxiv.org/abs/1912.11390}{{\ttfamily
  1912.11390}}].

\bibitem{Li:2019hyv}
S.~W. Li, P.~Lowdon, O.~Oliveira and P.~J. Silva, \emph{{The generalised
  infrared structure of the gluon propagator}},
  \href{https://doi.org/10.1016/j.physletb.2020.135329}{\emph{Phys. Lett. B}
  {\bfseries 803} (2020) 135329}
  [\href{https://arxiv.org/abs/1907.10073}{{\ttfamily 1907.10073}}].

\bibitem{Dudal:2020uwb}
D.~Dudal, D.~M. van Egmond, M.~S. Guimaraes, L.~F. Palhares, G.~Peruzzo and
  S.~P. Sorella, \emph{{Spectral properties of local gauge invariant composite
  operators in the $SU(2)$ Yang\textendash{}Mills\textendash{}Higgs model}},
  \href{https://doi.org/10.1140/epjc/s10052-021-09008-9}{\emph{Eur. Phys. J. C}
  {\bfseries 81} (2021) 222}
  [\href{https://arxiv.org/abs/2008.07813}{{\ttfamily 2008.07813}}].

\bibitem{Bonanno:2021squ}
A.~Bonanno, T.~Denz, J.~M. Pawlowski and M.~Reichert, \emph{{Reconstructing the
  graviton}},
  \href{https://doi.org/10.21468/SciPostPhys.12.1.001}{\emph{SciPost Phys.}
  {\bfseries 12} (2022) 001}
  [\href{https://arxiv.org/abs/2102.02217}{{\ttfamily 2102.02217}}].

\bibitem{Maas:2020kda}
A.~Maas and R.~Sondenheimer, \emph{{Gauge-invariant description of the Higgs
  resonance and its phenomenological implications}},
  \href{https://doi.org/10.1103/PhysRevD.102.113001}{\emph{Phys. Rev. D}
  {\bfseries 102} (2020) 113001}
  [\href{https://arxiv.org/abs/2009.06671}{{\ttfamily 2009.06671}}].

\bibitem{Binosi:2019ecz}
D.~Binosi and R.-A. Tripolt, \emph{{Spectral functions of confined particles}},
  \href{https://doi.org/10.1016/j.physletb.2019.135171}{\emph{Phys. Lett. B}
  {\bfseries 801} (2020) 135171}
  [\href{https://arxiv.org/abs/1904.08172}{{\ttfamily 1904.08172}}].

\bibitem{Dudal:2019aew}
D.~Dudal, D.~M. van Egmond, M.~S. Guimar\~aes, O.~Holanda, B.~W. Mintz, L.~F.
  Palhares et~al., \emph{{Some remarks on the spectral functions of the Abelian
  Higgs Model}}, \href{https://doi.org/10.1103/PhysRevD.100.065009}{\emph{Phys.
  Rev. D} {\bfseries 100} (2019) 065009}
  [\href{https://arxiv.org/abs/1905.10422}{{\ttfamily 1905.10422}}].

\bibitem{Hayashi:2021nnj}
Y.~Hayashi and K.-I. Kondo, \emph{{Reconstructing confined particles with
  complex singularities}},
  \href{https://doi.org/10.1103/PhysRevD.103.L111504}{\emph{Phys. Rev. D}
  {\bfseries 103} (2021) L111504}
  [\href{https://arxiv.org/abs/2103.14322}{{\ttfamily 2103.14322}}].

\bibitem{Hayashi:2021jju}
Y.~Hayashi and K.-I. Kondo, \emph{{Reconstructing propagators of confined
  particles in the presence of complex singularities}},
  \href{https://doi.org/10.1103/PhysRevD.104.074024}{\emph{Phys. Rev. D}
  {\bfseries 104} (2021) 074024}
  [\href{https://arxiv.org/abs/2105.07487}{{\ttfamily 2105.07487}}].

\bibitem{Horak:2020eng}
J.~Horak, J.~M. Pawlowski and N.~Wink, \emph{{Spectral functions in the
  $\phi^4$-theory from the spectral DSE}},
  \href{https://doi.org/10.1103/PhysRevD.102.125016}{\emph{Phys. Rev. D}
  {\bfseries 102} (2020) 125016}
  [\href{https://arxiv.org/abs/2006.09778}{{\ttfamily 2006.09778}}].

\bibitem{Roth:2021nrd}
J.~V. Roth, D.~Schweitzer, L.~J. Sieke and L.~von Smekal, \emph{{Real-time
  methods for spectral functions}},
  \href{https://doi.org/10.1103/PhysRevD.105.116017}{\emph{Phys. Rev. D}
  {\bfseries 105} (2022) 116017}
  [\href{https://arxiv.org/abs/2112.12568}{{\ttfamily 2112.12568}}].

\bibitem{Horak:2021pfr}
J.~Horak, J.~Papavassiliou, J.~M. Pawlowski and N.~Wink, \emph{{Ghost spectral
  function from the spectral Dyson-Schwinger equation}},
  \href{https://doi.org/10.1103/PhysRevD.104.074017}{\emph{Phys. Rev. D}
  {\bfseries 104} (2021) } [\href{https://arxiv.org/abs/2103.16175}{{\ttfamily
  2103.16175}}].

\bibitem{Horak:2022myj}
J.~Horak, J.~M. Pawlowski and N.~Wink, \emph{{On the Complex Structure of
  Yang-Mills Theory}},  \href{https://arxiv.org/abs/2202.09333}{{\ttfamily
  2202.09333}}.

\bibitem{Braun:2022mgx}
J.~Braun et~al., \emph{{Renormalised spectral flows}},
  \href{https://arxiv.org/abs/2206.10232}{{\ttfamily 2206.10232}}.

\bibitem{Platania:2022gtt}
A.~Platania, \emph{{Causality, unitarity and stability in quantum gravity: a
  non-perturbative perspective}},
  \href{https://arxiv.org/abs/2206.04072}{{\ttfamily 2206.04072}}.

\bibitem{Ilgenfritz:2017kkp}
E.-M. Ilgenfritz, J.~M. Pawlowski, A.~Rothkopf and A.~Trunin, \emph{{Finite
  temperature gluon spectral functions from $N_f=2+1+1$ lattice QCD}},
  \href{https://doi.org/10.1140/epjc/s10052-018-5593-7}{\emph{Eur. Phys. J. C}
  {\bfseries 78} (2018) 127}
  [\href{https://arxiv.org/abs/1701.08610}{{\ttfamily 1701.08610}}].

\bibitem{Fischer:2017kbq}
C.~S. Fischer, J.~M. Pawlowski, A.~Rothkopf and C.~A. Welzbacher,
  \emph{{Bayesian analysis of quark spectral properties from the
  Dyson-Schwinger equation}},
  \href{https://doi.org/10.1103/PhysRevD.98.014009}{\emph{Phys. Rev. D}
  {\bfseries 98} (2018) 014009}
  [\href{https://arxiv.org/abs/1705.03207}{{\ttfamily 1705.03207}}].

\bibitem{Horak:2021syv}
J.~Horak, J.~M. Pawlowski, J.~Rodr\'\i{}guez-Quintero, J.~Turnwald, J.~M.
  Urban, N.~Wink et~al., \emph{{Reconstructing QCD spectral functions with
  Gaussian processes}},
  \href{https://doi.org/10.1103/PhysRevD.105.036014}{\emph{Phys. Rev. D}
  {\bfseries 105} (2022) 036014}
  [\href{https://arxiv.org/abs/2107.13464}{{\ttfamily 2107.13464}}].

\bibitem{Capri:2016aqq}
M.~A.~L. Capri, D.~Dudal, D.~Fiorentini, M.~S. Guimaraes, I.~F. Justo, A.~D.
  Pereira et~al., \emph{{Local and BRST-invariant Yang-Mills theory within the
  Gribov horizon}},
  \href{https://doi.org/10.1103/PhysRevD.94.025035}{\emph{Phys. Rev. D}
  {\bfseries 94} (2016) 025035}
  [\href{https://arxiv.org/abs/1605.02610}{{\ttfamily 1605.02610}}].

\bibitem{Capri:2016gut}
M.~A.~L. Capri, D.~Dudal, A.~D. Pereira, D.~Fiorentini, M.~S. Guimaraes, B.~W.
  Mintz et~al., \emph{{Nonperturbative aspects of Euclidean Yang-Mills theories
  in linear covariant gauges: Nielsen identities and a BRST-invariant two-point
  correlation function}},
  \href{https://doi.org/10.1103/PhysRevD.95.045011}{\emph{Phys. Rev. D}
  {\bfseries 95} (2017) 045011}
  [\href{https://arxiv.org/abs/1611.10077}{{\ttfamily 1611.10077}}].

\bibitem{Capri:2017abz}
M.~A.~L. Capri, D.~Fiorentini, A.~D. Pereira and S.~P. Sorella, \emph{{A
  non-perturbative study of matter field propagators in Euclidean
  Yang\textendash{}Mills theory in linear covariant, Curci\textendash{}Ferrari
  and maximal Abelian gauges}},
  \href{https://doi.org/10.1140/epjc/s10052-017-5107-z}{\emph{Eur. Phys. J. C}
  {\bfseries 77} (2017) 546}
  [\href{https://arxiv.org/abs/1703.03264}{{\ttfamily 1703.03264}}].

\bibitem{Wetterich:1992yh}
C.~Wetterich, \emph{{Exact evolution equation for the effective potential}},
  \href{https://doi.org/10.1016/0370-2693(93)90726-X}{\emph{Phys. Lett.}
  {\bfseries B301} (1993) 90}
  [\href{https://arxiv.org/abs/1710.05815}{{\ttfamily 1710.05815}}].

\bibitem{Ellwanger:1993mw}
U.~Ellwanger, \emph{{Flow equations for N point functions and bound states}},
  \href{https://doi.org/10.1007/BF01555911}{\emph{Z. Phys.} {\bfseries C62}
  (1994) 503} [\href{https://arxiv.org/abs/hep-ph/9308260}{{\ttfamily
  hep-ph/9308260}}].

\bibitem{Morris:1993qb}
T.~R. Morris, \emph{{The Exact renormalization group and approximate
  solutions}}, \href{https://doi.org/10.1142/S0217751X94000972}{\emph{Int. J.
  Mod. Phys.} {\bfseries A9} (1994) 2411}
  [\href{https://arxiv.org/abs/hep-ph/9308265}{{\ttfamily hep-ph/9308265}}].

\bibitem{Litim:2000ci}
D.~F. Litim, \emph{{Optimization of the exact renormalization group}},
  \href{https://doi.org/10.1016/S0370-2693(00)00748-6}{\emph{Phys.Lett.}
  {\bfseries B486} (2000) 92}
  [\href{https://arxiv.org/abs/hep-th/0005245}{{\ttfamily hep-th/0005245}}].

\bibitem{Litim:2001up}
D.~F. Litim, \emph{{Optimized renormalization group flows}},
  \href{https://doi.org/10.1103/PhysRevD.64.105007}{\emph{Phys.Rev.} {\bfseries
  D64} (2001) 105007} [\href{https://arxiv.org/abs/hep-th/0103195}{{\ttfamily
  hep-th/0103195}}].

\bibitem{Wilson:1971dc}
K.~G. Wilson and M.~E. Fisher, \emph{{Critical exponents in 3.99 dimensions}},
  \href{https://doi.org/10.1103/PhysRevLett.28.240}{\emph{Phys. Rev. Lett.}
  {\bfseries 28} (1972) 240}.

\bibitem{Lovelace:1982hz}
C.~Lovelace, \emph{{Universality at Large N}},
  \href{https://doi.org/10.1016/0550-3213(82)90435-7}{\emph{Nucl. Phys. B}
  {\bfseries 201} (1982) 333}.

\bibitem{Benini:2019dfy}
F.~Benini, C.~Iossa and M.~Serone, \emph{{Conformality Loss, Walking, and 4D
  Complex Conformal Field Theories at Weak Coupling}},
  \href{https://doi.org/10.1103/PhysRevLett.124.051602}{\emph{Phys. Rev. Lett.}
  {\bfseries 124} (2020) 051602}
  [\href{https://arxiv.org/abs/1908.04325}{{\ttfamily 1908.04325}}].

\end{thebibliography}\endgroup

\end{document}